\documentclass[journal]{IEEEtran}
\usepackage{times, verbatim}
\usepackage[cmex10]{amsmath}
\usepackage{amssymb,amsfonts}

\usepackage{graphicx}
\graphicspath{{./figures/}}
\usepackage{epsfig}
\usepackage{epstopdf}

\usepackage{color}
\usepackage{float}
\usepackage{algorithm}
\usepackage{algorithmic}

\usepackage{enumitem}
\usepackage{tablefootnote}
\usepackage{multirow}
\usepackage{array}

\begin{document}
% single column
% \onecolumn
\title{Optimized Data Representation for Interactive Multiview Navigation}

\author{Rui~Ma,~\IEEEmembership{Student Member,~IEEE,}
        Thomas~Maugey,~\IEEEmembership{Member,~IEEE,}
        and~Pascal~Frossard,~\IEEEmembership{Senior~Member,~IEEE}% <-this % stops a space
\thanks{R. Ma is with the Department of Electrical and Computer Engineering, The Hong Kong University of Science and Technology, Hong Kong (e-mail: rmaaa@connect.ust.hk).}% <-this % stops a space
\thanks{T. Maugey is with Inria Rennes-Bretagne Atlantique Research Centre, Rennes Cedex 35042, France (e-mail: thomas.maugey@inria.fr).}% <-this % stops a space
\thanks{P. Frossard is with Signal Processing Laboratory LTS4, \'{E}cole Polytechnique F\'{e}d\'{e}rale de Lausanne CH-1015, Switzerland (e-mail: pascal.frossard@epfl.ch).}
%\thanks{Manuscript received XX xx, 2016; revised XX xx, 2016.}
}

\maketitle

% page numbering
\thispagestyle{plain}
\pagestyle{plain}

\begin{abstract}
In contrary to traditional media streaming services where a unique media content is delivered to different users, interactive multiview navigation applications enable users to choose their own viewpoints and freely navigate in a 3-D scene. 
The interactivity brings new challenges in addition to the classical rate-distortion trade-off, which considers only the compression performance and viewing quality.
On the one hand, interactivity necessitates sufficient viewpoints for richer navigation; on the other hand, it requires to provide low bandwidth and delay costs for smooth navigation during view transitions.
In this paper, we formally describe the novel trade-offs posed by the navigation interactivity and classical rate-distortion criterion. 
Based on an original formulation, we look for the optimal design of the data representation by introducing novel rate and distortion models and practical solving algorithms. 
Experiments show that the proposed data representation method outperforms the baseline solution by providing lower resource consumptions and higher visual quality in all navigation configurations, which certainly confirms the potential of the proposed data representation in practical interactive navigation systems. 
\end{abstract}

\begin{IEEEkeywords}
Multiview navigation, interactivity, navigation segment, multiview image compression
\end{IEEEkeywords}

% \fi
\IEEEpeerreviewmaketitle

\section{Introduction}
\label{sec:intro}
\begin{figure}[tb]
 \centering
\begin{minipage}[b]{.95\linewidth}
  \centering
  \centerline{\includegraphics[width=\linewidth]{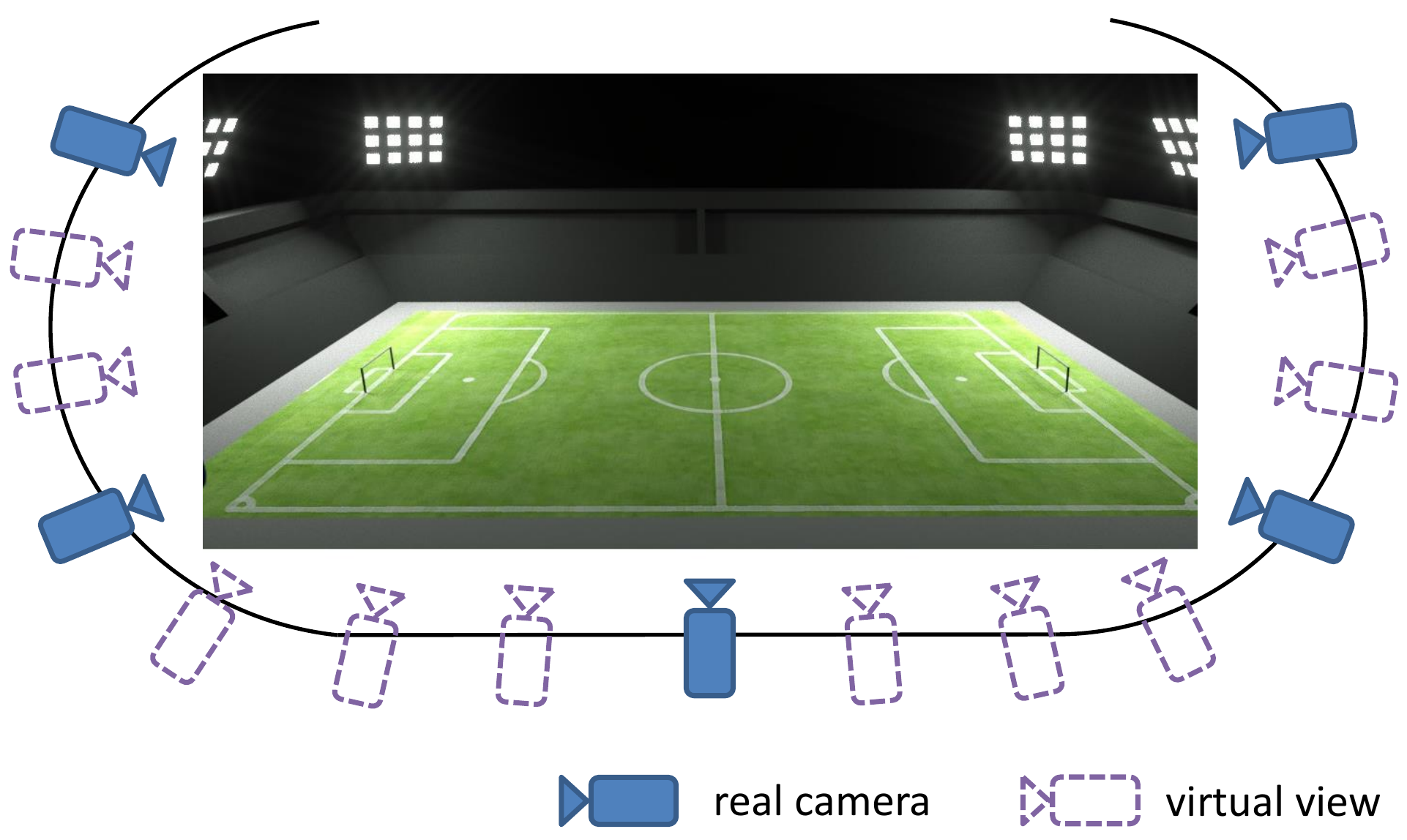}}
 % \vspace{1.5cm}
 % \centerline{(b) }%\medskip
\end{minipage}
\caption{ A typical multiview navigation scenario. Users are able to freely navigate along virtual views that are synthesized from views captured at different camera positions. }
\label{fig:ftv}
%\vspace{10pt}
\end{figure}

\IEEEPARstart{W}{ith} the development of multiview imaging techniques, there has been a lot of interest in interactive multiview navigation \cite{spm2007multiview, tanimoto2006overview}.
Differently from traditional media streaming systems where a unique media content is streamed to all users, interactive multiview navigation systems provide users with different media data depending on their interactions with the server.
In particular, each user watches a specific 2-D image corresponding to his own choice of viewing position and orientation (called a \emph{viewpoint}) and is able to navigate in the scene by freely changing this viewpoint (see Fig. \ref{fig:ftv}).
These virtual views are synthesized from the content of different cameras positioned in the 3-D scene.

\begin{figure}[tb]
 \centering
\begin{minipage}[b]{.98\linewidth}
  \centering
  \centerline{\includegraphics[width=\linewidth]{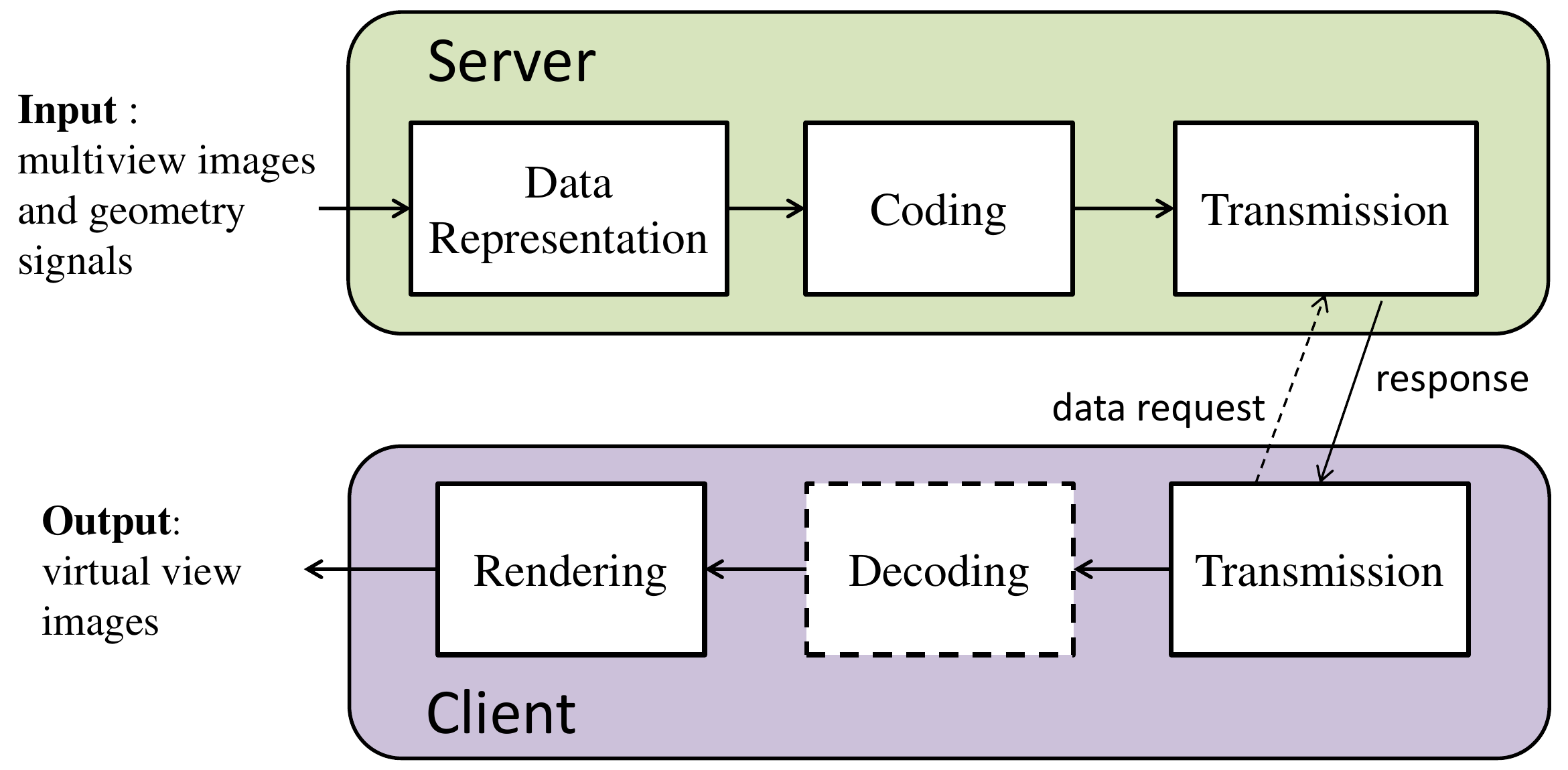}}
 % \vspace{1.5cm}
 % \centerline{(b) }%\medskip
\end{minipage}
\caption{ System architecture for interactive multiview navigation. }
\label{fig:nv_sys}
%\vspace{10pt}
\end{figure}

In order to achieve interactive navigation, it is necessary to consider a complete processing chain consisting of different connected components, including data representation, coding, transmission and view rendering (see Fig. \ref{fig:nv_sys}).
Indeed, the consideration of every component in isolation can only lead to suboptimal performance.
In the literature, the individual components of the navigation system have been extensively studied, e.g., 3-D scene representation \cite{muller2011_3D_depth_repr, tanimoto2011free}, multiview video compression \cite{chen2009emerging, muller2013hevc3d}, multiview data streaming \cite{de_abreu2015layered_repr, toni2016packet_scheduling} and view synthesis \cite{xiu2011optimal, xiu2012delay}. However, there is clearly a lack of fully integrated frameworks that incorporate these techniques in an end-to-end system and jointly optimize them.

The end-to-end system optimization involves complex design trade-offs.
As \emph{interactivity} denotes the users' flexibility to choose arbitrary viewpoints during navigation, it first requires a sufficiently large navigation range, i.e., a large set of achievable viewpoints. 
Second, as users’ are willing to watch only a subpart of this navigation range, the system must transmit \emph{only what is useful} due to bandwidth limitations in practice.
This original trade-off between bandwidth limitations, visual quality and navigation flexibility has not been solved in the literature. Indeed, the good compression performance of traditional multiview schemes \cite{chen2009emerging, muller2013hevc3d} is obtained at a price of possibly long coding prediction paths, which however prevent independent view decoding (an analogue problem is posed by random access in monoview video \cite{sullivan2012overview}). A careful redesign of the whole system is thus needed, starting for example from the representation of the multiview data itself. In that spirit, the work in \cite{maugey2013navi_dom_repr} has proposed to organize the achievable viewpoints in independently decodable partitions, namely the \emph{navigation segments}. Indeed, the navigation segment can be regarded as the spatial analogue to the temporal GOP (group of pictures) in monoview video transmissions. This approach has however left some important questions opened, such as the optimal design of the navigation segments.

\begin{figure}[tb]
 \centering
\begin{minipage}[htb]{.5\linewidth}
  \centering
  \centerline{\includegraphics[width=\linewidth]{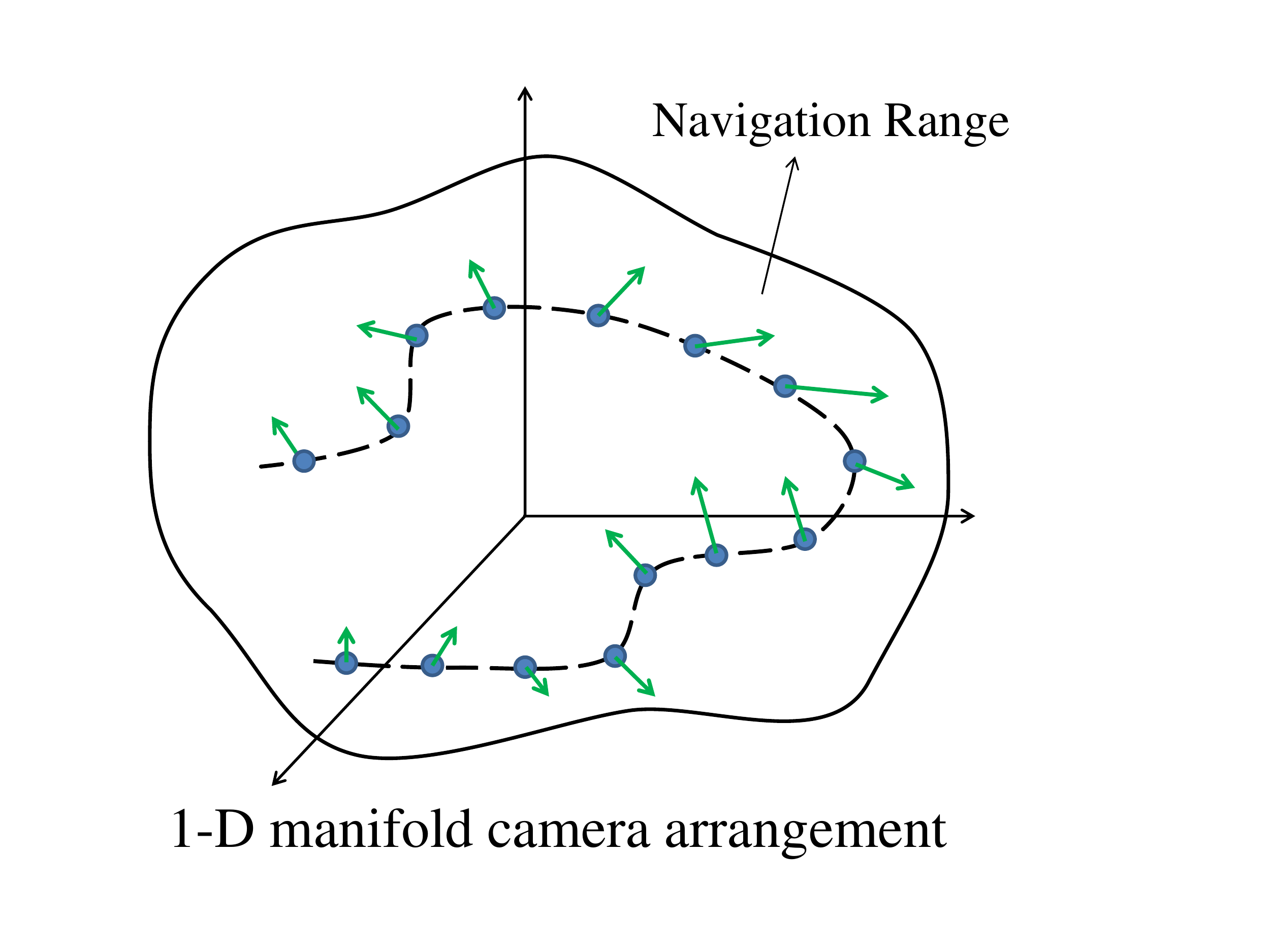}}
 % \vspace{1.5cm}
  \centerline{(a) }%\medskip
\end{minipage}
\hspace{10pt}
\begin{minipage}[htb]{.36\linewidth}
  \centering
  \vspace{.7cm}
  \centerline{\includegraphics[width=\linewidth]{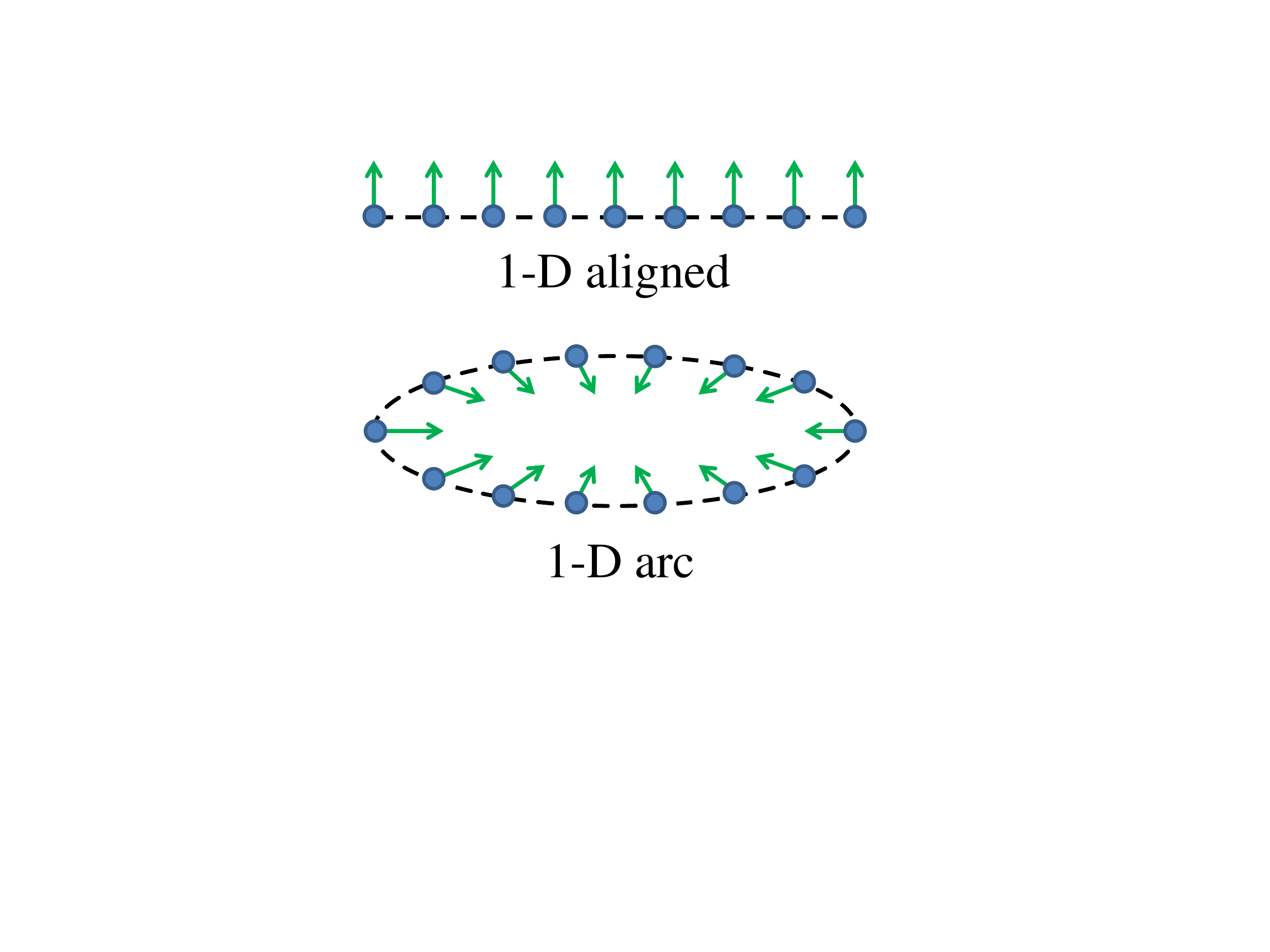}}
  \vspace{.45cm}
  \centerline{(b) }%\medskip
\end{minipage}
\caption{ An illustration of camera arrangements: (a) 1-D manifold camera arrangement (b) typical camera arrangements. The solid circle and arrow denote the camera position and orientation respectively, while the dash line is the camera trajectory. }
\label{fig:nv_manifold_camera}
\end{figure}

In this paper, we formally describe the novel trade-offs posed by interactive schemes between bandwidth limitations, visual quality and navigation interactivity, in the context of the \emph{navigation segment representation}. Based on this original formulation, we propose to study \emph{the optimal design of the navigation segments}. 
We further take into account the delay between user requests and actual data receiving, by introducing the concept of \emph{navigation ball}. 
We conduct our study in the challenging scenario of wide navigation range like the 1-D manifold camera arrangement depicted in Fig. \ref{fig:nv_manifold_camera}.
Experiments on the New Tsukuba Dataset \cite{peris2012stereo, martull2012realistic} show that the proposed navigation segment representation outperforms the baseline equidistant solution (where navigation segments are equally divided). Our approach offers lower resource consumptions and higher visual quality in all different navigation configurations, due to its high adaptability to various navigation parameters, like the navigation speed and the view popularity. 

The rest of the paper is organized as follows.
Section \ref{sec:relatedwork} introduces the related work.
Section \ref{sec:navisys} describes the proposed navigation segment representation under the navigation environment.
Section \ref{sec:naviopt} proposes the optimization framework for the navigation system, and Section \ref{sec:navimodel} further elaborates on the problem formulation with novel rate and distortion models.
Section \ref{sec:navisol} investigates practical solutions and analyses their complexity.
Experimental results are demonstrated in Section \ref{sec:expr}, and Section \ref{sec:concl} draws the conclusions.

\section{Related Work}
\label{sec:relatedwork}
The problem of interactive multiview navigation \cite{spm2007multiview} has recently gained interest in the research community.
A first category of works provides navigation interactivity by switching views between a predefined set of real camera viewpoints.
The H.264 SP/SI-frames \cite{karczewicz2003sp}, for example, is able to increase the interactivity between view switchings by avoiding transmission of previous frames in the new view. A SP-frame can be inserted at view switching point, which is able to be identically decoded from a cross-view reference instead of a reference in the same view \cite{chen2009emerging}.
The distributed source coding (DSC) can also be utilized for interactive streaming \cite{cheung2009distributed, petrazzuoli2011using}, since a DSC frame can be identically reconstructed from different predictors.
Another way to increase the interactivity is to produce multiple decoded versions of the media subset.
In \cite{cheung2009generation, cheung2009bandwidth}, redundant P-frames are used to support multiple decoding process. In \cite{liu2010rd}, the multiple encoding versions are stored in the server for diverse user requests. 
These methods, however, require large server storage.
The interactive navigation also needs to consider the user behaviors.
In \cite{kurutepe2007client, tekalp2007overip}, for example, the prediction structure is adapted to the user position estimated by Kalman filtering.
Although the above methods can increase the interactivity, it is limited to actual camera viewpoints, resulting in abrupt and unnatural view switchings. Also, they do not consider the viewing delays incurred in data transmission and processing.

Some approaches propose to extend the navigation interactivity beyond the camera viewpoints by utilizing the virtual view synthesis techniques \cite{shum2003survey, na2008multi}.
In \cite{xiu2011optimal, xiu2012delay}, for example the users can access to any virtual views that are rendered using the two nearest coded camera views.
The rendering can also be performed on the server side.
In \cite{maugey2011interactive}, the virtual views are encoded using predictive coding and stored in the server, before being streamed directly to the users. However the storage burden is largely increased because every accessible virtual view must be stored.

In order to support the high-quality rendering of virtual views, appropriate data representations are therefore extensively studied.
In \cite{tanimoto2006overview, tanimoto2011free}, the light-field representation \cite{levoy1996light} is adopted for view synthesis due to its efficient and high-quality rendering. However the dense representation of light-field is heavily redundant and poses additional challenges in data compression and transmission.
Some other data representation methods are considered to remove the data redundancy in the representation stage. 
In \cite{maugey2013navi_dom_repr, maugey2012consistent}, for example, the scene is represented using only one texture and one depth map, plus some auxiliary information that helps the view synthesis. However the choice of the appropriate auxiliary information is still an open question.
In \cite{takyar2014extended_layered_depth}, the layered depth image format is used for data representation, where multiple images are constructed in layers corresponding to different levels of occlusion.
Although data redundancy can be mostly removed from these representations, additional efforts are required to convert the captured data into the specific representation formats.

The increasing interactivity also brings challenges in data transmission, where the transmission policy needs to react to different requests of multiple users. 
In \cite{de_abreu2015layered_repr, toni2016packet_scheduling}, for example, the streaming of multiview video content in a navigation environment is studied, where the optimal transmission strategies are designed to provide high-quality content to heterogenous users under limited resources.
All the above examples indicate that designing an interactive navigation system relates to many issues, including data representation, compression, transmission and rendering. 
While most existing works focus on a particular part of the system, only a few approaches investigate an end-to-end system design in the literature. 
In \cite{bauermann2008rdtc1, bauermann2008rdtc2}, the classical rate-distortion optimization is extended to the interactive streaming scenario by considering the transmission rate and decoding complexity. 
However it mainly focuses on the coding aspect and does not consider the data representation at first. 
The design of effective solutions for multiview data representation and coding in the interactive navigation scenario is still an open problem.

Compared with the previous approaches, our work has the following contributions.
First, we consider an end-to-end interactive navigation system design from data representation to rendering and propose to jointly optimize the novel trade-offs between navigation interactivity, bandwidth limitation and visual quality.
Second, we show that the proper data representation plays an important role in optimizing the navigation system and we investigate practical solutions to find effective data representation strategies adapted to various navigation parameters (e.g., navigation speed, view popularity). 
Third, since the viewing delay caused by data transmission and processing is discussed in many works but often not properly handled, we propose a novel mechanism called navigation ball to prohibit the viewing delays and enable smooth user navigations.
Fourth, we consider a rich 1-D manifold camera arrangement with high degrees of freedom in camera translation and rotation for user navigation, which extends the classical camera array arrangements.

\section{Navigation Segment Representation for Low-Delay Navigations}
\label{sec:navisys}
In this section, we describe the proposed interactive multiview navigation system step by step. Based on that, we present our navigation segment representation. 

\begin{figure}[tb]
 \centering
\begin{minipage}[b]{.85\linewidth}
  \centering
  \centerline{ \quad \includegraphics[width=\linewidth]{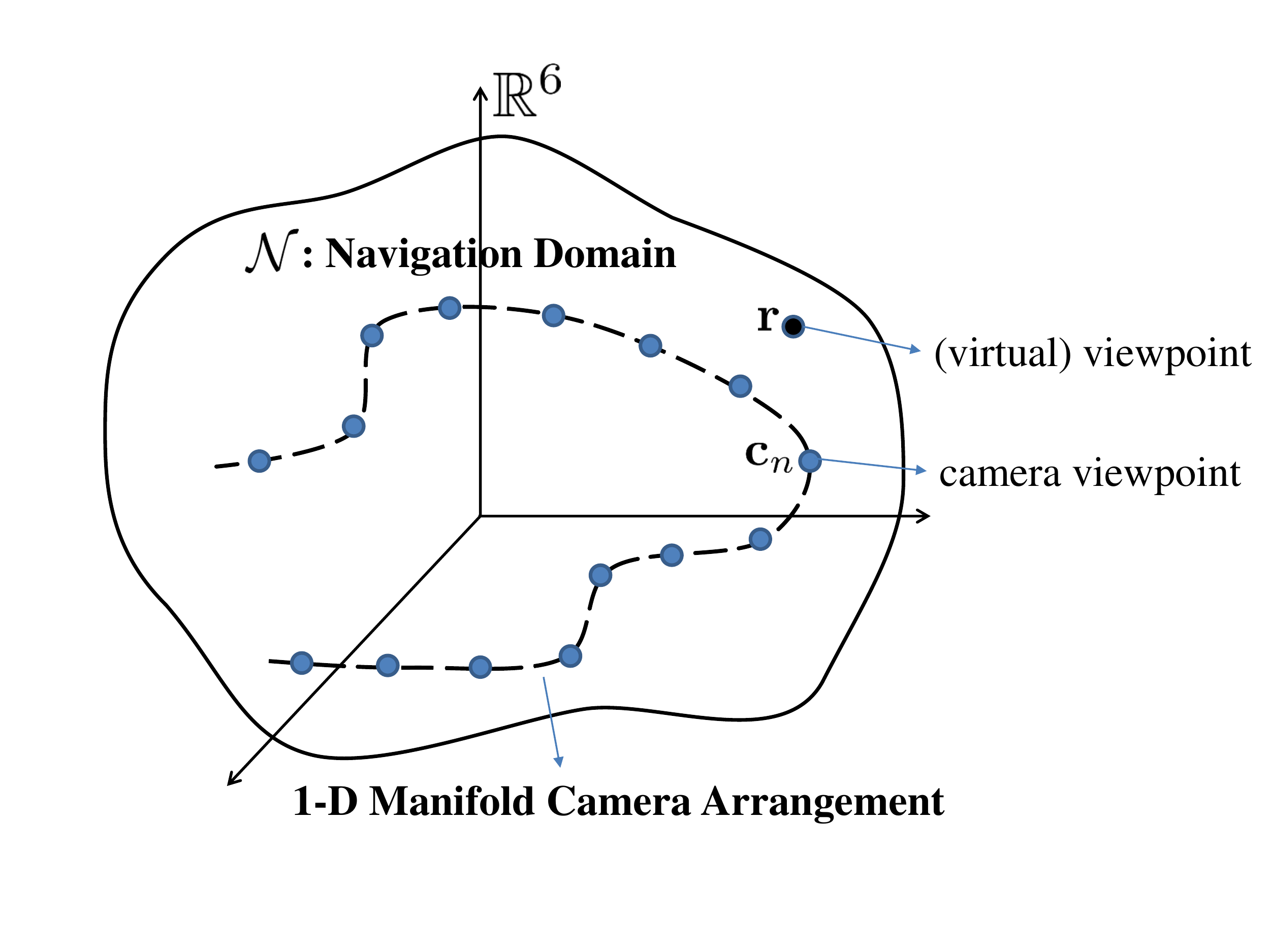}}
  \vspace{-5pt}
 % \centerline{(b) }%\medskip
\end{minipage}
\caption{ Navigation domain: A user is able to navigate in the whole navigation domain, based on the data captured by camera viewpoints that lie on a 1-D manifold. }
\label{fig:nv_domain}
%\vspace{-10pt}
\end{figure}

\subsection{1-D Manifold Camera Arrangement}
We are interested in a navigation scenario in a \emph{static} 3-D scene, which is captured by a set of cameras positioned in different locations and orientations. 
A \emph{camera viewpoint} in 3-D scene can be represented as a 6-D vector $\mathbf{c} = [x,y,z,\theta, \phi, \psi]^T $, where $ [x,y,z] $ denotes the position and $ [\theta, \phi, \psi] $ denotes the orientation.
In our work, we study the challenging camera arrangement depicted in Fig. \ref{fig:tsukuba_camera_arrange}, where all camera viewpoints lie in a 1-D manifold embedded in the 6-D space $ \mathbb{R}^6 $.
This camera arrangement greatly extends the navigation interactivity in terms of navigation range, where multiple degrees of freedom for camera motion can be replicated, including translation and rotation. 
For simplicity, we index each camera viewpoint along the 1-D manifold and denote it as $ \mathbf{c}_n $, where $ n \in [1, N_V]$ and $N_V$ is the number of cameras.
We assume that all the cameras provide both images and depth maps of the 3-D scene. We use $ Y_n $ to represent the image and the depth map captured at camera viewpoint $ \mathbf{c}_n $.

\subsection{Navigation Domain and Navigation Path}
\label{sec:navi-desc}
Similarly to the camera viewpoints, a (virtual) \emph{viewpoint} in the 3-D scene can also be represented as a 6-D vector $\mathbf{r} \in \mathbb{R}^6$. 
The virtual views are rendered using a depth-image-based rendering (DIBR) technique \cite{fehn2004dibr} with data from the closest camera views.
In a navigation scenario, the set of all accessible viewpoints within the navigation range forms the \emph{navigation domain}, and it is denoted as $\mathcal{N} \subset \mathbb{R}^6 $.
Fig. \ref{fig:nv_domain} shows the navigation domain, which can be much larger than the camera set.

The user's navigation process is associated with a path traveling through all viewpoints visited by this user in $ \mathcal{N} $.
We call this path a \emph{navigation path}.
In practice, the navigation path is discrete and finite, due to finite frame rate $ f $ and bounded navigation period $ T $.
Then the total number of visited viewpoints in one navigation path is $ N_f = T f $.
We define the navigation path $ P $ as the set that sequentially contains all visited viewpoints within the navigation period, i.e., $P = \{ \mathbf{r}_1, \mathbf{r}_2, \cdots, \mathbf{r}_{N_f} \}$, with $ \mathbf{r}_i $ the $ i $-th viewpoint in $ P $.

\subsection{Navigation Ball}
When a user navigates along a path $ P $, he will repeatedly request data in order to render views at each $ \mathbf{r}_i $.
However, the data response time will lag behind the request time due to the system delay, which includes the transmission delay and other data processing delays (e.g., decoding, rendering).
Therefore, more data than the one required by the current viewpoint needs to be transmitted in order to compensate for system delays.

\begin{figure}[tb]
 \centering
\begin{minipage}[htb]{.95\linewidth}
  \centering
  \centerline{\includegraphics[width=\linewidth]{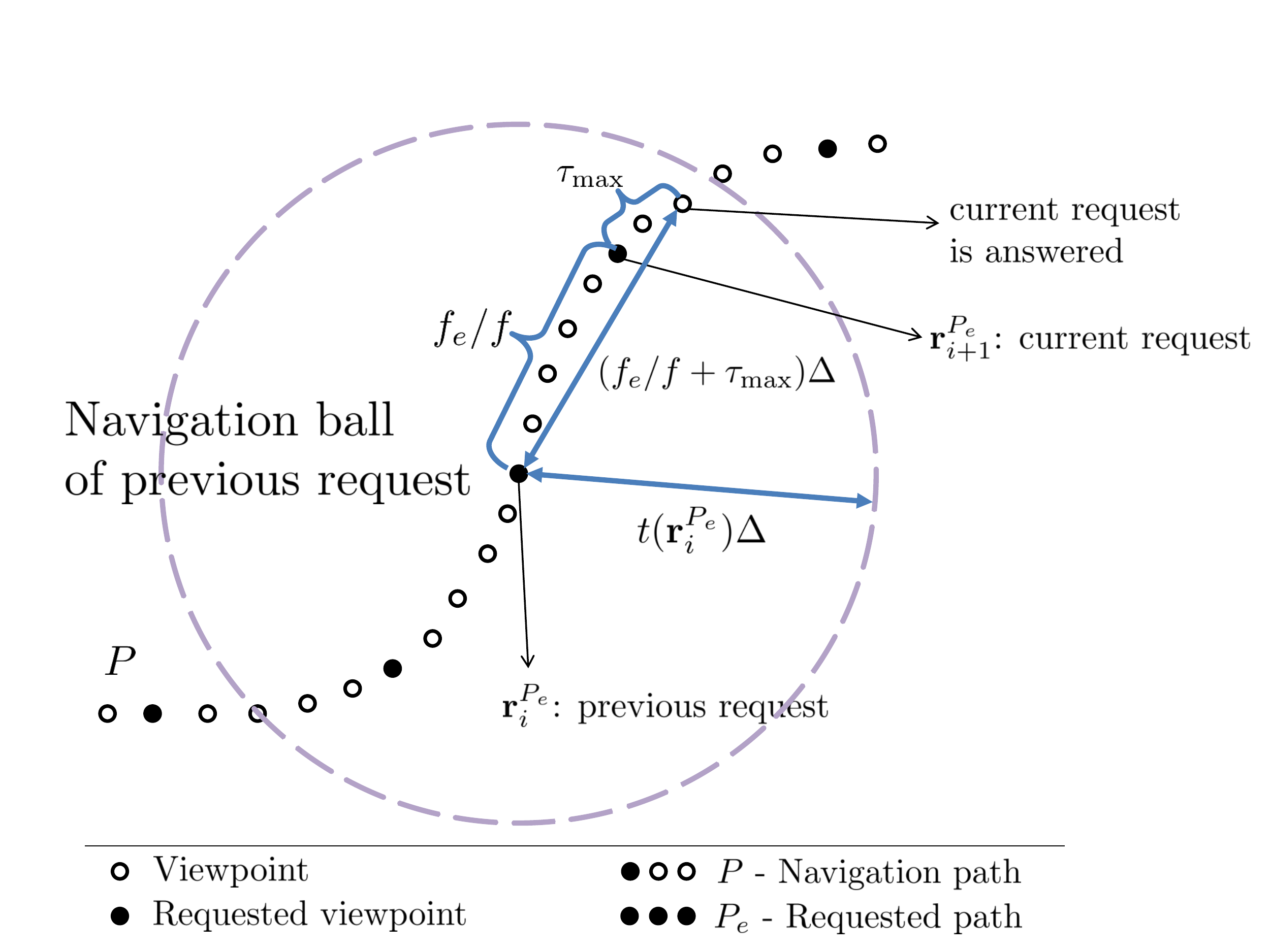}}
 % \vspace{1.5cm}
 % \centerline{(b) }%\medskip
\end{minipage}
\caption{Navigation ball for data buffering: A navigation ball gathers all viewpoints that will possibly be visited by the user before his next data request is handled by the server.}
\label{fig:nv_ball}
%\vspace{-10pt}
\end{figure}

For that purpose, we introduce the concept of \emph{navigation ball} as illustrated in Fig. \ref{fig:nv_ball}.
In more details, we assume that a data request is periodically sent by the user to the server every $ f_e $ frames, i.e., data request is sent at viewpoints $ \mathbf{r}_1, \mathbf{r}_{1+f_e}, \mathbf{r}_{1+2f_e}$, etc. These viewpoints are called \emph{requested viewpoints}. 
The set of all requested viewpoints forms a special subset of $P$, called the \emph{requested path} $ P_e = \{ \mathbf{r}_1, \mathbf{r}_{1+f_e} \cdots \mathbf{r}_{1+(N_e-1) f_e} \} $,
where $ N_e = N_f / f_e $ is the number of requested viewpoints within a single path. 
Different from $ P $ that is purely related to user navigation, the requested path $ P_e $ is associated to actual data sent to the user, i.e., the data to be transmitted depends on the location of each requested viewpoint.
For each $ \mathbf{r} \in P_e $, we target to transmit data that enables the user to render any views in a neighborhood around $ \mathbf{r} $. 
This neighborhood is called the navigation ball, and it is defined as
\begin{equation} \label{eq:naviball}
\mathcal{N}_B(\mathbf{r}) = \{ \mathbf{r}' \in \mathcal{N} | d (\mathbf{r}', \mathbf{r}) \leq t({\mathbf{r}}) \Delta \},
\end{equation}
where $ d(\cdot) $ is a distance function and $ t({\mathbf{r}}) \Delta $ measures the size of the ball.
The parameter $ \Delta $ is the navigation speed describing the maximum velocity of the user in the navigation domain, and $ t({\mathbf{r}}) $ is the tolerable delay of the navigation ball.
By increasing $ t({\mathbf{r}}) $, longer delay can be tolerated, and therefore more viewpoints can be visited without additional data from the server.

When $ t({\mathbf{r}}) $ increases beyond a certain value, the effect of the system delay can be eliminated and the entire navigation becomes smooth.
For the sake of simplicity, we assume a maximum system delay $\tau_{\max} $ for any data request. 
As illustrated in Fig. \ref{fig:nv_ball}, the tolerable delay $ t({ \mathbf{r}}) $ requires to compensate for the overall delay consisting of the time interval between consecutive requests $ f_e/f $ ($f$ is the frame rate) and the system delay $ \tau_{ \max } $, i.e.,
\begin{equation} \label{eq:navi_ball_smooth}
t({\mathbf{r}}) \geq f_e / f + \tau_{\max},\quad \forall \mathbf{r} \in P_e, ~\forall P_e.
\end{equation}
When this inequality is satisfied at all requested viewpoints, the entire user navigation is smooth and there is no data starvation at the client side.

\subsection{Navigation Segment Representation}
\label{sec:sub:navi_seg}
\begin{figure}[t]
 \centering
\begin{minipage}[htb]{\linewidth}
  \centering
  \centerline{\includegraphics[width=.95\linewidth]{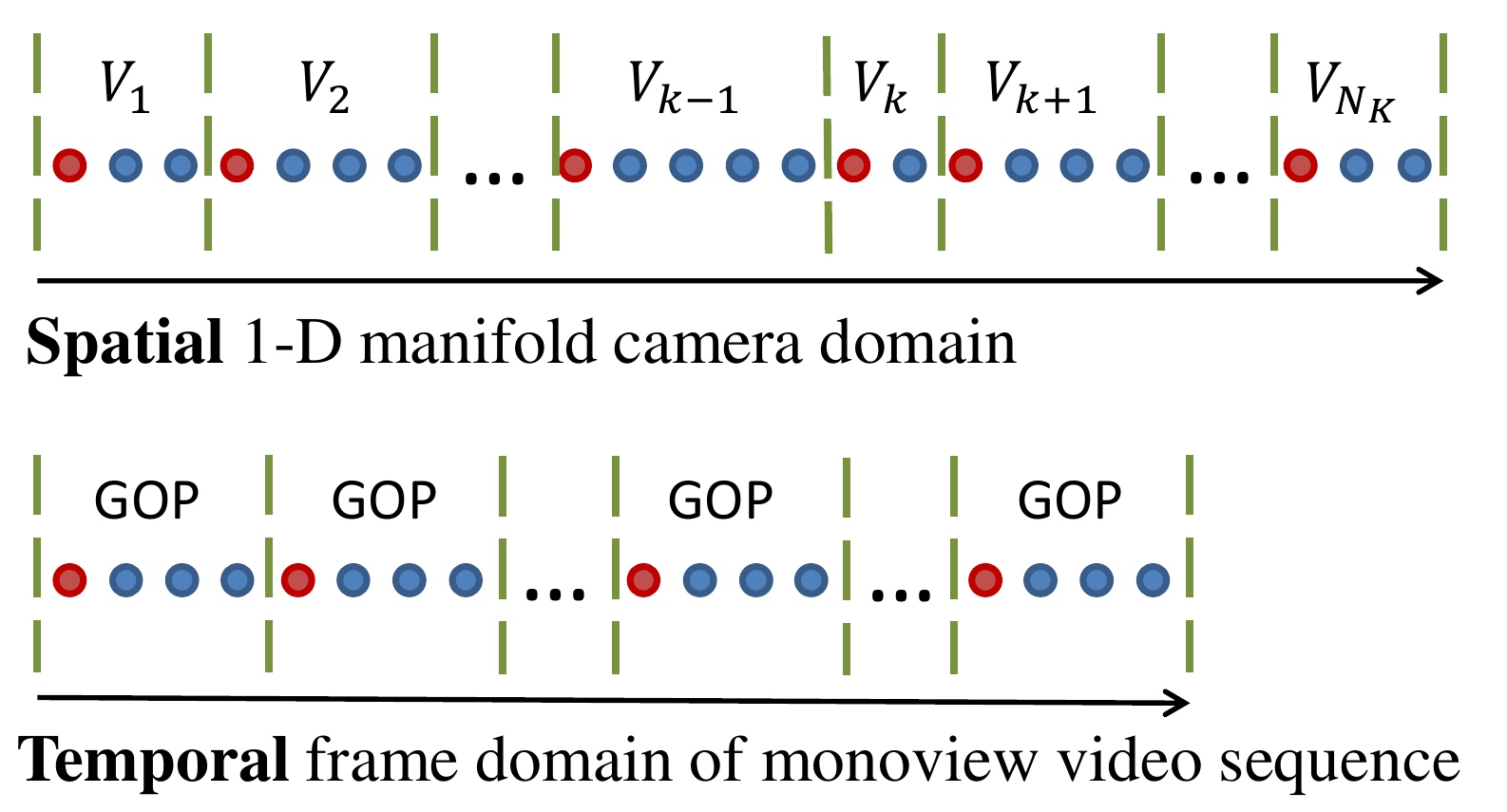}}
 % \vspace{1.5cm}
 % \centerline{(b) }%\medskip
\end{minipage}
\caption{ Navigation segment representation as a spatial analogue to temporal GOP structure in monoview video sequence.}
\label{fig:navi_seg}
%\vspace{10pt}
\end{figure}

An appropriate data representation format is crucial to the efficiency of data transmission and compression in the navigation system. For each data request, the system needs to \emph{transmit only the data that is sufficient to cover the navigation ball of $ \mathcal{N}_B(\mathbf{r}) $}. Thus, the design of data representation should allow for certain flexibility to choose any potential subset of the whole multiview data.
Similarly to \cite{maugey2013navi_dom_repr}, we investigate a data representation based on navigation segments.
Basically, a \emph{navigation segment} is a set of camera views $ Y_n $, which is coded independently of the rest data. 
Suppose all camera views are divided into $ N_K $ navigation segments. The $k$-th segment is denoted by
\[ V_k = \{ n \in [1,N_V] ~|~ Y_n \text{~is~in~} k \text{-th~segment} \}, ~ \forall k \in [1,N_K], \]
and it corresponds to the set of indices of the camera views included in this segment.
We further assume that the navigation segments are non-overlapping and connected along the underlying 1-D manifold of the camera views. 
%Therefore, we have $ {V}_k \cap {V}_l = \emptyset, \forall k \neq l $. 
In this case, the camera views in the left segment is always to the left of camera views in the right segment.
An illustration of our navigation segment representation is shown in Fig. \ref{fig:navi_seg}.

In fact, navigation segments can be regarded as an spatial analogue to the temporal GOP structure in monoview video coding (see Fig. \ref{fig:navi_seg}). The GOP structure supports temporal random access to frames, and the navigation segment structure supports the spatial random access to viewpoints.
Since each segment $V_k$ is independently decodable, the user only requires the necessary segments to enable the current navigation. The bandwidth and delay costs are thus reduced by avoiding the transmission of unnecessary segments.
By further adjusting the shape of the navigation segments, the navigation system is able to quantitatively control the interactivity in terms of bandwidth and delay during the data transmission. 

\subsection{Compressing Navigation Segments}
\label{subsec:code_strategy}
We compress each navigation segment independently using predictive coding in order to remove data redundancy and save bit rate.
Basically, in each navigation segment, one of the camera views is chosen as the anchor frame (i.e. I-frame), which is intra coded. The other camera views are predicted frames (i.e. P-frames) and each P-frame uses one of its neighboring views as the reference view for prediction.
The images and depth maps are coded using the same prediction structure.

\begin{figure}[tb]
 \centering
\begin{minipage}[htb]{.65\linewidth}
  \centering
  \vspace{4pt}
  \centerline{\includegraphics[width=\linewidth]{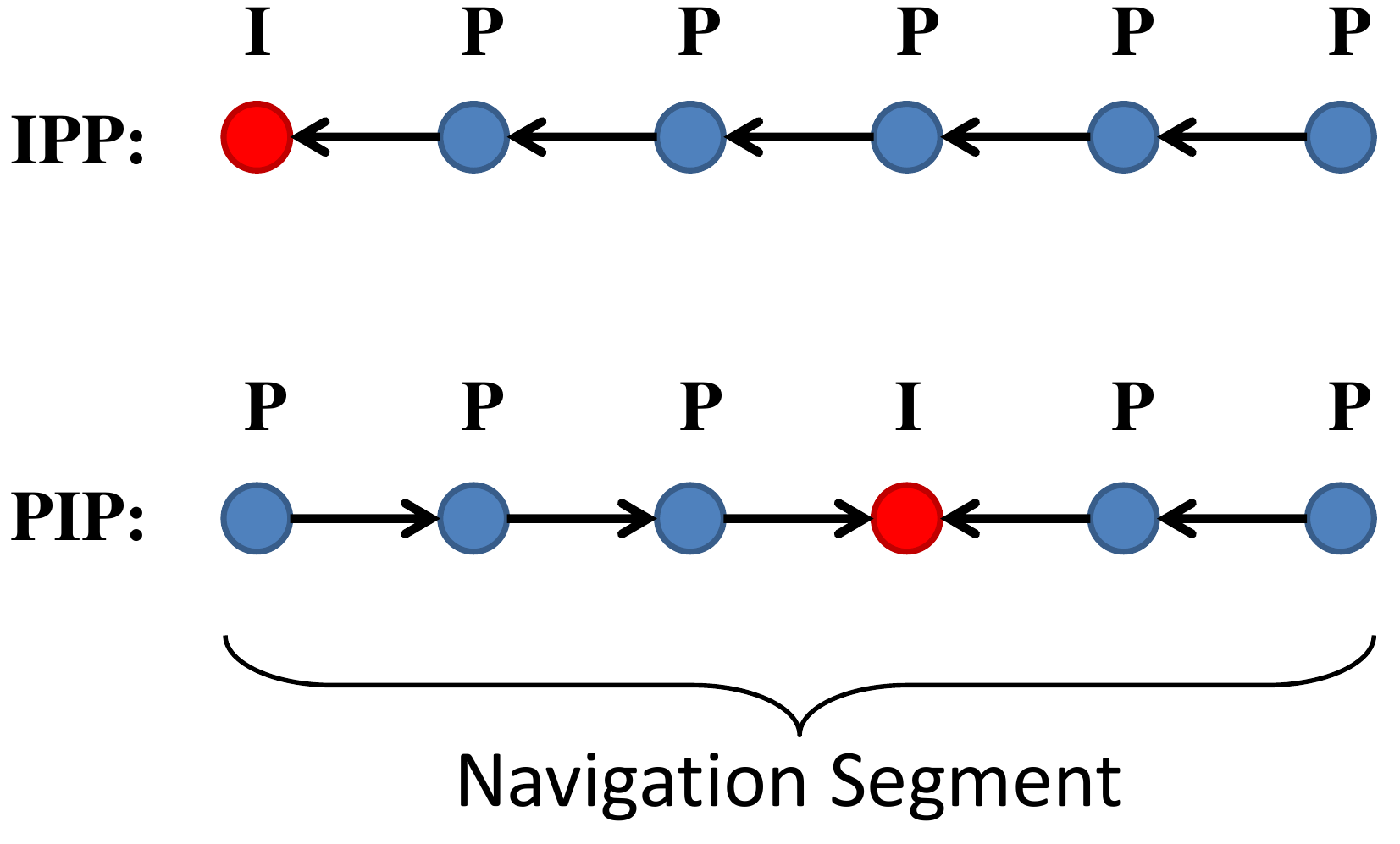}}
  \vspace{3pt}
 \centerline{ (a) } \medskip
\end{minipage} \\
\begin{minipage}[htb]{.65\linewidth}
  \centering
  \centerline{\includegraphics[width=\linewidth]{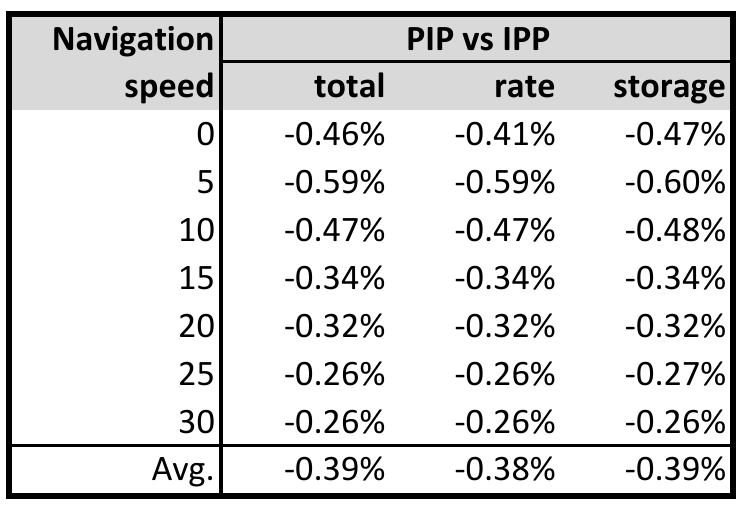}}
  \centerline{ (b) }
\end{minipage}
\caption{ Prediction structures. (a) Illustration of two typical prediction structures. (b) Performance comparison in terms of navigation cost reduction of `PIP' against `IPP'. }
\label{fig:expr_coding_struct}
\end{figure}

Two typical prediction structures can be chosen as illustrated in Fig. \ref{fig:expr_coding_struct}a. In the IPP structure, the position of I-frame is fixed to be the first view in each navigation segment, and each P-frame takes its previous view as the reference frame.
While in the PIP structure, the position of I-frame is float and each P-frame takes its left / right view as the reference frame when it is on the right / left side of the I-frame. The position of I-frame in the PIP structure is further optimized in order to achieve the minimum encoding bit rate for each navigation segment.
We compare the performance of the two prediction structures in terms of file size of compressed navigation segments and the average transmission rate by conducting experiments on the datasets \cite{peris2012stereo, martull2012realistic}. The details of the experimental setups can be referred in Sec. \ref{sec:expr}.
Fig. \ref{fig:expr_coding_struct}b illustrates the comparison results, where the two prediction structures provide very close performances.
We also notice that, with the IPP structure, the resulting problem can be efficiently solved (see Sec. \ref{sec:navisol}), which is important for a practical navigation system. However the resulting problem with PIP structure is difficult to handle on the other hand. 
Therefore, considering the trade-offs between performance and complexity, we choose the IPP prediction structure in our work.

There are several possible solutions for the actual compression of the navigation segments. 
We use the state-of-the-art MV-HEVC standard \cite{tech2013mvhevc, sullivan2013hevcextension} as the compression engine, which is an extension of the HEVC standard \cite{sullivan2012overview} for coding multiview sequences.
It exploits inter-view data redundancies by enabling disparity-compensated prediction, where previously coded images at the same time instant in neighboring views are used as references for prediction. 

\section{Navigation System Optimization}
\label{sec:naviopt}

\subsection{Optimization Framework}
\label{sec:sub_prob_form}
Based on the navigation definition and the navigation segment representation, we now propose an optimization framework to optimize the full navigation system. 
In particular, we consider the following optimization problem
\begin{equation} \label{eq:nv_prob}
\begin{aligned}[l]
& \min_{\mathcal{V}, \mathcal{S}, \mathcal{T}} ~~ U_R (\mathcal{V}, \mathcal{S}) + \mu \cdot U_S (\mathcal{V}) + \nu \cdot U_D (\mathcal{V}, \mathcal{S}) \\
& ~~\mathrm{s.t} \quad t({\mathbf{r}}) \geq f_e / f + \tau_{\max},\quad \forall \mathbf{r} \in P_e, ~\forall P_e.
\end{aligned}
\end{equation}
In this problem, we jointly optimize various \emph{navigation costs} of the navigation system, including the compressed data size on the server $ U_S (\mathcal{V}) $, the transmission rate $ U_R (\mathcal{V}, \mathcal{S}) $ and the view synthesis distortion $ U_D (\mathcal{V}, \mathcal{S}) $.
The parameters $ \mu $ and $ \nu $ are weights for $ U_S (\mathcal{V}) $ and $ U_D (\mathcal{V}, \mathcal{S}) $ respectively.
The constraint enforces smooth navigation by using the navigation ball mechanism as indicated in Eq. (\ref{eq:navi_ball_smooth}).

The optimization variables denote the optimal design of navigation segments and navigation balls. 
In particular, $ \mathcal{V} $ denotes the partition (or division) of navigation segments, while $ \mathcal{S} $ denotes the allocation (or delivery) of navigation segments. 
The last optimization variable, $ \mathcal{T} = \{ t({\mathbf{r}})~|~\forall \mathbf{r} \in P_e, \forall P_e \} $, controls the size of the navigation balls.
The solution to this problem deals with the optimal partition and allocation of the navigation segments and the optimal choice of every navigation ball, which provides the best trade-offs between the navigation quality and the resource consumption for the system.
We next study each term of the cost function. 

\subsection{Navigation Costs}

\vspace{5pt}
\begin{flushleft}
\textbf{Storage cost}
\end{flushleft}
\vspace{-5pt}
The storage cost $ U_S $ denotes the size of the compressed multiview data stored in the server.
As we compress each navigation segment independently, the overall storage is the sum of the segment size, i.e.,
\begin{equation} \label{eq:cost_storage}
U_S(\mathcal{V}) = \sum\limits_{k=1}^{N_K} h^{(Q)}(V_k),
\end{equation}
where $ \mathcal{V} = \{ V_1, \cdots, V_{N_K} \} $ is the partitions of navigation segments.
The function $ h^{(Q)}(\cdot) $ is the generic compression function.
When the navigation segment is predictively coded using the prediction structure as described Section \ref{sec:sub:navi_seg}, it is written as
\begingroup\makeatletter\def\f@size{10}\check@mathfonts
\begin{equation} \label{eq:ipp_coding_struct}
\begin{aligned}
h^{(Q)}(V_k) = h_I^{(Q)}(Y_{i_k}) + \sum_{j \in \{{V}_k \backslash i_k \}} h_P^{(Q)}(Y_j|\hat{Y}_{j-1}),
\end{aligned}
\end{equation}
\endgroup
where $ h_I^{(Q)}(\cdot) $ and $ h_P^{(Q)}(\cdot) $ represent the compression functions of I-frame and P-frame respectively with quantization step size $ Q $. 
The notation $ i_k $ denotes the index of the first camera view in segment $V_k$, and $ \hat{Y}_n $ is the reconstruction of $ Y_n $.

We consider that the quantization step size $ Q $ is constant for all segments 
in order to stabilize the quality of all frames, which is important for a pleasant navigation with steady viewing quality.
We further assume that the function $ h_P^{(Q)}(Y_j|\hat{Y}_{j-1}) $ is independent of the segment partition $ \mathcal{V} $, because the quality of the reference view $ \hat{Y}_{j-1} $ is steady in different partition choices given a fixed $ Q $ value.

\vspace{5pt}
\begin{flushleft}
\textbf{Rate cost}
\end{flushleft}
\vspace{-5pt}
The rate cost $ U_R $ denotes the transmission rate and it measures the navigation interactivity in terms of system bandwidth.
In our work, we define the transmission rate as the size of total transmitted data per data request.
We first express the transmission rate at a single requested viewpoint $ \mathbf{r} \in P_e $ as
\begin{equation} \label{eq:cost_rate_request}
u_R (\mathbf{r}; \mathcal{V}, \mathcal{S}) = \sum_{k=1}^{N_K} h^{(Q)}(V_k) \cdot s(\mathbf{r}, V_k; t(\mathbf{r}) ),
\end{equation}
where $ s(\mathbf{r}, V_k; t(\mathbf{r}) ) $ is the indicator function for segment allocation. Its value is $ 1 $ if segment $ V_k $ is required for view rendering considering the navigation ball at $ \mathbf{r} $ with size $ t(\mathbf{r}) $. Otherwise it is $ 0 $. 
The set $ \mathcal{S} = \{ s(\mathbf{r}, V_k; t(\mathbf{r}) )~|~ \forall \mathbf{r} \in P_e, P_e, k \} $ contains all indicator functions.
Note that we treat the navigation segments as the minimum unseparated unit for data transmission. 

From the perspective of a system, the definition of $ U_R $ should consider the navigation of different users, as they will have different navigation paths and accordingly different data transmission instances. Therefore we define $ U_R $ as the expected average transmission rate over all possible navigation paths:
\begin{equation} \label{eq:cost_rate}
U_R (\mathcal{V}, \mathcal{S}) = E_{P_e} \left[ \frac{1}{N_e} \sum_{\mathbf{r} \in P_e} u_R (\mathbf{r}; \mathcal{V}, \mathcal{S}) \right],
\end{equation}
where we first compute the average transmission rate per requested path, and then take the expected value over all possible requested paths of the users.

It should be pointed out that, we assume a memoryless transmission scheme, where we do not consider the client's memory capacity. This means that the user does not reuse the data received at the previous requests. 
Therefore the transmission rate of a path is simply the sum of individual rates of each request.

\vspace{5pt}
\begin{flushleft}
\textbf{View synthesis distortion}
\end{flushleft}
\vspace{-5pt}
The view synthesis distortion $ U_D $ is the distortion in the rendered views and it represents the quality of navigation. 
We first denote the view synthesis distortion at a single viewpoint as $ u_D (\mathbf{r}; \mathcal{V}, \mathcal{S}) $, because both the partition and allocation of navigation segments influence this distortion.
Similar to the rate cost, the distortion term $ U_D $ also requires to consider the navigation of different users. Therefore we represent it as the expected sum of distortion over all navigation paths:
\begin{equation} \label{eq:cost_dist}
U_D (\mathcal{V}, \mathcal{S}) = E_{P} \left[ \sum_{\mathbf{r} \in P} u_D (\mathbf{r}; \mathcal{V}, \mathcal{S}) \right].
\end{equation}
We first compute the sum of view synthesis distortion in a single navigation path, and then calculate the expected distortion over all possible navigation paths of the users.

\subsection{Influencing Navigation Parameters}
There exist many navigation parameters influencing the navigation system, like the quantization step size $Q$, the system delay $\tau_{\max}$, the weights $\mu$ and $ \nu$, etc.
In our work, these parameters are not treated as optimization variables, but are regarded as input parameters of the optimization framework, since we focus our study on the data representation using the navigation segments.

\section{Model-Based Problem Formulation}
\label{sec:navimodel}

\subsection{Overview}
{
The above navigation problem in Eq. (\ref{eq:nv_prob}) is difficult to handle.
First, we need to introduce rate and distortion models in order to properly deal with the distortion function $ u_D (\mathbf{r}; \mathcal{V}, \mathcal{S}) $ and the expectation operator $ E_{P_e}[\cdot] $ and $ E_{P}[\cdot] $ in the rate and distortion terms respectively.
Second, it is difficult to solve the segment partition $ \mathcal{V} $ and the segment allocation $ \mathcal{S} $ simultaneously. 
In our work, we propose to study the navigation problem by considering the following two subproblems.
\begin{itemize}[leftmargin=12pt]
\item [1)]We first consider a fixed allocation solution, namely $ \mathcal{S}_0 $. Given $ \mathcal{S}_0 $, we solve for the optimal segment partition $ \mathcal{V}^\star $ and the optimal size of navigation balls $ \mathcal{T}^\star $ in Eq. (\ref{eq:nv_prob}).
\item [2)] With the derived optimal $ \mathcal{V}^\star $ and $ \mathcal{T}^\star $ in 1), we further solve for the optimal segment allocation $ \mathcal{S}^\star $ in Eq. (\ref{eq:nv_prob}).
\end{itemize}
This approach guarantees the optimal solution for users with fixed allocation solution $ \mathcal{S}_0 $, and provides suboptimal solution to users with other allocation solutions.
We next present how we formulate these two subproblems using our rate and distortion models.
}

\subsection{The Partitioning Problem with Fixed Segment Allocation}

\begin{flushleft}
\textbf{Fixed segment allocation $ \mathcal{S}_0 $}
\vspace{-5pt}
\end{flushleft}
{We consider a fixed allocation solution $ \mathcal{S}_0 $ that targets a low-distortion rendering.
In order to define $ \mathcal{S}_0 $, we need to consider the reference views in DIBR.}
In many existing approaches, people use two or more reference views for DIBR in order to reduce the view synthesis distortion. 
However, in our work, we assume a single reference view due to the following reasons. 
First, the rendering quality is already satisfying with a single reference, because the virtual view is mostly derived from one reference while the rest of the references mainly provide side information for occlusion handling.
Second, under the single reference assumption, the subsequent modeling process is much simplified and it becomes easier to solve the navigation problem in Eq. (\ref{eq:nv_prob}).

Under this single reference assumption, we define a fixed allocation solution $ \mathcal{S}_0 $ as follows. 
For any virtual viewpoint within the navigation ball, namely $ \mathbf{r}' \in \mathcal{N}_B(\mathbf{r}) $, we choose the camera view that is closest to $ \mathbf{r}' $ as the reference view for rendering, and the index of this camera view is denoted as $ l_0(\mathbf{r}') $.
We then transmit the corresponding navigation segments that contain the camera view $ l_0(\mathbf{r}') $ for all $ \mathbf{r}' \in \mathcal{N}_B(\mathbf{r}) $.
Since any virtual view is assigned with its closest camera view for rendering, the solution $ \mathcal{S}_0 $ generally provides a low view synthesis distortion very close to the minimum value. However it does not guarantee the minimum transmission rate. 
The definition of $ \mathcal{S}_0 $ is consistent with the purpose of having a high quality rendering at the price of a potential suboptimal transmission rate.

\begin{flushleft}
\textbf{Optimal size of navigation balls $ \mathcal{T}^{\star} $}
\vspace{-5pt}
\end{flushleft}
Under the allocation solution $ \mathcal{S}_0 $, the optimal value of $ t(\mathbf{r}) \in \mathcal{T} $ can actually be inferred.
As $ t(\mathbf{r}) $ grows, more virtual viewpoints are included in the navigation ball. As a result, each navigation segment will be requested more often, and consequently the rate term $ U_R (\mathcal{V}, \mathcal{S}_0) $ in Eq. (\ref{eq:nv_prob}) will keep increasing.
On the other hand, the distortion term $ U_D (\mathcal{V}, \mathcal{S}_0) $ is not affected by $ t(\mathbf{r}) $, because the view synthesis distortion of each viewpoint $ \mathbf{r}' $ is fixed due to its unique and determined reference view $ l_0(\mathbf{r}') $ in $ \mathcal{S}_0 $. 
As a result, the objective function of Eq. (\ref{eq:nv_prob}) will keep increasing as $ t(\mathbf{r}) $ grows. 
Then the optimal $ t^{\star}(\mathbf{r}) $ is obtained when the equality holds in the smooth navigation constraint:
\begin{equation} \label{eq:opt_t}
t^{\star}(\mathbf{r}) = f_e/f + \tau_{\max} \equiv t^{\star}, ~~\forall t(\mathbf{r}) \in \mathcal{T}.
\end{equation}
In other words, all navigation balls will have the identical optimal size indicated by $ t^{\star} $.

\begin{flushleft}
\vspace{5pt}
\textbf{Modeling process: rate model}
%\vspace{-5pt}
\end{flushleft}
Based on $ \mathcal{S}_0 $ and $ t^{\star} $ defined above, we now propose rate and distortion models in order to convert the original navigation problem in Eq. (\ref{eq:nv_prob}) into a solvable problem.
We first rewrite the rate cost $ U_R (\mathcal{V}, \mathcal{S}_0) $ using Eq. (\ref{eq:cost_rate_request}) and (\ref{eq:cost_rate}) as follows.
\begingroup\makeatletter\def\f@size{10}\check@mathfonts
\begin{equation} \label{eq:model_rate2alpha}
\begin{aligned}
U_R (\mathcal{V}, \mathcal{S}_0) 
&= \frac{1}{N_e} \sum_{k=1}^{N_K} h^{(Q)}(V_k) \cdot \alpha(V_k, \mathcal{S}_0),
\end{aligned}
\end{equation}
\endgroup
where $ \alpha(V_k, \mathcal{S}_0) = E_{P_e} \left[  \sum_{\mathbf{r} \in P_e} s_0(\mathbf{r}, V_k; t(\mathbf{r}) ) \right] $ denotes the expected number of requests of segment $ V_k $ per navigation path.
The notation $ \alpha(V_k, \mathcal{S}_0) $ allows us to consider the global influence of the allocation solution $ \mathcal{S}_0 $ instead of looking into individual indicator functions.

We first study $ \alpha_{0}(V_k, \mathcal{S}_0) $ with no navigation ball, i.e. $ t^{\star} = 0$.
In that case, the navigation ball shrinks to a single viewpoint $ \mathcal{N}_B(\mathbf{r}) = \mathbf{r} $, and the indicator function is degraded to $ s_0(\mathbf{r}, V_k) $.
We then use the following approximation
\begin{equation*}
\begin{aligned}
\alpha_{0}(V_k, \mathcal{S}_0) &= E_{P_e} \left[  \sum_{\mathbf{r} \in P_e}  s_0(\mathbf{r}, V_k ) \right] \approx  N_e \cdot E_{\mathbf{r}} [s_0(\mathbf{r}, V_k) ],
\end{aligned}
\end{equation*}
\textcolor{black}{where we assume that each $ \mathbf{r} $ is considered independently from the path $ P_e $. 
This is justified in the case of a large number of user accesses, since the expectation related to one single path tends to a global expectation evaluated over the whole navigation domain.}
We further represent $ s_0(\mathbf{r}, V_k) $ as
\begin{equation*}
s_0(\mathbf{r}, V_k) = \left\lbrace
\begin{array}{l}
 1, \quad \mathbf{r} \in \bigcup\limits_{ n \in {V}_k} \mathcal{N}(Y_n) \\
0, \quad \text{otherwise}
\end{array} \right. .
\end{equation*}
The subset $ \mathcal{N}(Y_n) = \{ \mathbf{r}~|~l_0(\mathbf{r}) = n \} $ is the set of viewpoints which require camera view $ Y_n $ for view rendering.
Recall that we consider the case where a single reference view is used for rendering in $ \mathcal{S}_0 $.
Therefore each $ \mathbf{r} $ requires only one $ Y_n $, resulting in a non-overlapping division of $ \mathcal{N}(Y_n) $, i.e. $ \mathcal{N}(Y_n) \cap \mathcal{N}(Y_m) = \emptyset, \forall n \neq m $.
Hence, we can rewrite the expectation term in $ \alpha_{0}(V_k, \mathcal{S}_0) $ as 
\begin{equation*}
\begin{aligned}
&E_{\mathbf{r}} [s_0(\mathbf{r},V_k) ] 
= \int_{\mathcal{N}} p_r(\mathbf{r}) \cdot s_0(\mathbf{r},V_k) \mathrm{d} \mathbf{r} \\
&= \int_{\bigcup\limits_{ n \in {V}_k} \mathcal{N}(Y_n)} p_r(\mathbf{r}) \mathrm{d} \mathbf{r} 
= \sum\limits_{ n \in {V}_k} \int_{\mathcal{N}(Y_n)} p_r(\mathbf{r}) \mathrm{d} \mathbf{r},
\end{aligned}
\end{equation*}
where $ p_r(\mathbf{r}) $ is the density function of $ \mathbf{r} $. We further define the \emph{view popularity} by converting the density function of virtual viewpoints into the popularity function of camera viewpoints:
\begin{equation}
p_n \equiv \int_{\mathcal{N}(Y_n)} p_r(\mathbf{r}) \mathrm{d} \mathbf{r}.
\end{equation}
The popularity $ p_n $ represents the popularity of the camera view $ Y_n $ being required by users for view rendering.
Finally we can approximate $ \alpha_{0}(V_k, \mathcal{S}_0) $ using the view popularity as follows.
\begin{equation} \label{eq:model_alpha_base}
\alpha_{0}(V_k, \mathcal{S}_0) \approx N_e \sum\limits_{n \in {V}_k} p_n.
\end{equation}

\begin{figure}[tb]
 \centering
\begin{minipage}[htb]{.8\linewidth}
  \centering
  \centerline{\hspace{-20pt}\includegraphics[width=\linewidth]{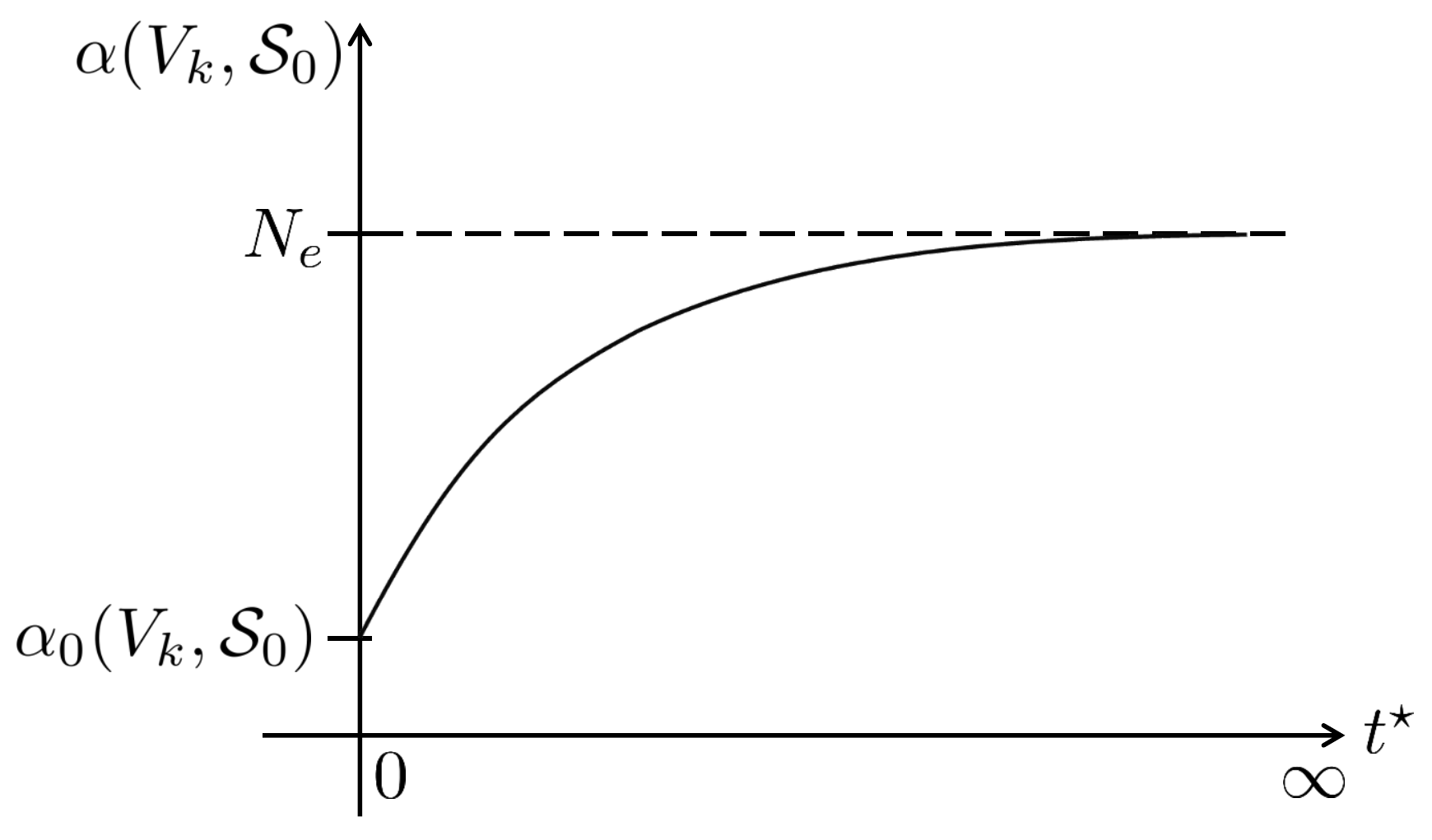}}
 % \vspace{1.5cm}
  \centerline{(a) } \medskip
\end{minipage} \\
\begin{minipage}[htb]{.75\linewidth}
  \centering
  \centerline{\includegraphics[width=\linewidth]{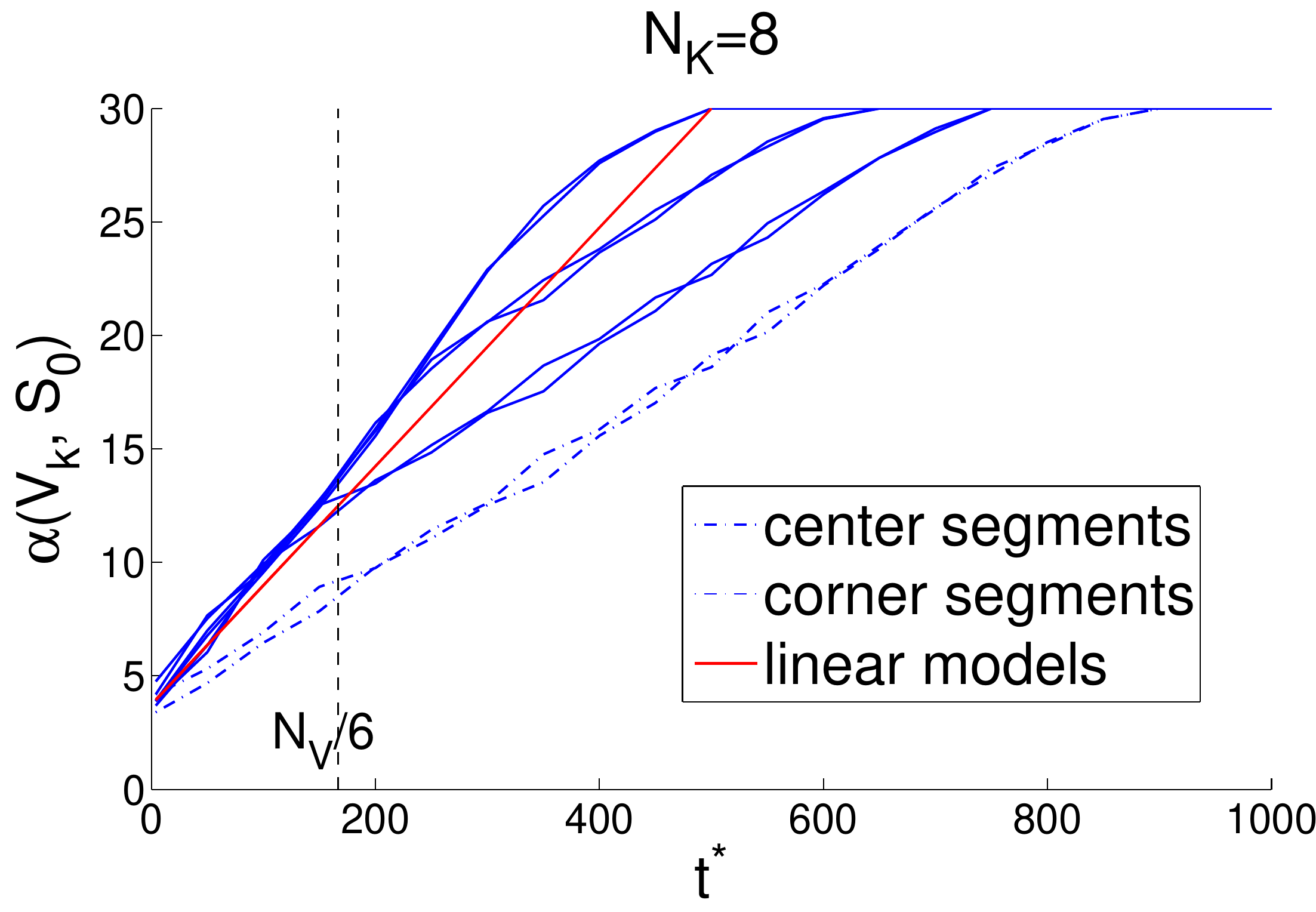}}
 % \vspace{1.5cm}
  \centerline{(b) }%\medskip
\end{minipage}
\caption{ Rate model with navigation ball: relationship between $ \alpha(V_k, \mathcal{S}_0) $ and $ t^{\star} $. (a) Qualitative relationship. (b) Quantitative result obtained by experiments on simulated navigation paths ($N_K = 8$). }
\label{fig:nv_naviball_model}
%\vspace{10pt}
\end{figure}

\begin{flushleft}
\textbf{Modeling process: rate model with navigation ball}
%\vspace{-5pt}
\end{flushleft}
We note that $ \alpha(V_k, \mathcal{S}_0) $ is influenced by the size of the navigation ball.
When $ t^{\star} = 0 $, $ \alpha(V_k, \mathcal{S}_0) $ has the minimum value $ \alpha_{0}(V_k, \mathcal{S}_0) $. As $ t^{\star} $ increases, each navigation segment will be requested more often, and therefore $ \alpha(V_k, \mathcal{S}_0) $ will keep increasing until the maximum value of $ N_e $, where the navigation ball expands to the entire navigation domain and segment $ V_k $ is always being requested in every requested viewpoint of a path.
This relationship is qualitatively illustrated in Fig. \ref{fig:nv_naviball_model}a.
We propose to model this relationship using a monotonic decreasing function $ g(t^{\star}) $ as follows,
\begin{equation} \label{eq:model_alpha_nb}
\begin{aligned}
\alpha(V_k, \mathcal{S}_0) &= ( 1 - g(t^{\star})) N_e + g(t^{\star}) \cdot \alpha_{0}(V_k, \mathcal{S}_0). 
\end{aligned}
\end{equation}

In order to derive an appropriate expression of $g(t^{\star})$, we carry out experiments on simulated navigation paths.
In particular, we generate $500$ navigation paths based on $1000$ imaginary camera viewpoints lying in a 1-D manifold. A detailed explanation of the generation process can be referred in Section \ref{subsec:eval_part+strm}.
We adopt a baseline equidistance partitioning of navigation segments, where the navigation segments are equally divided with the same width (see Section \ref{subsec:expr_setup}).
We apply the segment allocation solution $ \mathcal{S}_0 $, and compute $ \alpha(V_k, \mathcal{S}_0) $ by averaging the number of request for segment $ V_k $ over all navigation paths at different values of $ t^{\star} $.
Different navigation configurations are tested by adjusting the navigation speed $ \Delta $ and the number of navigation segments $ N_K $.
Fig. \ref{fig:nv_naviball_model}b plots the results when $N_K = 8$.
It is observed that different navigation segments reach the maximum value $N_e$ at different values of $ t^{\star} $, because the corresponding navigation ball of each navigation segment takes different time to expand the whole navigation domain with the same navigation speed $\Delta$.
However when the navigation ball is small ($ t^{\star} \Delta \leq N_V / 6 $), almost all navigation segments (except the two corner ones) have the same linear behavior. This condition is satisfied in our problem, because we are dealing with hundreds or even thousands of camera viewpoints and in general $ N_V \gg t^{\star} \Delta $. We also ignore the influence of two corner navigation segments, as typically $N_K \gg 2$.
Then we obtain an identical linear function of $ g(t^{\star}) $ for all navigation segments, i.e.
\begin{equation}
\label{eq:g_func}
g(t^{\star}) = \max ( 1 - 2 t^{\star} \Delta / N_V, ~0).
\end{equation}
The truncation is to ensure that the range of $ g(t^{\star}) $ is in $ [0,1] $.
We plot this function in Fig. \ref{fig:nv_naviball_model}b, which fits the curves well when $ t^{\star} \Delta \leq N_V / 6 $.
Note that, although we illustrate a particular example, extensive experiments under different navigation configurations verify the effectiveness of the derived model.

Finally we derive our rate model using Eq. (\ref{eq:model_rate2alpha}) (\ref{eq:model_alpha_base}) (\ref{eq:model_alpha_nb}) as follows.
\begin{equation} \label{eq:model_rate}
\begin{aligned}
%U_R (\mathcal{V}, \mathcal{S}_0) = \sum_{k=1}^{N_K} h^{(Q)}(V_k) \left( 1 - g(t^{\star}) + g(t^{\star}) \sum\limits_{ n \in {V}_k} p_n  \right).
U_R (\mathcal{V}, \mathcal{S}_0) &= \big(1 - g(t^{\star})\big) \sum_{k=1}^{N_K} h^{(Q)}(V_k) + \\
&\hspace{30pt} g(t^{\star}) \sum_{k=1}^{N_K} \bigg( h^{(Q)}(V_k) \sum\limits_{ n \in {V}_k} p_n \bigg) .
\end{aligned}
\end{equation}
\textcolor{black}{This rate model contains two terms. The first term is a storage-like term with weight $ 1 - g(t^{\star}) $, and the second term is a basic rate term with weight $ g(t^{\star}) $. When we neglect the navigation ball, we have $ g(t^{\star}) = 1 $, and the rate cost is entirely determined by the basic rate term, which prefers smaller sizes of navigation segments. As the navigation ball increases, $ g(t^{\star}) $ decreases to $ 0 $, and the storage-like term will gradually dominate the overall rate cost, resulting in larger sizes of navigation segments. This phenomenon is further presented in our experiments in Sec. \ref{sec:expr}. }
In this model, the rate term is influenced by the partitions of navigation segments, the size of navigation balls and the view popularity.

\begin{flushleft}
\textbf{Modeling process: distortion model}
\vspace{-5pt}
\end{flushleft}
We next investigate the modeling of the distortion term $ U_D (\mathcal{V}, \mathcal{S}_0) $ in the navigation problem of (\ref{eq:nv_prob}).
We first look at the view synthesis distortion at a single viewpoint, namely $ u_D (\mathbf{r}; \mathcal{V}, \mathcal{S}) $, which is in general difficult to estimate \cite{ma2013distmetric}.
However, under the allocation solution $ \mathcal{S}_0 $, we can derive
\begin{equation*}
\begin{aligned}
& u_D (\mathbf{r}; \mathcal{V}, \mathcal{S}_0) = u_D (\mathbf{r}, \hat{Y}_{l_0(\mathbf{r})} ),
\end{aligned}
\end{equation*}
where we use the nearest camera viewpoint indexed by $ l_0(\mathbf{r}) $ as the reference view for rendering.
With this, we can rewrite the distortion term in Eq. (\ref{eq:cost_dist}) as
\begingroup\makeatletter\def\f@size{10}\check@mathfonts
\begin{equation} \label{eq:model_dist_s1}
\begin{aligned}
U_D (\mathcal{V}, \mathcal{S}_0) &= E_{P} \left[ \sum_{\mathbf{r} \in P} u_D (\mathbf{r}, \hat{Y}_{l_0(\mathbf{r})} ) \right] \\
&\approx N_f E_{\mathbf{r}} \left[ u_D(\mathbf{r}, \hat{Y}_{l_0(\mathbf{r})}) \right].
\end{aligned}
\end{equation}
\endgroup
Similar to the rate model, here we assume that each $ \mathbf{r} $ is considered independently from the navigation path $ P $.
We then further rewrite the expectation term using the density function $ p_r(\mathbf{r}) $ and the non-overlapping subset $ \mathcal{N}(Y_n) $ introduced previously, and we derive
\begingroup\makeatletter\def\f@size{10}\check@mathfonts
\begin{equation} \label{eq:model_dist_s2}
\begin{aligned}
E_{\mathbf{r}} \left[ u_D(\mathbf{r}, \hat{Y}_{l_0(\mathbf{r})}) \right] 
&= \int_{\mathcal{N}} p_r(\mathbf{r}) u_D(\mathbf{r}, \hat{Y}_{l_0(\mathbf{r})}) \mathrm{d} \mathbf{r} \\
&= \sum\limits_{n=1}^{N_V} \int_{\mathcal{N}(Y_n)} p_r(\mathbf{r}) u_D(\mathbf{r}, \hat{Y}_n) \mathrm{d} \mathbf{r}.
\end{aligned}
\end{equation}
\endgroup
This is very similar to the derivation in the rate model, except that we have the distortion function $ u_D(\mathbf{r}, \hat Y_n) $, which computes the view synthesis distortion given a single reference view.

As aforementioned, the distortion function $ u_D(\mathbf{r}, \hat Y_n) $ depends on the rendering method and also the quantization level of the camera views.
In DIBR, a virtual view image is first generated by warping the reference view image according to the corresponding depth map, and then inpainting is applied for hole filling \cite{maugey2015guided}. Therefore we can separate the virtual view image into two parts: hole regions and non-hole regions \cite{maugey2015GBR}.
For pixels in hole regions, the distortion mostly comes from inpainting. If we assume a constant inpainting distortion $ D_{inp} $ at each pixel location, then the overall distortion in hole regions is $ D_{inp} \cdot \Omega(\mathbf{r}, \mathbf{c}_n) $, where $ \Omega(\mathbf{r}, \mathbf{c}_n) $ is the number of pixels in the hole regions at viewpoint $ \mathbf{r} $ given reference view at $ \mathbf{c}_n $.
For pixels in non-hole regions, the distortion mostly comes from rendering.
We assume that an integer location wrapping is performed during rendering process, and we ignore the mismatching of pixels due to a depth distortion.
Then the pixels in non-hole regions are exactly copied from the reference image, and therefore they have the same reconstruction distortion as pixels in the reference image. 
We further assume that the reconstruction distortion depends only on the quantization step size $ Q $, and is denoted as $ D_{rec}^{( Q )} $. 
The distortion in the non-hole regions is thus $ D_{rec}^{( Q )} \cdot (WH - \Omega(\mathbf{r}, \mathbf{c}_n)) $, where $ W $ and $ H $ are the width and height of the image respectively.
Finally we derive the distortion function of a single virtual view by aggregating the distortions in both regions, i.e.
\begingroup\makeatletter\def\f@size{10}\check@mathfonts
\begin{equation} \label{eq:model_dist_dibr}
\begin{aligned}
u_D(\mathbf{r}, \hat Y_n) = D_{inp} \Omega(\mathbf{r}, \mathbf{c}_n) + D_{rec}^{( Q )} (WH - \Omega(\mathbf{r}, \mathbf{c}_n)),
\end{aligned}
\end{equation}
\endgroup
where we further provide the estimation of $ \Omega(\mathbf{r}, \mathbf{c}_n) $ in detail in Appendix.
Finally, with Eq. (\ref{eq:model_dist_s1}) (\ref{eq:model_dist_s2}) and (\ref{eq:model_dist_dibr}), we derive the distortion model as 
\begingroup\makeatletter\def\f@size{10}\check@mathfonts
\begin{equation} \label{eq:model_dist}
\begin{aligned}
& U_D (\mathcal{V}, \mathcal{S}_0) \approx N_f D_{rec}^{(Q)} W H + \\
& \qquad N_f (D_{inp} - D_{rec}^{(Q)} ) \cdot \sum_{n=1}^{N_V} \int_{\mathcal{N}(Y_n)} p_r(\mathbf{r}) \Omega(\mathbf{r}, \mathbf{c}_n) \mathrm{d} \mathbf{r} .
\end{aligned}
\end{equation}
\endgroup
In this equation, the right hand side only contains the navigation parameters and does not have the optimization variables, because under $ \mathcal{S}_0 $ the view synthesis distortion does not depend on the segment partition $ \mathcal{V} $.
In particular, all components in the above expression clearly have determined values, except for the integral. 
In the integral, the size of the hole regions $ \Omega(\mathbf{r}, \mathbf{c}_n) $ is uniquely determined by the viewpoints $ \mathbf{r} $ and $ \mathbf{c}_n $. 
The density function $ p_r(\mathbf{r}) $ and the subset $ \mathcal{N}(Y_n) $ are both determined by the navigation domain. 
Therefore the integral also has a determined value though it is hard to compute. 
We finally note that with the distortion model in Eq. (\ref{eq:model_dist}), the distortion term is not influenced by the optimization variables.

\begin{flushleft}
\textbf{The partitioning problem given $ \mathcal{S}_0 $}
\vspace{-5pt}
\end{flushleft}
We are now able to derive a solvable navigation problem by substituting the rate and distortion models in Eq. (\ref{eq:model_rate}) and (\ref{eq:model_dist}) into the original navigation problem in Eq. (\ref{eq:nv_prob}), and we have
\begingroup \makeatletter\def\f@size{10}\check@mathfonts
\begin{equation} \label{eq:nv_prob_part}
\begin{aligned}
\mathcal{V}^{\star} 
&= \arg\min_{\mathcal{V}} \sum\limits_{k = 1}^{N_K} \Big( h_I^{(Q)}(Y_{i_k}) + \sum_{j \in \{{V}_k \backslash i_k \}} h_P^{(Q)}(Y_j|\hat{Y}_{j-1}) \Big) \\
& \qquad \cdot \Big( \mu + 1 - g(t^{\star}) + g(t^{\star}) \sum_{n \in {V}_k} p_n \Big),
\end{aligned}
\end{equation}
\endgroup
where the compression function is expanded using Eq. (\ref{eq:ipp_coding_struct}).
Since the allocation solution is already given by $ \mathcal{S}_0 $ and the optimal size of navigation balls $ t^{\star} $ is provided in Eq. (\ref{eq:opt_t}), the only optimization variable in this problem is $ \mathcal{V} $, which represents the partition of navigation segments. 
Thus this problem is called \emph{the partitioning problem} in our work.
The goal is to find the optimal $ \mathcal{V}^{\star} $ that minimizes the rate and storage costs of the navigation system (note that the distortion cost is discarded using Eq. (\ref{eq:model_dist})).

\subsection{Complementary Allocation Problem}
Given the optimal size of navigation balls $ t^\star $ and the optimal segment partition $ \mathcal{V}^\star $ derived in Case 1, we further solve for the optimal segment allocation in the navigation problem of Eq. (\ref{eq:nv_prob}), and we derive the following problem.
\begin{equation}
\begin{aligned}
\mathcal{S}^\star &= \arg \min_{\mathcal{S}} ~~ U_R (\mathcal{V}^\star, \mathcal{S}) + \nu \cdot U_D (\mathcal{V}^\star, \mathcal{S}) \\
\end{aligned}
\end{equation}
Note that the storage cost is discarded in this problem, because it is fixed with the segment partition $ \mathcal{V}^\star $ and it is not influenced by the segment allocation $ \mathcal{S} $. 

It is realized that we are able to further separate this problem into smaller problems for each data request at $ \mathbf{r} \in P_e $ using the definition of rate and distortion costs in Eq. (\ref{eq:cost_rate}) and (\ref{eq:cost_dist}) as follows.
\begin{equation} \label{eq:nv_comple_alloc}
\begin{aligned}
&\{ s^{\star}(\mathbf{r}, V_k; t^\star )~|~\forall k \} = \arg \min ~~ \sum_{k=1}^{N_K} h^{(Q)}(V_k) \cdot s(\mathbf{r}, V_k; t^\star ) \\
& \qquad \qquad + \nu \cdot E_{\mathbf{r} } \left[ u_D(\mathbf{r}', \hat{Y}_{l(\mathbf{r}')})~|~\mathbf{r}' \in \mathcal{N}_B( \mathbf{r}) \right].
\end{aligned}
\end{equation}
In the cost function, the first term is the rate cost, which is further expanded using Eq. (\ref{eq:cost_rate_request}). The second term is the distortion term, which is the expected view synthesis distortion within the navigation ball $ \mathcal{N}_B( \mathbf{r}) $.
The optimization variable are the streaming indicator functions $ s(\mathbf{r}, V_k; t^\star ), \forall k \in [1, N_K] $. 

Here we still take the assumption of single reference rendering, but the camera index $ l(\mathbf{r}') $ denotes the closest camera viewpoint to $ \mathbf{r}' $ \emph{given only the transmitted segments} indicated by $ s^{\star}(\mathbf{r}, V_k; t^\star ) $. Note that it is different from the rendering in $ l_0(\mathbf{r}') $, which always uses the closest camera view as reference.
With this, the solution allows to use farther camera views as references for rendering, which leads to possibly larger view synthesis distortion.
In the extreme case when $ \nu \rightarrow \infty $, the solution converges to $ \mathcal{S}_0 $ that provides the minimum distortion. 
As $ \nu $ decreases, the solution gradually provides larger distortion than $ \mathcal{S}_0 $, but with correspondingly lower transmission rate.
Therefore, the derived allocation solution complements the fixed allocation solution $ \mathcal{S}_0 $ for practical users that might not be able to afford the amount of data required by $ \mathcal{S}_0 $ due to bandwidth limitations. It provides a flexible trade-off between the transmission rate and the viewing quality. 
We call this problem \emph{the complementary allocation problem} in our work.

\section{Optimization Algorithm}
\label{sec:navisol}
In this section we propose solving algorithms for the partitioning problem in Eq. (\ref{eq:nv_prob_part}) and the complementary allocation problem in Eq. (\ref{eq:nv_comple_alloc}).

\subsection{The Partitioning Problem}
\label{sec:sol_dijkstra}
The partitioning problem in Eq. (\ref{eq:nv_prob_part}) can be solved using the Dijkstra shortest path algorithm, similar to a key view selection problem in \cite{maugey2016refview}.
The Dijkstra's algorithm works for all partitioning parameters, i.e., $ i_k, \mathcal{V}_k, N_K $, simultaneously. Moreover, it is a fast algorithm that provides a globally optimal solution.

\begin{figure}[tb]
 \centering
\begin{minipage}[htb]{.9\linewidth}
  \centering
  \centerline{\includegraphics[width=\linewidth]{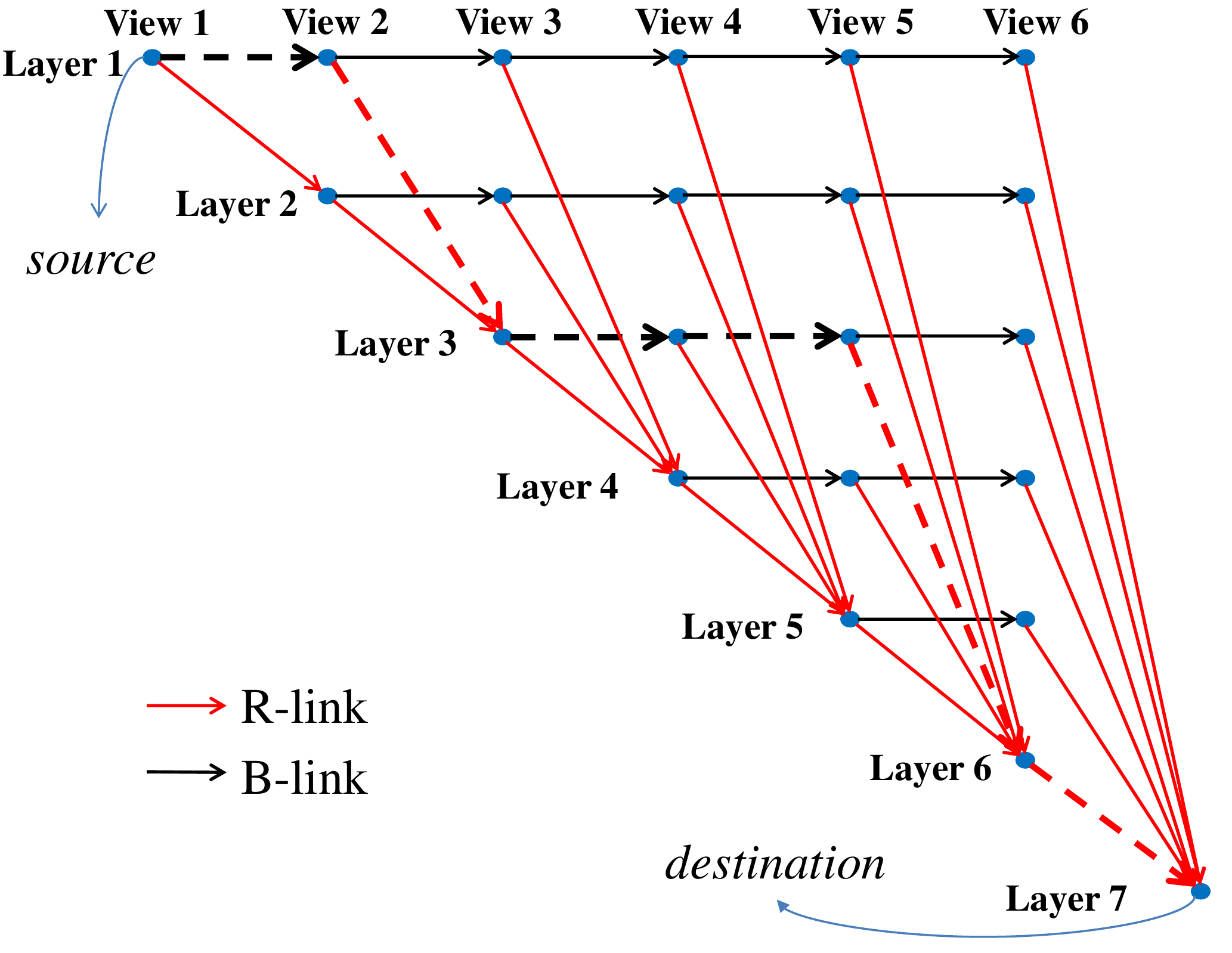}}
 % \vspace{1.5cm}
 % \centerline{(b) }%\medskip
\end{minipage}
\caption{ The graph constructed for the partitioning problem (\ref{eq:nv_prob_part}): an example of 6 camera viewpoints. The dashed path represents a navigation segment structure of \{view 1, 2\}, \{view 3, 4, 5\} and \{view 6\}. }
\label{fig:nv_dijk_graph}
%\vspace{-10pt}
\end{figure}

We first construct a graph where the vertices represents the views in the navigation domain. The views are organized in different layers that represent the potential navigation segments. An example of such a graph is given in Fig. \ref{fig:nv_dijk_graph}.
The nodes aligned in the same vertical line across different layers represent the same camera view.
In each layer, the first node is encoded as an I-frame and the other nodes are encoded as P-frames according to our coding structure in each navigation segment.
The last layer is the destination node, which is not a real camera view.
The source node is the first node in the first layer.
Two kinds of directed links, namely R-link and B-link, are assigned in the graph. Their costs are defined as follows.
The link cost between neighbouring views in different layers (R-link) is the cost of starting a new navigation segment by adding the first view of the former layer as an I-frame.
While the link cost between neighbouring views of the same layer (B-link) is the cost of adding the latter view to the current navigation segment as a P-frame.

We derive the link costs from the cost function of Eq. (\ref{eq:nv_prob_part}) as follows.
First of all, the original cost function can be separated into the unary term and the pairwise term:
\begingroup \makeatletter\def\f@size{9}\check@mathfonts
\begin{equation*}
(\mu + 1 - g(t^\star)) \sum_{k=1}^{N_K} \sum_{n \in V_k} h_n  + g(t^\star) \sum_{k=1}^{N_K} \sum_{n \in V_k} \sum_{m \in V_k} h_n p_m.
\end{equation*}
\endgroup
For simplicity, here we use $ h_n $ to denote the encoding bits of an arbitrary I-/P- frame for view $n$.
The unary term aggregates only the encoding bits $h_n$ of each view, while the pairwise term computes a pairwise cost of $h_n p_m$ for each ordered pair of $(n,m)$ in segment $V_k$. 
Based on this, we are able to write the link costs as
\begingroup \makeatletter\def\f@size{9}\check@mathfonts
\begin{equation*}
\begin{aligned}
&\textrm{R-link}: e(v_i^l, v_{i+1}^m) = h_I^{(Q)}(Y_{l}) \cdot \big( \mu + 1 - g(t^{\star}) + g(t^{\star}) p_l \big), m > i  \\
&\textrm{B-link}: e(v_i^l, v_{i+1}^l) = \\
& \left\lbrace
\begin{array}{l}
 h_I^{(Q)}(Y_{l}) \cdot g(t^{\star}) \cdot p_{l+1} + \\ \vspace{5pt}
 \hspace{2pt} h_P^{(Q)}(Y_{l+1}|\hat{Y}_{l}) \cdot \big(\mu + 1-g(t^{\star}) + g(t^{\star}) (p_l + p_{l+1}) \big), ~i=l \\
\big( h_I^{(Q)}(Y_{l}) + \sum_{j=l}^{i-1} h_P^{(Q)}(Y_{j+1}|\hat{Y}_{j}) \big) \cdot g(t^{\star}) \cdot p_{i+1} + \\
 \hspace{2pt} h_P^{(Q)}(Y_{i+1}|\hat{Y}_{i}) \cdot \big(\mu + 1-g(t^{\star}) + g(t^{\star}) \sum_{j=l}^{i+1}p_j \big), ~i \geq l+1
\end{array} \right.
\end{aligned}
\end{equation*}
\endgroup
where we let $ e(v_i^l, v_j^m) $ to be the link cost between view $i$ in layer $l$ and view $j$ in layer $m$.
The R-link $ e(v_i^l, v_{i+1}^m) $ has the cost of starting a new segment with the first view of layer $l$, which is view $l$. Therefore, its cost is the unary term of view $l$ plus the pairwise term of itself $(l,l)$.
The B-link $ e(v_i^l, v_{i+1}^l) $ has the cost of appending view $i+1$ to the current segment of layer $l$. Therefore, its cost contains the pairwise term of $(j,i+1), \forall j<i+1$ (the first term of B-link), plus the pairwise term of $(i+1, j), \forall j \leq i+1$ and the unary term of $i+1$ (the second term of B-link). 
Moreover, the design of the graph in Fig. \ref{fig:nv_dijk_graph} guarantees that each unary and pairwise term is aggregated only once along any solution from the source to the destination. As a result, the sum of the link costs of any solution exactly corresponds to the value of the cost function in Eq. (\ref{eq:nv_prob_part}). Therefore, the shortest path solution of this graph is exactly the solution to the partitioning problem in Eq. (\ref{eq:nv_prob_part}).

Once the graph is constructed, we can apply the Dijkstra's algorithm to find the shortest path from the source to the destination.
The resulting shortest path yields the optimal $ \mathcal{V}^{\star} $, where each passed layer represents a navigation segment. For example, the dashed path in Fig. \ref{fig:nv_dijk_graph} represents three navigation segments: $ V_1 = \{1, 2\} $, $ V_2 = \{3, 4, 5\} $ and $ V_3 = \{6\} $.

\subsection{The Complementary Allocation Problem} \label{sec:solve_strm}
In the complementary allocation problem of Eq. (\ref{eq:nv_comple_alloc}), the rate term is computed by adding up the sizes of the compressed navigation segments to be transmitted.
The distortion term is estimated using Eq. (\ref{eq:model_dist_dibr}).
Differently from the partitioning problem, which can be pre-computed offline before the actual user navigation, the allocation problem has a real-time requirement, where the system needs to react immediately and selects the best navigation segments to be transmitted for each data request.
For this purpose, instead of finding the optimal solution, we adopt an efficient heuristic algorithm for real-time processing.
It contains three steps as follows.
\vspace{10pt}
\begin{itemize}[leftmargin=0pt]
\item[] \textbf{Step 1}: For each requested viewpoint $ \mathbf{r} \in P_e $, determine the subset of its navigation ball, namely $  \mathcal{N}_B^{S}( \mathbf{r}) \subseteq \mathcal{N}_B( \mathbf{r}) $, using Eq. (\ref{eq:naviball}) but with a smaller tolerable delay $ t_S \leq t^{\star }$.
\item[] \textbf{Step 2}: For each $ \mathbf{r}' \in \mathcal{N}_B^{S}( \mathbf{r}) $, find the nearest camera viewpoint indexed by $ l_0(\mathbf{r}') $ and determine the corresponding navigation segment $ V_k $ it belongs to, i.e., $ l_0(\mathbf{r}') \in V_k $.
\item[] \textbf{Step 3}: The streaming indicator function $ s^{\star}(\mathbf{r}, V_k; t^\star ) $ for all navigation segments is then determined by:
\begin{equation} \label{eq:nv_comple_alloc_sol}
s^{\star}(\mathbf{r}, V_k; t^\star ) = \left\lbrace
\begin{array}{l}
 1, \quad \exists~ \mathbf{r}' \in \mathcal{N}_B^{S}( \mathbf{r}), ~l_0(\mathbf{r}') \in V_k \\
 0, \quad \text{otherwise}
\end{array} \right. , ~\forall k.
\end{equation}
\end{itemize}
In this solution, we always guarantee the best rendering quality for the virtual viewpoints that are closest to the requested viewpoint, i.e., the viewpoints within $ \mathcal{N}_B^{S}( \mathbf{r}) $, because the user will more likely visit these viewpoints than the ones farther away from the requested viewpoint $ \mathbf{r} $.
The rate-distortion trade-off, which is originally controlled by the weight $ \nu $ in Eq. (\ref{eq:nv_comple_alloc}), can be alternatively achieved by changing the size of $ \mathcal{N}_B^{S}( \mathbf{r}) $, namely $ t_S $.
When $ t_S = t^\star $, the solution is exactly $ S_0 $.
As $ t_S $ decreases from $ t^\star $ to a lower value, the algorithm will gradually request a smaller amount of navigation segments with smaller rate cost. The distortion will however increase because more and more viewpoints outside of $ \mathcal{N}_B^{S}( \mathbf{r}) $ will no longer have the closest camera view for rendering.
Note that, in practice it is impossible to run the above algorithm for all viewpoints in  $ \mathcal{N}_B^{S} ( \mathbf{r}) $, simply because there is infinite number of them. Instead, we sample $ \mathcal{N}_B^{S} ( \mathbf{r}) $ with equal distance, and run the algorithm only on the sampled viewpoints.

It should be pointed out that, although this solution is designed for our problem that assumes a memoryless transmission scheme, it can be extended to the case that considers the client's memory by simply avoiding the transmission of the repeated navigation segments received in the previous requests.

\begin{figure*}[tb]
 \centering
\begin{minipage}[htb]{.95\linewidth}
  \centering
  \centerline{\includegraphics[width=\linewidth]{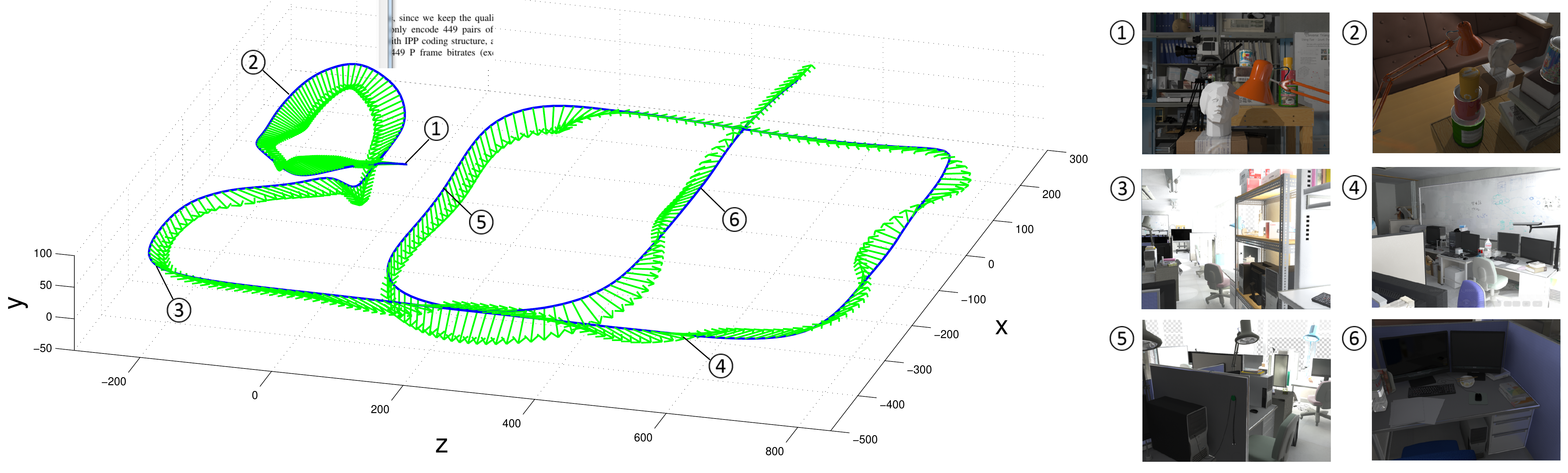}}
 % \vspace{1.5cm}
 % \centerline{(b) }%\medskip
\end{minipage}
\caption{Illustration of 1-D manifold camera arrangement using the New Tsukuba Stereo Dataset \cite{peris2012stereo, martull2012realistic}. The solid line denotes the camera trajectory while the arrows point out the orientations. We also show some camera views at different locations.}
\label{fig:tsukuba_camera_arrange}
%\vspace{-10pt}
\end{figure*}

\subsection{Complexity Analysis}
We briefly analyze the complexity of our algorithms here.
For the offline partitioning problem, the Dijkstra's algorithm runs in time $ \mathcal{O}(|E| + |V| \log |V|) $ when a min-priority queue is used \cite{fredman1987fibonacci}, where $ |E| $ and $ |V| $ represent the number of edges and nodes respectively.
In our problem (Fig. \ref{fig:nv_dijk_graph}), $ |E| =  N_V^2$ and $ |V| = \frac{1}{2} N_V^2 + \frac{1}{2} N_V + 1 $,
where $ N_V $ is the number of camera viewpoints. 
The computational complexity is thus $ \mathcal{O}(N_V^2 \log N_V) $.
This complexity is tolerable when $ N_V $ is not huge. In our experiment, when $ N_V = 450 $, it costs around 120 seconds on average in Matlab on a Inter(R) Core(TM) i5-3320M PC, which suits for an offline solution.
However it might still be necessary to speed up the running time especially when $N_V$ keeps growing. 
One simple way to reduce the complexity is to assume a maximum number of camera viewpoints in each navigation segment.
It can be verified that in this case the computational complexity reduces to $ \mathcal{O}(N_V \log N_V) $.

The computational complexity of the complementary allocation problem is more critical, as this has to be solved during the real-time data transmissions. 
In the solutions of Eq. (\ref{eq:nv_comple_alloc_sol}), the main computation lies in the search of the nearest camera viewpoint. 
We suppose that $ M $ points are sampled from the subset $ \mathcal{N}_B^{S} ( \mathbf{r}) $. 
For each sampled viewpoint, the worst case complexity for finding its nearest camera viewpoint is $ \mathcal{O}(N_V) $.
Then the worst case complexity for solving the allocation problem is $ \mathcal{O}(M N_V) $.
Therefore we can adjust the complexity by changing the sampling rate in the solutions.
In real implementations, when $ N_V = 450, M \approx 200, t_S = 4s, \Delta = 20$ (a dense sampling for a considerable size of the navigation ball), the run time is around 1 second in Matlab on the same Inter(R) Core(TM) i5-3320M PC. This complexity is suitable for the real-time computation in the streaming session.

\section{Experimental Results}
\label{sec:expr}
In this section, we evaluate the performance of the proposed navigation segment representation. In particular, we compute the resource consumptions in terms of storage and rate costs, and the navigation quality in terms of view synthesis distortion under different navigation configurations. We compare the proposed method with a baseline method, where the navigation segments are equally divided.

\subsection{Experimental Setup}
\label{subsec:expr_setup}
\begin{flushleft}
\textbf{Dataset}
\vspace{-5pt}
\end{flushleft}
We perform experiments on the New Tsukuba Stereo Dataset \cite{peris2012stereo} \cite{martull2012realistic}, which provides groundtruth stereo image and depth pairs for 1800 camera viewpoints along a 1-D manifold trajectory.
We further uniformly sample the 1800 camera viewpoints and obtain 450 of them, because practical navigation systems generally could not afford too many camera viewpoints due to resource limitations.
Fig. \ref{fig:tsukuba_camera_arrange} illustrates this camera arrangement.

\begin{flushleft}
\textbf{Navigation parameters}
\vspace{-5pt}
\end{flushleft}
We set the navigation parameters as shown in Table \ref{tab:expr_para_setting}. 
As mentioned in Sec. \ref{sec:sub_prob_form}, these navigation parameters are not optimized in our problem, but are treated as input parameters of the optimization, i.e., their values are set beforehand.
Note that the navigation speed $\Delta$ and the view popularity $p_n$ are to be determined according to different navigation configurations in our experiments.
The 450 image and depth pairs are encoded by the MV-HEVC engine \cite{tech2013mvhevc} \cite{sullivan2013hevcextension}. The quantization step size $Q$ is controlled by the QP (quantization parameter) in the MV-HEVC engine.
The higher the QP value, the larger the quantization step size.
As aforementioned, we assume that the compression function $ h_P^{(Q)}(Y_n|\hat{Y}_{n-1})$ does not depend on navigation segment partitions under the constant $Q$ value.
Therefore we only encode the pair of neighboring camera viewpoints and obtain 450 values of $h_I^{(Q)}(Y_n)$ and 449 values of $ h_P^{(Q)}(Y_n|\hat{Y}_{n-1})$ (excluding the first camera viewpoint) in order to estimate the compression function in the optimization.
%Note that these pre-encoded bit rates are only used for the optimization process; after obtaining the optimal navigation segment representation, each navigation segment requires to be re-encoded for actual navigation.

%
\begin{table}
\caption{ Navigation Parameter Setting }
\label{tab:expr_para_setting}
%\begingroup\makeatletter\def\f@size{10}\check@mathfonts
\begin{center}
\vspace{-10pt}
\begin{tabular}{c|c|c}
\hline
\textbf{Notation} & \textbf{Value} & \textbf{Description}\\ \hline
$N_V$ & 450 & \# camera viewpoints \\
$T$ & 90s & navigation period \\
$f$ & 30fps & frame rate \\
$N_f$ & 2700 & \# total frames ($N_f= Tf$) \\
%$t_e$ & 3s & request interval \\
$f_e$ & 90 & request interval in frames \\
$N_e$ & 30 & \# requested viewpoints ($N_e = N_f / f_e$) \\
$\tau_{\max}$ & 1s & system delay \\
$\mu$ & 0.05 & weight for storage cost \\
$\Delta$ & TBD & navigation speed \\
$p_n$ & TBD & view popularity \\
\hline
QP & 25 & quantization parameter for I-frames\tablefootnote{QP values for P-frames and depth maps are assigned automatically using the default settings of the MV-HEVC engine.} \\
$h_I^{(Q)}(Y_n)$ & \multirow{2}{*}{-} & encoding bit rates of I- and P-frames: \\
$ h_P^{(Q)}(Y_n|\hat{Y}_{n-1})$ &  & determined by the MV-HEVC engine \\
\hline
\end{tabular}
\end{center}
%\endgroup
\vspace{-10pt}
\end{table}

\begin{flushleft}
\textbf{Comparison algorithms}
\vspace{-5pt}
\end{flushleft}
In our experiments, we use ``NBPA'' (navigation ball and popularity-aware) to denote the proposed partitioning algorithm, and ``NBPU'' (navigation ball and popularity-unaware) to denote its popularity-unaware version where a uniform view popularity is assigned. 
We compare our algorithms with a baseline method that corresponds to a blind $N_K$-equidistant partitioning, where the $N_V$ camera views are equally divided into $N_K$ non-overlapping navigation segments (when $N_V$ is not divisible by $N_K$, a rounding is performed).
We consider two types of baseline method. 
The first one is denoted as ``Baseline'', which always uses fixed value of $N_K$ for different navigation configurations. The value of $N_K$ is determined by evaluating the proposed cost function in the partitioning problem of Eq. (\ref{eq:nv_prob_part}) with the baseline equidistant partitions. Because it does not consider the navigation configurations, we set $ \Delta = 0 $, and $ p_n $ to be uniform respectively.
%A full search is performed to find the value of $N_K$ that reveals the minimum function cost.
The second one is denoted as ``Baseline-NB'', which further considers the usage of navigation ball, and the value of $N_K$ is flexible for different values of $\Delta$. We determine $N_K$ by changing the value of $\Delta$ when we evaluate the cost function in Eq. (\ref{eq:nv_prob_part}) using the baseline equidistant partitions. 

\begin{figure}[tb]
 \centering
\begin{minipage}[htb]{.492\linewidth}
  \centering
  \centerline{\includegraphics[width=\linewidth]{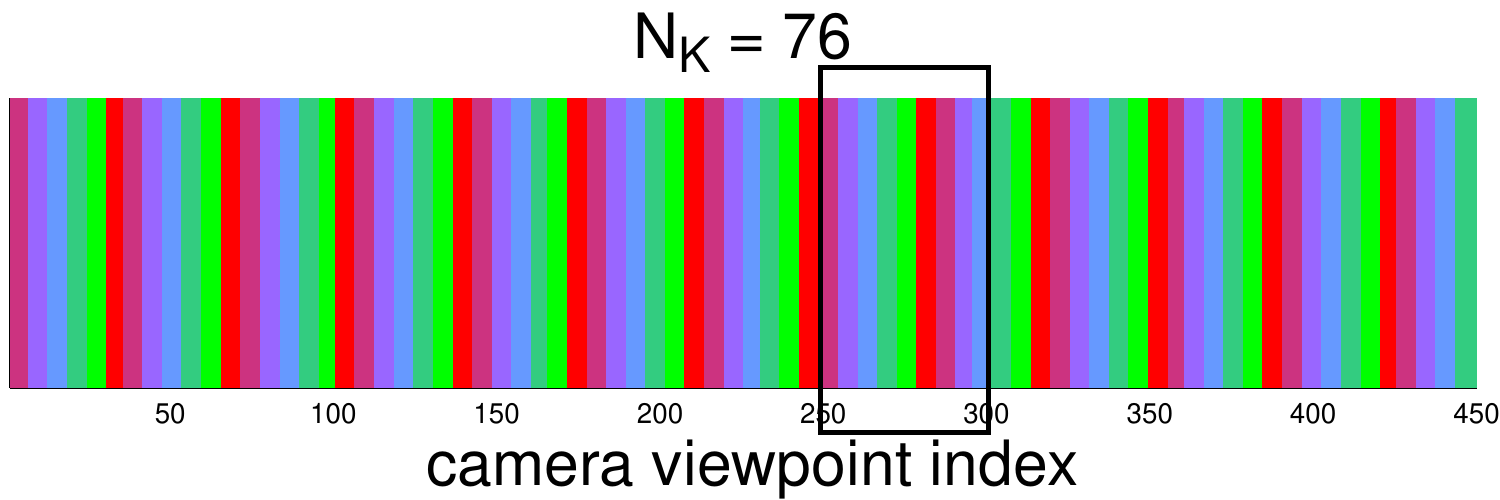}} 
 % \vspace{1.5cm}
  %\centerline{(a) } %\medskip
\end{minipage} %\hspace{10pt}
\begin{minipage}[htb]{.492\linewidth}
  \centering
  \centerline{\includegraphics[width=\linewidth]{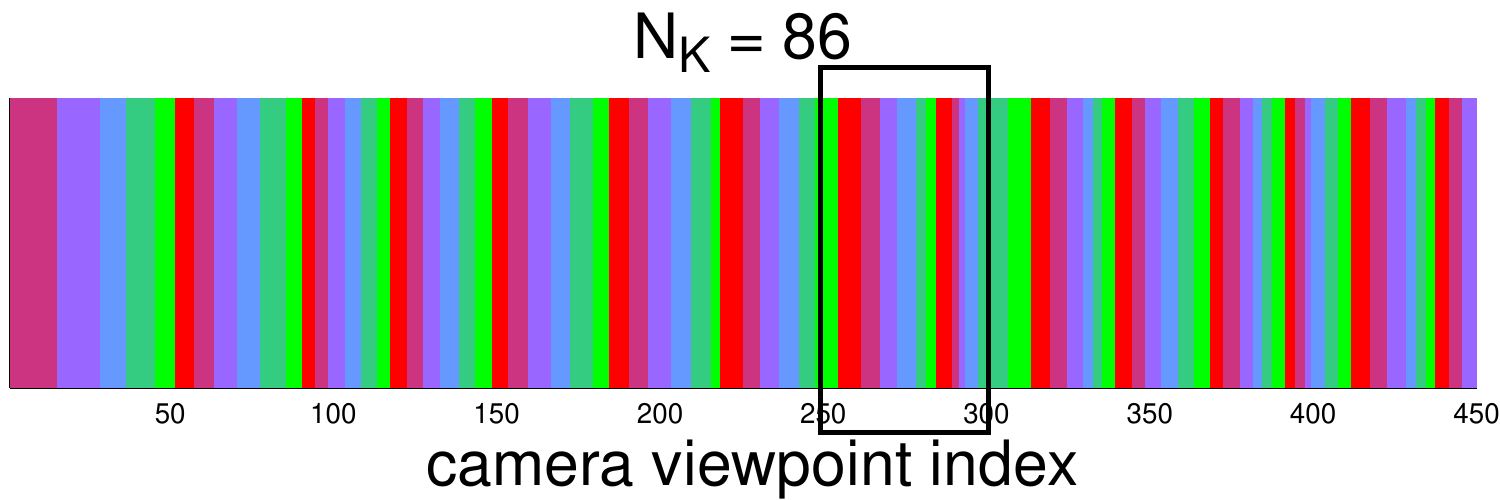}}
 % \vspace{1.5cm}
  %\centerline{(b) $\Delta = 0$} %\medskip
\end{minipage} \\
\vspace{8pt}
\begin{minipage}[htb]{.38\linewidth}
  \centering
  \centerline{\includegraphics[width=\linewidth]{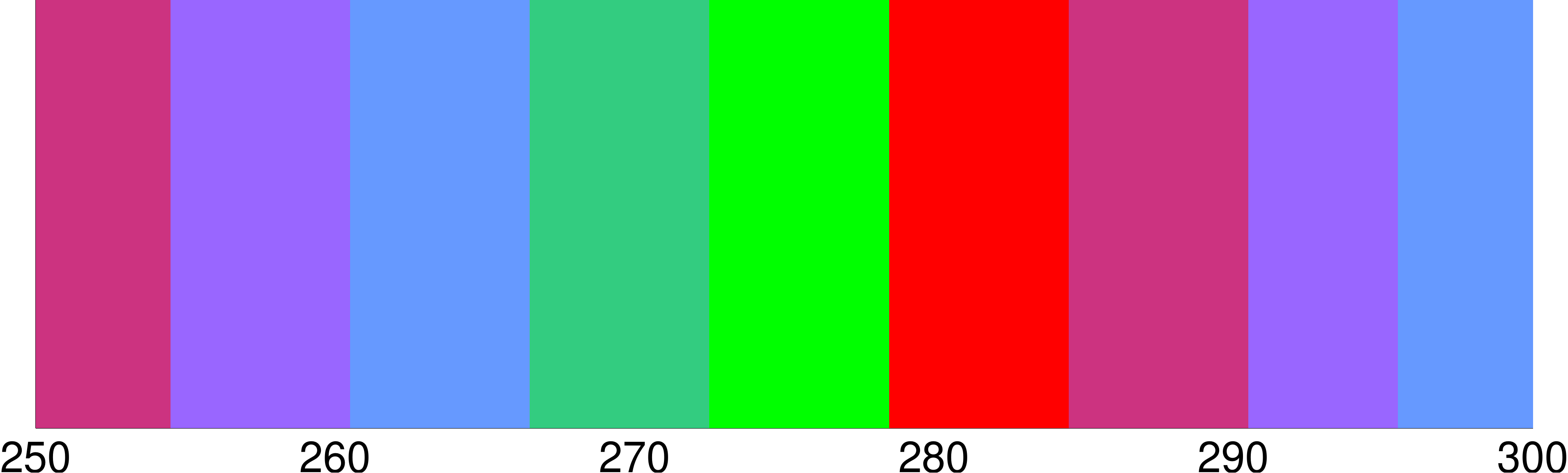}}
 % \vspace{1.5cm}
  \centerline{(a) baseline equidistant } %\medskip
\end{minipage} \hspace{25pt}
\begin{minipage}[htb]{.38\linewidth}
  \centering
  \centerline{\includegraphics[width=\linewidth]{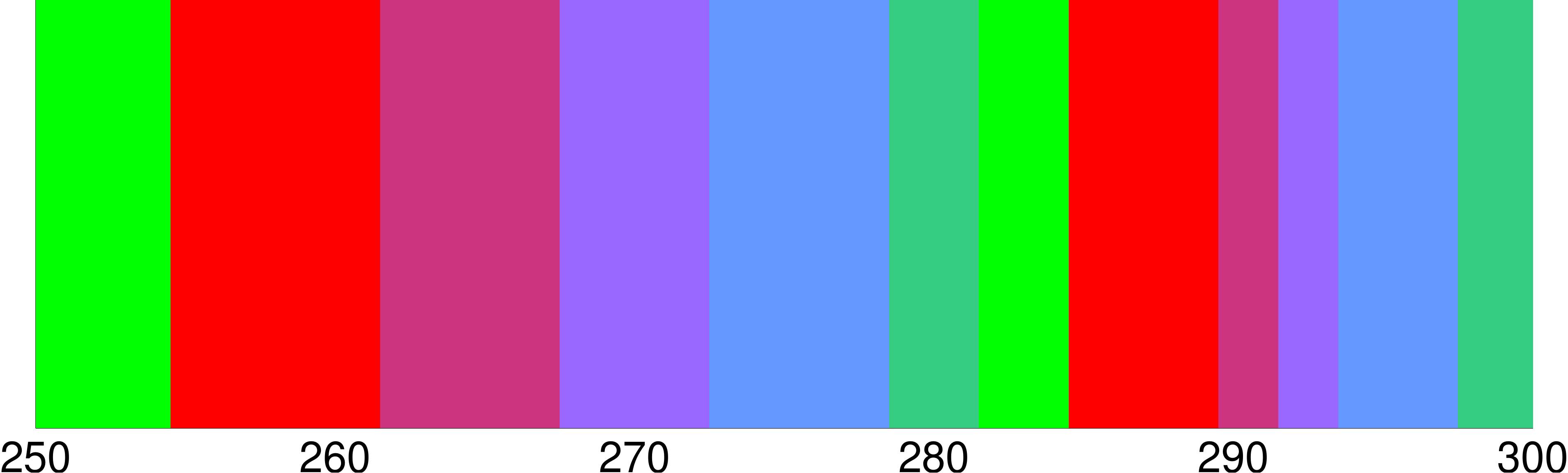}}
 % \vspace{1.5cm}
  \centerline{(b) NBPU, $ \Delta = 0 $ } %\medskip
\end{minipage}
\caption{ Partitions of navigation segments for different methods. Top: full partition patterns. Bottom: enlarged patterns for camera indices from $ 200 $ to $ 350 $. }
\label{fig:expr_naviseg_comp}
%\vspace{-5pt}
\end{figure}
\begin{table}[t]
\caption{ Width of navigation segments in proposed NBPU ($\Delta=0$) }
\label{tab:expr_naviseg_bitrates}
\begin{center}
\vspace{-10pt}
\begin{tabular}{|>{\centering\arraybackslash}m{2.2cm}||>{\centering\arraybackslash}m{.8cm}|>{\centering\arraybackslash}m{.8cm}|>{\centering\arraybackslash}m{.8cm}|>{\centering\arraybackslash}m{1cm}|}
\hline
& \multicolumn{4}{c|}{\textbf{Width of navigation segments}} \\
& \multicolumn{4}{c|}{\textbf{(in \# camera views)}} \\ \cline{2-5}
& mean & max & min & max/min \\ \hline
\textbf{Real encoding ~~~bit rates} & {5.23} & {15} & {2} & {7.50} \\ \hline
\textbf{Artificial constant encoding bit rates} & 6 & 6 & 6 & 1\\
\hline
\end{tabular}
%\vspace{-10pt}
\end{center}
\end{table}

\subsection{Partitioning Evaluation}
\label{sec:sub_expr_part}
We first evaluate only the partitioning results, where we compare different partitioning methods in terms of storage and rate costs under different navigation configurations of navigation speeds and view popularities.

\begin{flushleft}
\textbf{Irregular partitioning}
\vspace{-5pt}
\end{flushleft}
We compare the visual results of proposed method and the baseline method in Fig. \ref{fig:expr_naviseg_comp}. For fair comparison, we neglect the navigation speed ($\Delta=0$) and the view popularity (uniform distribution) for the propose method, i.e., we use NBPU with $\Delta=0$.
We align the camera views along the 1-D manifold and use color bars to represent different navigation segments.
It is clearly seen that, the proposed method provides an irregular partitioning compared to the baseline equidistant method. 
We further show the corresponding width of the navigation segments (in number of camera views) in Table \ref{tab:expr_naviseg_bitrates}, and we observe that the widths of the navigation segments are quite unbalanced with a $7.5$ max/min ratio.
A comparison experiment is conducted using constant encoding bit rates, where we first manually assign two sets of artificial constant bit rates to I- and P-frames respectively, and then we run the proposed partitioning algorithm.
We provide this result in the bottom row of Table \ref{tab:expr_naviseg_bitrates}. We observe that all navigation segments turn to have the same width and the proposed partitioning method is actually degraded to the baseline $N_K$-equidistance partitioning.
It indicate that the unbalanced irregular partitions of navigation segments is due to the variation of the encoding bit rates of I- and P-frames. 

\begin{figure}[tb]
 \centering
\begin{minipage}[htb]{.492\linewidth}
  \centering
  \centerline{\includegraphics[width=\linewidth]{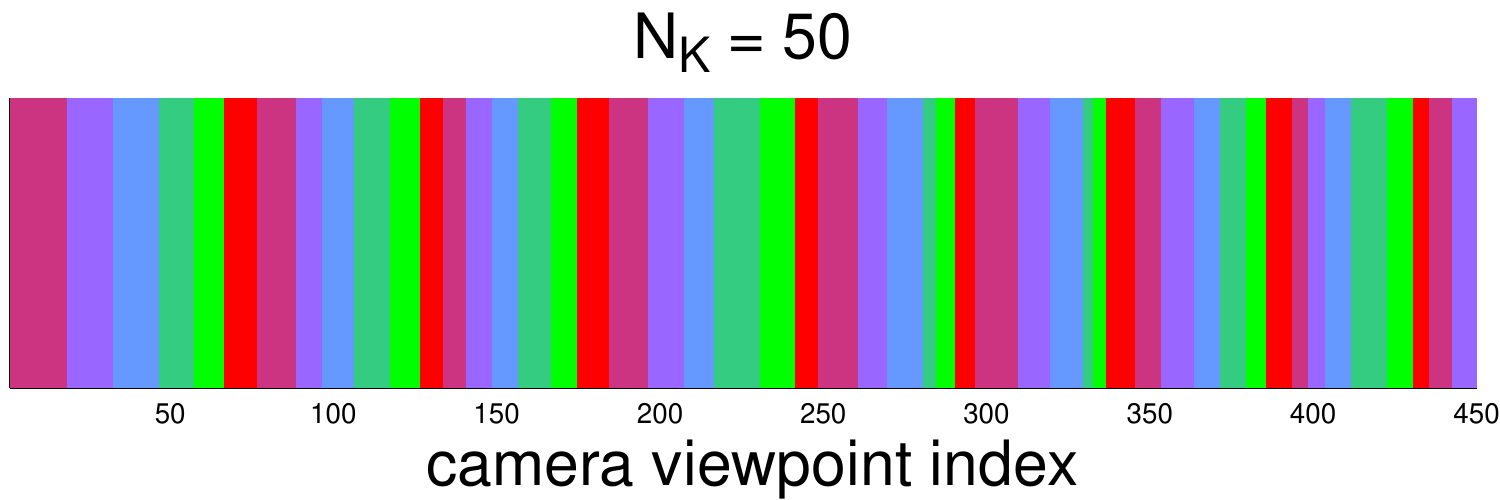}}
 % \vspace{1.5cm}
  \centerline{ (a) NBPU, $\Delta = 5$ } %\medskip
\end{minipage} %\hspace{10pt}
\begin{minipage}[htb]{.492\linewidth}
  \centering
  \centerline{\includegraphics[width=\linewidth]{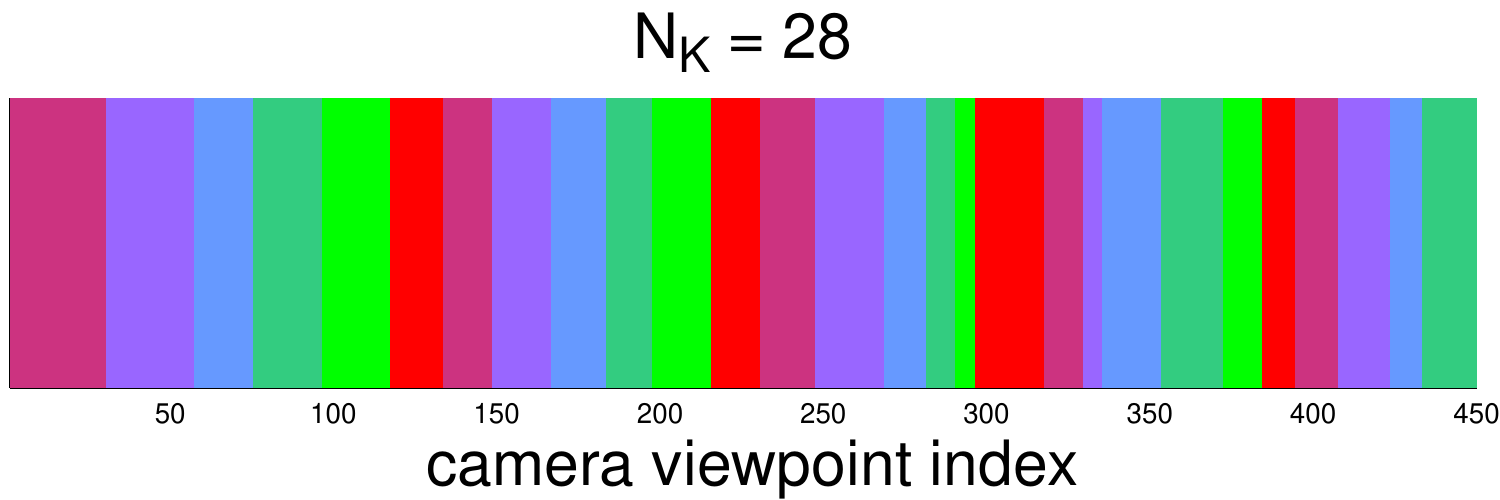}}
 % \vspace{1.5cm}
  \centerline{ (b) NBPU, $\Delta = 20$ } %\medskip
\end{minipage}
\caption{ Partitions of navigation segments for proposed NBPU with different navigation speeds. }
\label{fig:expr_part_speed}
%\vspace{-5pt}
\end{figure}
\begin{figure}[t]
 \centering
\begin{minipage}[htb]{.75\linewidth}
  \centering
  \centerline{\includegraphics[width=\linewidth, height=.62\linewidth]{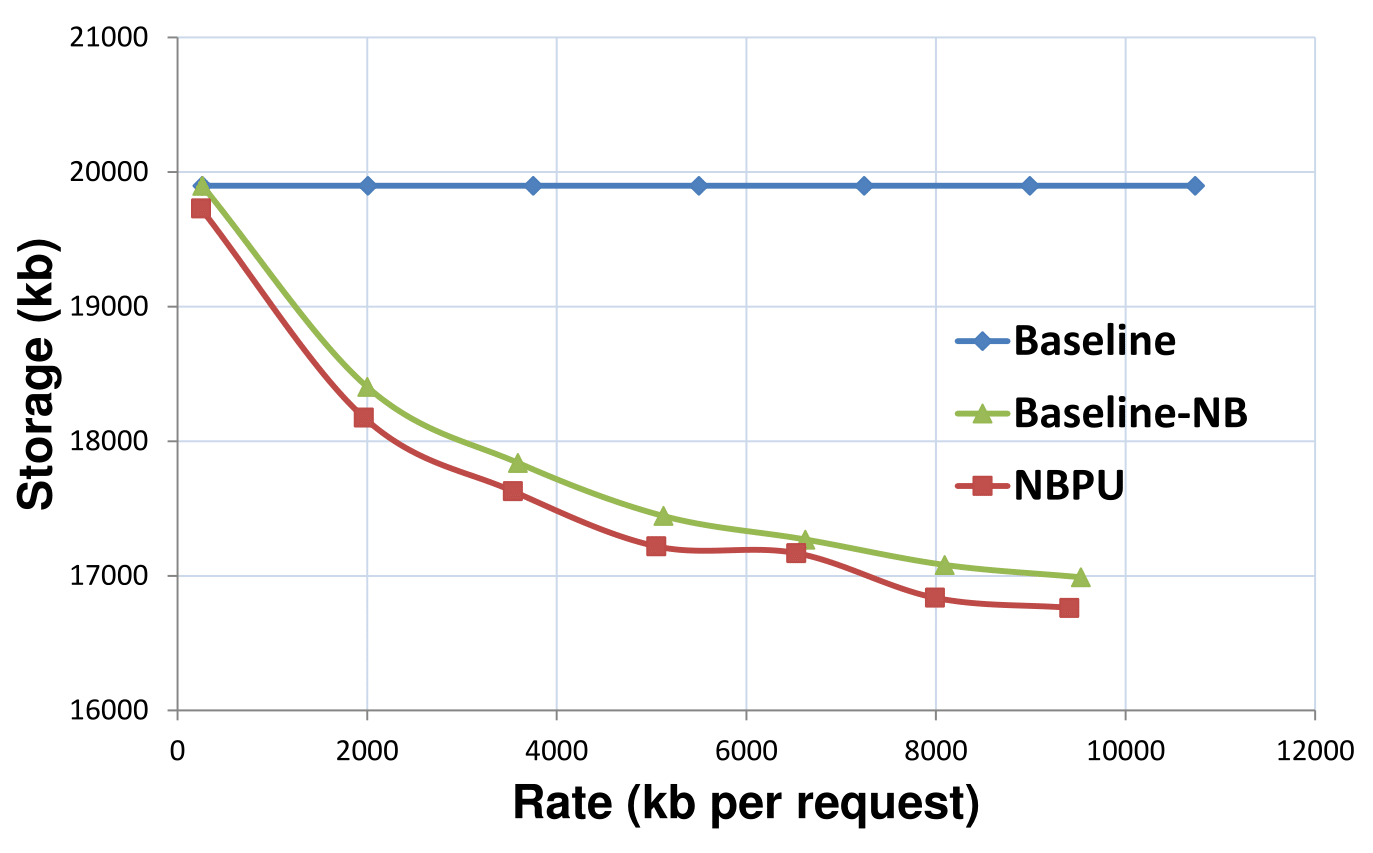}}
  \centerline{  } %\medskip
\end{minipage}
\caption{ Rate-storage curves of different partitioning methods under various navigation speeds for a uniform view popularity. }
\label{fig:expr_part_costcomp}
%\vspace{-5pt}
\end{figure}
\begin{table*}[t]
\caption{ Performance comparison between proposed NBPU and the baseline method under various navigation speeds for a uniform view popularity }
\label{tab:expr_part_cost_table}
 \centering
\begin{minipage}[tb]{.7\linewidth}
  \centering
  \centerline{\includegraphics[width=\linewidth]{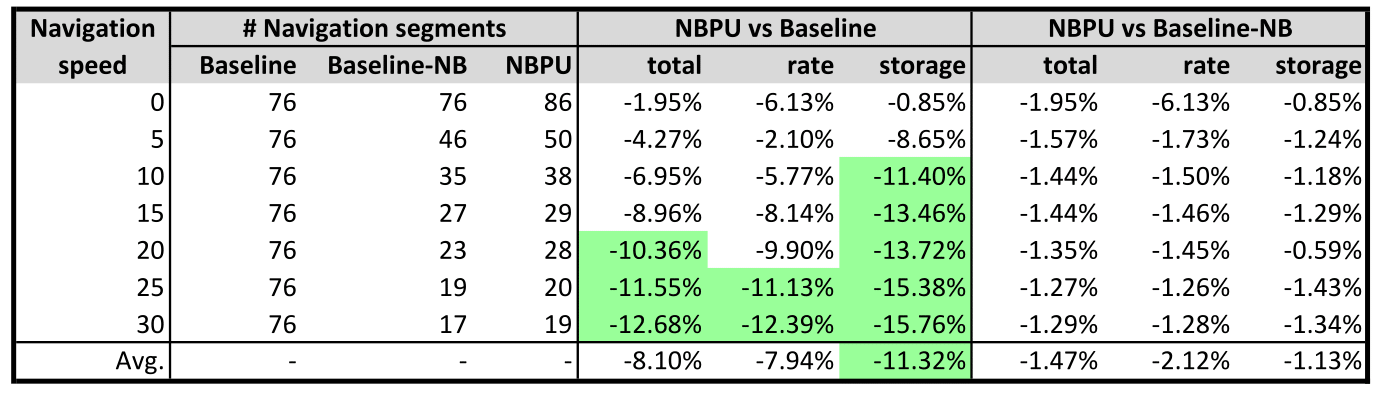}}
\end{minipage}
%
%\vspace{-10pt}
\end{table*}

\begin{flushleft}
\textbf{Influence of navigation speed}
\vspace{-5pt}
\end{flushleft}
We study next the influence of the navigation speed $\Delta$ on the navigation segment partitioning, and we use NBPU as we only focus on the navigation speed and we neglect the effect of the view popularity.
Since the cameras in our dataset are almost equally placed along the 1-D manifold, the distances between neighboring cameras are similar so that they are denoted by the unit distance $1$ for simplicity, and we measure $\Delta$ accordingly.

Fig. \ref{fig:expr_part_speed} illustrates the partitions of navigation segments of our NBPU under different values of $\Delta$.
It is seen that, as $ \Delta $ increases, the navigation segments grow wider and sparser.
This is because, with a larger $ \Delta $, more camera views are requested to support the view rendering within the increasing navigation ball, and therefore it becomes more efficient to compress and transmit a larger number of camera views together. As a result, the navigation segments tend to contain more camera views and therefore become wider.

Table \ref{tab:expr_part_cost_table} summarizes the comparison results of our NBPU and the baseline method, where we gradually increase $ \Delta $ and compute the relative reduction of the
partitioning function cost (``total''), the rate cost (``rate'') and the storage cost (``storage'') respectively according to Eq. (\ref{eq:nv_prob_part}). We highlight large performance improvements in green.
It is observed that our NBPU outperforms the baseline methods in all aspects. Compared to the Baseline with fixed value of $N_K$ , our NBPU achieves larger rate and storage cost reductions as $ \Delta $ increases. This shows the effectiveness of considering the influence of navigation speed. When compared to Baseline-NB, the reduction is less as expected, because Baseline-NB is more flexible as $N_K$ is further optimized for different values of $ \Delta $. The cost reduction against Baseline-NB is due to the irregular partitions of NBPU.

We also plot the corresponding rate-storage curves in Fig. \ref{fig:expr_part_costcomp}.
The navigation speed $\Delta$ increases from left to right. For all methods, the rate costs increase as $ \Delta $ grows, because a larger $ \Delta $ indicates a larger size of navigation balls and therefore more data is transmitted for data buffering at each request. 
The storage cost of Baseline is constant due to its fixed partition pattern. The storage costs of the other methods decrease as $ \Delta $ grows, because a larger $ \Delta $ leads to wider navigation segments and therefore more redundancies between camera views can be exploited and eliminated when encoding wider segments. 
In this figure, the gap between Baseline and Baseline-NB shows the cost reduction obtained by adapting the segment partitions to the navigation speed, and the gap between
Baseline-NB and NBPU shows the additional cost reduction induced by applying the irregular partitions. 

\begin{figure}[t]
 \centering
\begin{minipage}[htb]{.492\linewidth}
  \centering
  \centerline{\includegraphics[width=\linewidth]{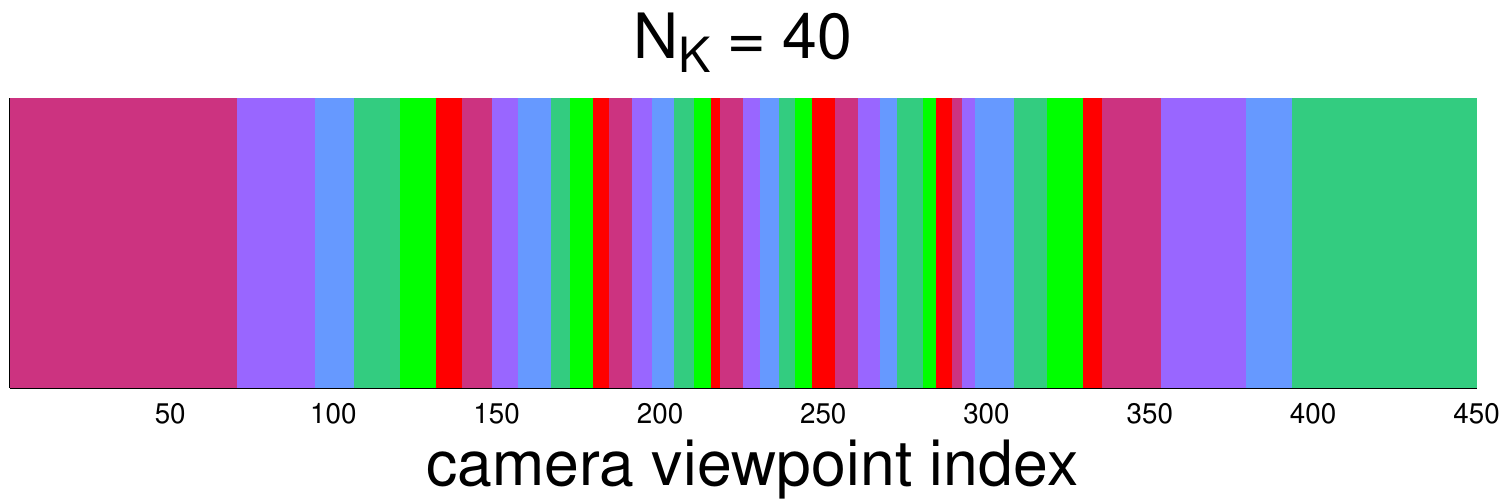}}
 % \vspace{1.5cm}
 %\centerline{ (a) center view preferred, $ \Delta = 5 $ } %\medskip
\end{minipage} %\hspace{10pt}
\begin{minipage}[htb]{.492\linewidth}
  \centering
  \centerline{\includegraphics[width=\linewidth]{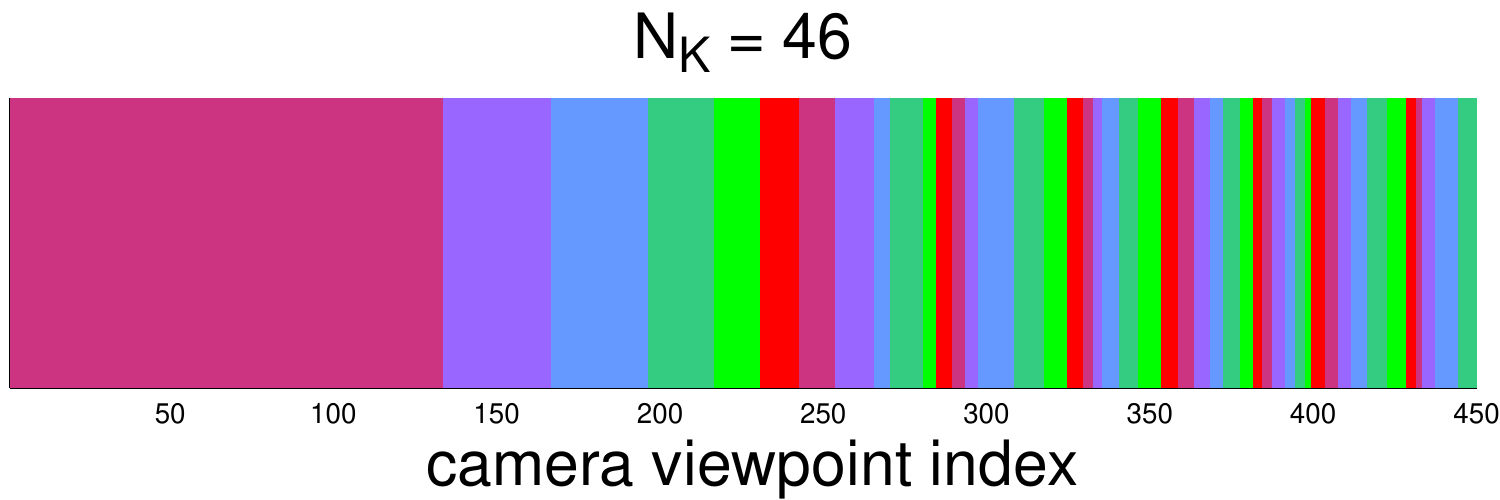}}
 % \vspace{1.5cm}
 %\centerline{ (b) center view preferred, $ \Delta = 20 $ } %\medskip
\end{minipage} \\ \vspace{5pt}
\begin{minipage}[htb]{.492\linewidth}
  \centering
  \centerline{\includegraphics[width=\linewidth]{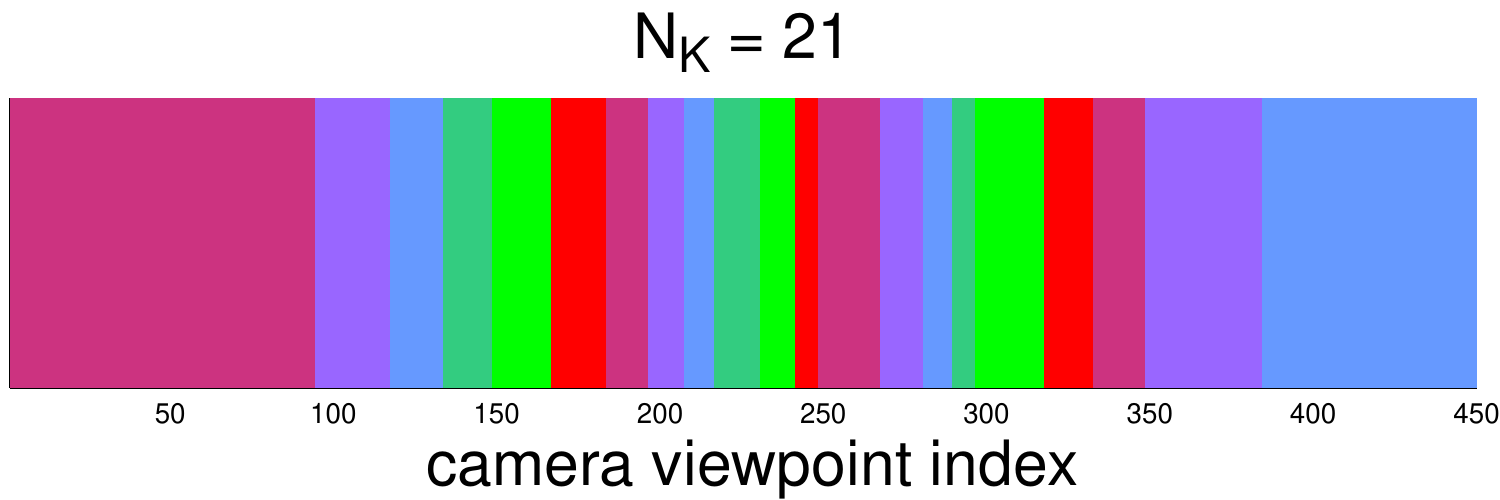}}
 % \vspace{.5cm}
 \centerline{ (a) center view preferred } %\medskip
\end{minipage} %\hspace{10pt}
\begin{minipage}[htb]{.492\linewidth}
  \centering
  \centerline{\includegraphics[width=\linewidth]{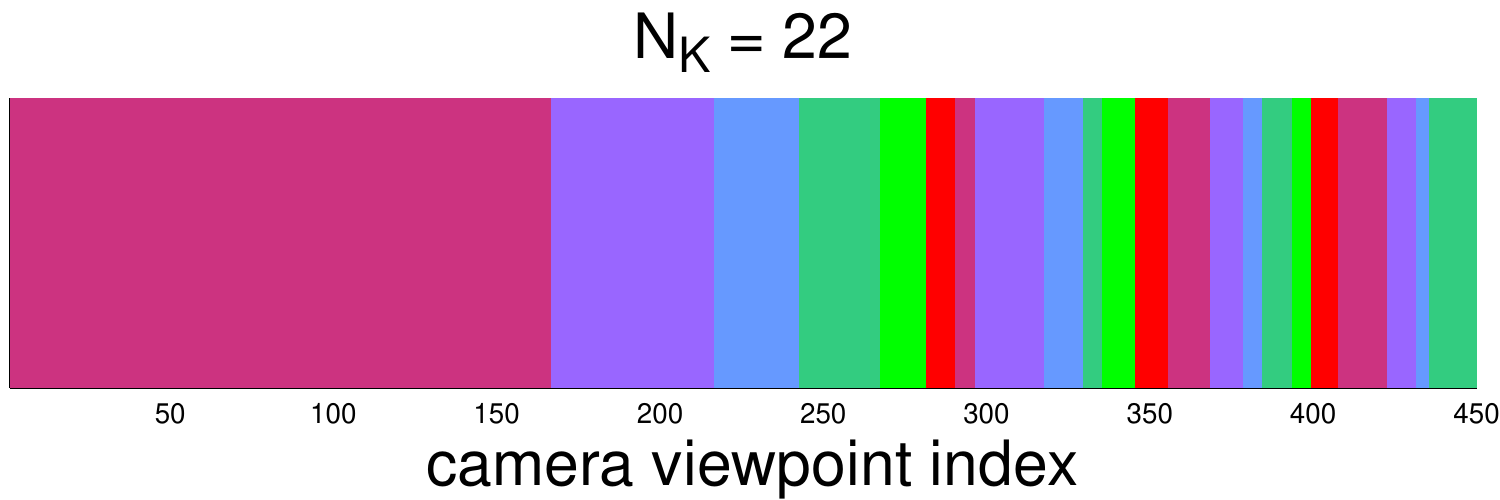}}
 % \vspace{.5cm}
 \centerline{ (b) right view preferred } %\medskip
\end{minipage}
\caption{ Partitions of navigation segments of proposed NBPA under pre-defined view popularity distributions at different navigation speeds. Top: low speed $ \Delta = 5 $. Down: high speed $ \Delta = 20 $.  }
\label{fig:expr_part_viewpop}
%\vspace{-5pt}
\end{figure}
\begin{table*}[tb]
\caption{ Performance comparison of different partitioning methods under various navigation speeds and pre-defined view popularity distributions. }
\label{tab:expr_part_cost_table_viewpop}
 \centering
\begin{minipage}[htb]{.88\linewidth}
  \centering
  \centerline{\includegraphics[width=\linewidth]{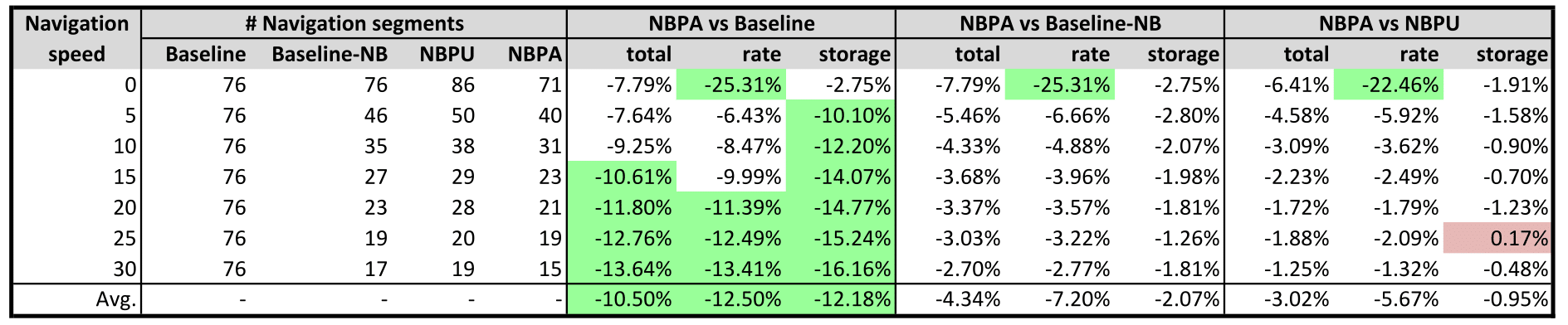}}
  \centerline{ (a) center view preferred (see Fig. \ref{fig:expr_part_viewpop}a) }
\end{minipage} \\
\begin{minipage}[htb]{.88\linewidth}
  \centering
  \centerline{ \hspace{2pt} \includegraphics[width=\linewidth]{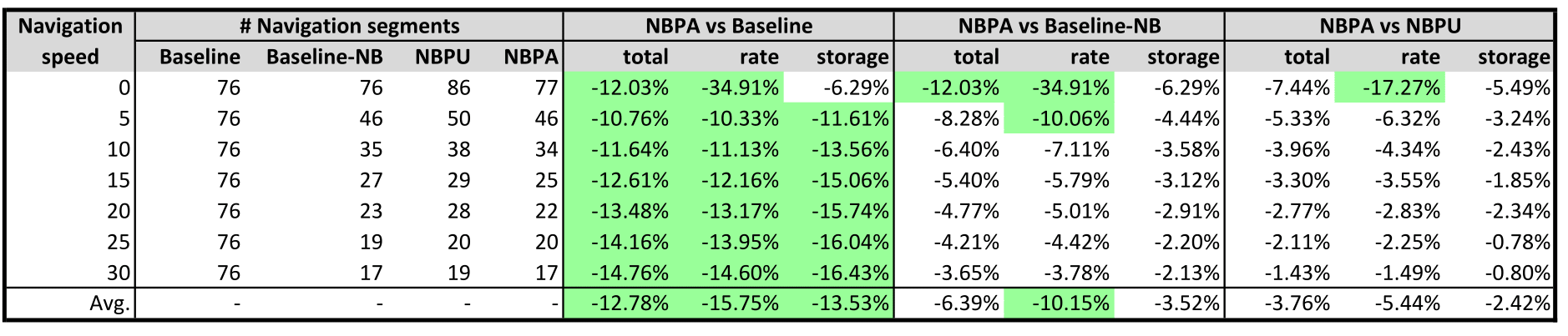}}
  \centerline{ (b) right view preferred (see Fig. \ref{fig:expr_part_viewpop}b)}
\end{minipage}
\vspace{-5pt}
\end{table*}

\begin{flushleft}
\textbf{Influence of view popularity}
\vspace{-5pt}
\end{flushleft}
We now study the influence of the view popularity that denotes the probability of each camera view being requested for view rendering during user navigation. It is generally non-uniformly distributed in practice due to the user preferences.
Fig. \ref{fig:expr_part_viewpop} illustrates the segment partitions of our NBPA under several pre-defined view popularity distributions.
A clear adaptation to the view popularity can be observed, where popular camera views have finer partitions and unpopular views have coarser ones.
Note that the finer partitions generally lead to lower transmission rate and higher navigation interactivity due to the less data dependencies between camera views.
The popular camera views are requested more often during navigation and consequently occupy a larger percentage of the transmission rate. Therefore it is more efficient to reduce the overall transmission rate by applying a finer partitioning to these camera views rather than the unpopular ones.

We present the performance of different partitioning methods under various view popularity distributions in Table \ref{tab:expr_part_cost_table_viewpop}, where the green box indicates large performance improvement and the red box indicates performance drop.
We first observe that the popularity-aware NBPA provides significant rate and storage cost reduction against the popularity-unaware NBPU when the navigation speed is low (more than $17\%$ rate reduction for $\Delta=0$). However, as $ \Delta $ increases, the performance gap becomes small.
This is because, for large $ \Delta $, the partitions become coarser and the partition patterns under different view popularity distributions become more similar to each other, leading to a closer performance between them.
Compare with the baseline methods, our NBPA achieves a further rate and storage reduction for non-uniform view popularities (referring to Table \ref{tab:expr_part_cost_table} for the uniform popularity).
These results indicate that, by adapting the segment partitions to the view popularity, the proposed method is able to provide different levels of navigation interactivity for camera views and further reduce the rate and storage consumptions of the system.

\begin{figure}[tb]
 \centering
\begin{minipage}[htb]{.75\linewidth}
  \centering
  \centerline{\includegraphics[width=\linewidth, height=.64\linewidth]{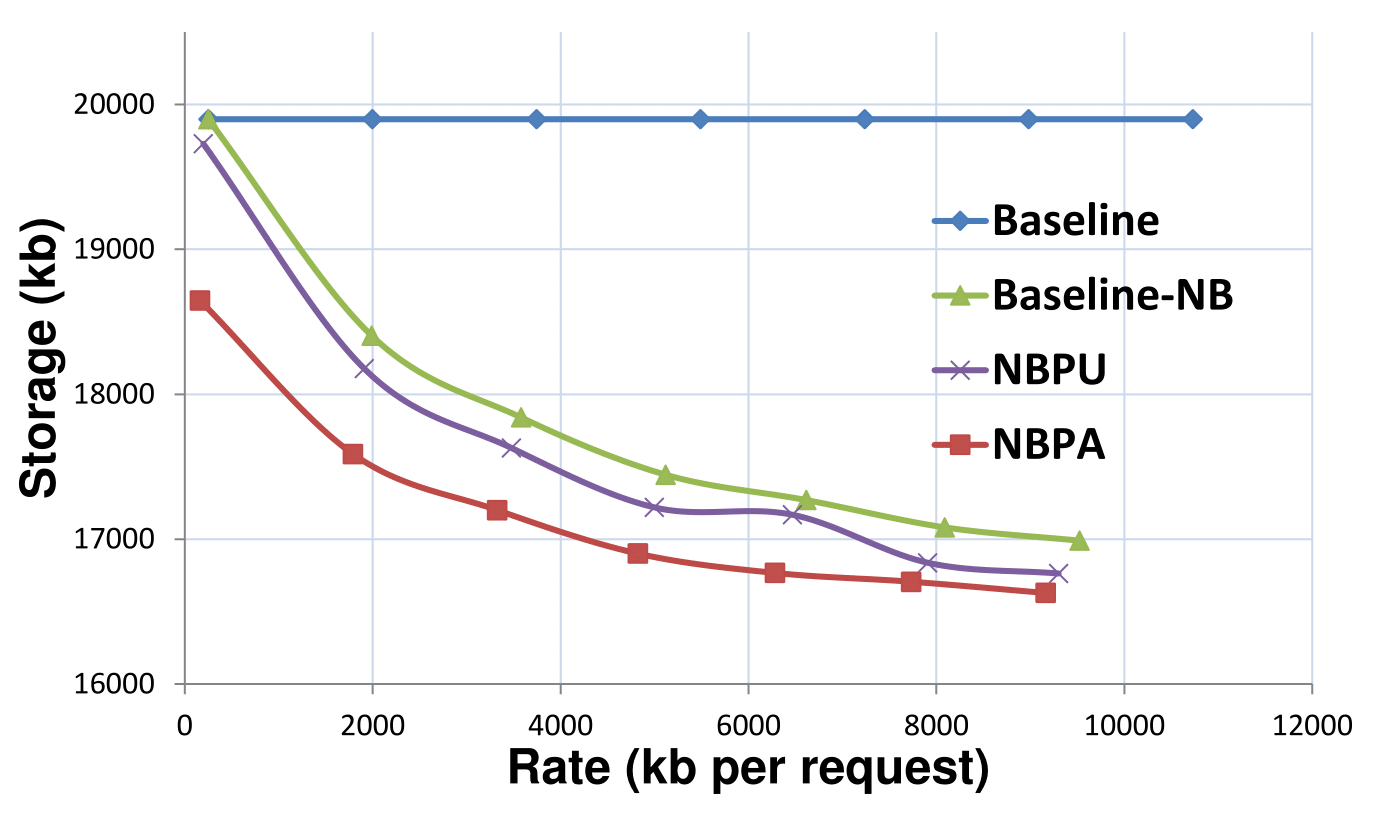}}
 % \vspace{1.5cm}
  \centerline{  } %\medskip
\end{minipage}
\caption{ Rate-storage curves of different partitioning methods under various navigation speeds for a popularity distribution that the right views are preferred. }
\label{fig:expr_part_costcomp_viewpop=r}
\vspace{-10pt}
\end{figure}

We further visualize the corresponding rate-storage curves in Fig. \ref{fig:expr_part_costcomp_viewpop=r} for the popularity distribution that the right views are preferred. The navigation speed $ \Delta $ increases from left to right.  The gap between NBPA and NBPU shows the additional rate and storage cost reductions of considering the influence of view popularity for the segment partitions. 

\begin{flushleft}
\textbf{Influence of other navigation parameters}
\vspace{-5pt}
\end{flushleft}
The partition of navigation segment is also influenced by other navigation parameters. We briefly analyze them here.

\begin{figure}[t]
 \centering
\begin{minipage}[htb]{.49\linewidth}
  \centering
  \centerline{\includegraphics[width=\linewidth]{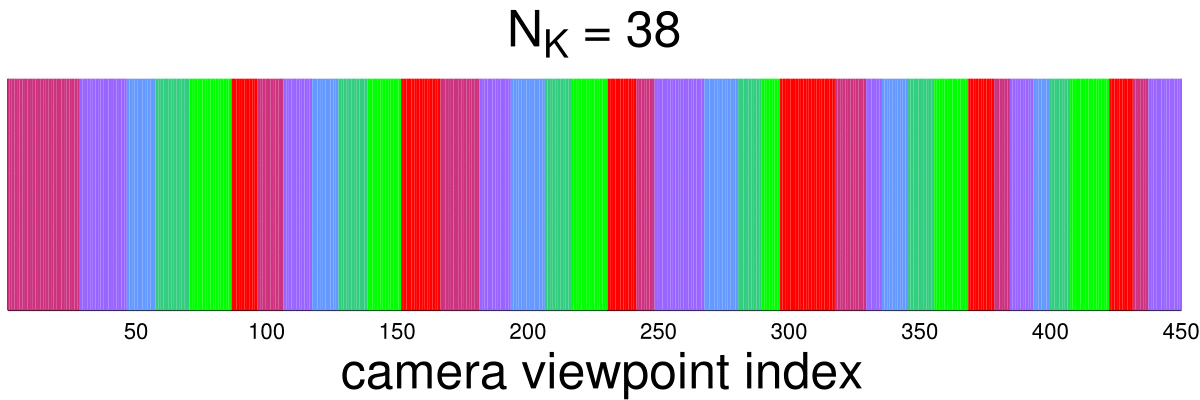}}
 % \vspace{1.5cm}
  \centerline{ (a) $\mu = 0.05$ } %\medskip
\end{minipage}
%
%\hspace{5pt}
\begin{minipage}[htb]{.49\linewidth}
  \centering
  \centerline{\includegraphics[width=\linewidth]{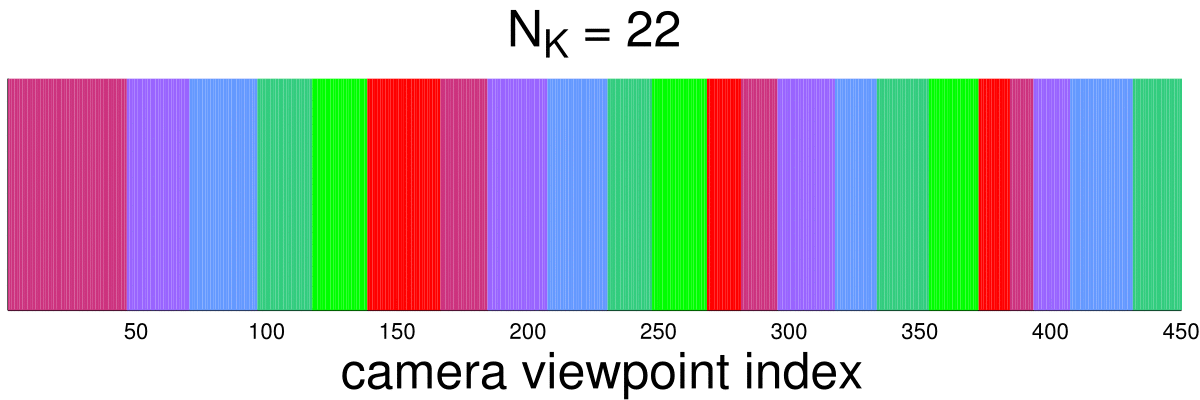}}
 % \vspace{1.5cm}
  \centerline{ (b) $\mu = 0.5$ } %\medskip
\end{minipage} \\
\caption{ Partitions of navigation segments for different $ \mu $ (NBPU with $\Delta = 10$). }
\label{fig:expr_part_mu}
%\vspace{-5pt}
\end{figure}
\begin{figure}[tb]
 \centering
\begin{minipage}[htb]{.75\linewidth}
  \centering
  \centerline{\includegraphics[width=\linewidth]{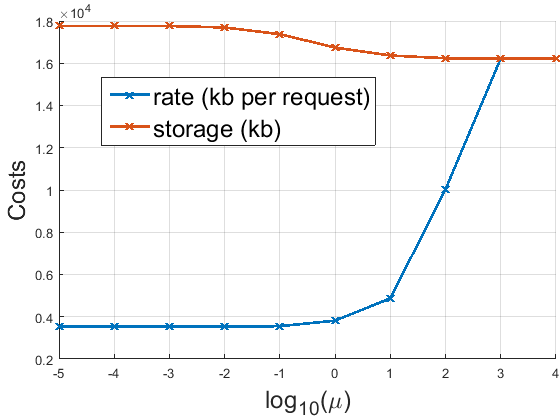}}
 % \vspace{1.5cm}
  \centerline{  } %\medskip
\end{minipage}
\caption{ Evolution of rate and storage costs versus storage weight $ \mu $ (NBPU with $\Delta = 10$). }
\label{fig:expr_rs_mu}
%\vspace{-5pt}
\end{figure}
\begin{figure}[t]
 \centering
\begin{minipage}[htb]{.85\linewidth}
  \centering
  \centerline{\includegraphics[width=\linewidth]{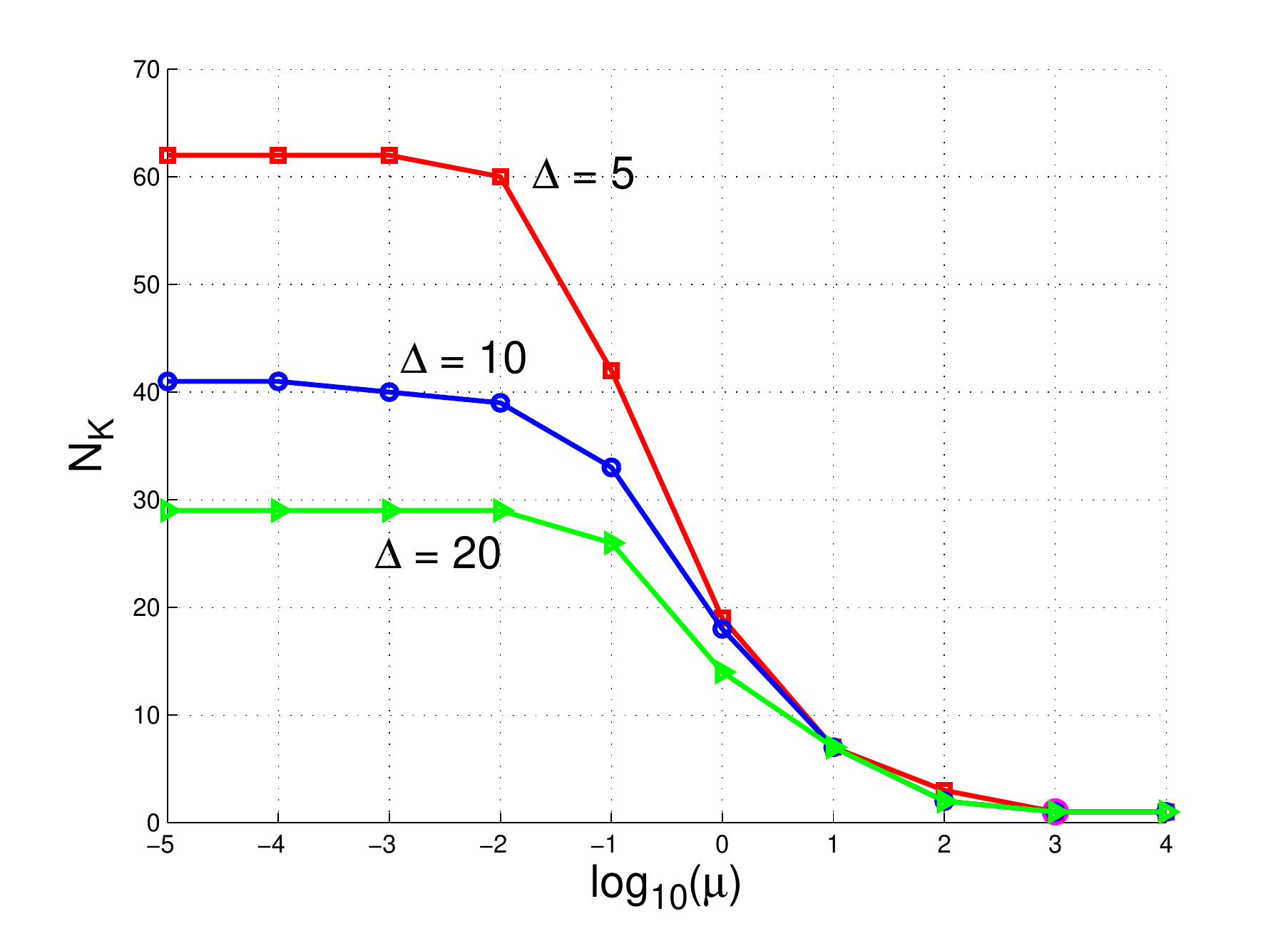}}
 % \vspace{1.5cm}
 % \centerline{ } %\medskip
\end{minipage}
\caption{ Evolution of the number of segments $N_K$ versus storage weight $ \mu $ (NBPU with different value of $\Delta$). }
\label{fig:expr_evol_mu}
%\vspace{10pt}
\end{figure}

\emph{\textbf{The storage weight $\mu$}} influence the relative weight between rate and storage costs, which further changes the optimal segment partitions. 
Fig. \ref{fig:expr_part_mu} shows the partition pattern of proposed NBPU for different $\mu$. As $\mu$ increases, more weight is given to the storage, and the segments become wider; wider segments increase the compression efficiency and decrease the storage cost. 
On the other hand, wider segments increase the transmission rate. This evolution of rate and storage costs is plotted in Fig. \ref{fig:expr_rs_mu} for the proposed NBPU. 
We also shows the evolution of the number of segments $ N_K $ versus $ \mu $ in Fig. \ref{fig:expr_evol_mu}.
In one extreme, as $\mu$ goes to infinity, the storage cost dominates and the entire multiview data is always represented by a single navigation segment, which is the classical multiview compression problem targeting for the best compression performance. In another extreme, as $\mu$ decreases to $0$, the system purely minimizes the transmission rate and does not consider the storage, and the partitions of navigation segments then converge to a certain pattern with the maximum $N_K$. 
Note that this pattern is different for different navigation speeds, and the width of navigation segment in this pattern does not shrink to $1$ because of the navigation ball with non-zero navigation speed.
Different values of $ \mu $ provide different combinations of rate and storage costs. The proper choice of $ \mu $ depends on whether the storage or the bandwidth resource is more limited in practical navigation scenarios.

\begin{table*}[tb]
\caption{ RD performance comparison of different methods averaged over 100 simulated navigation paths along camera viewpoints }
\label{tab:expr_part+strm_bd_cam}
 \centering
\begin{minipage}[htb]{.95\linewidth}
  \centering
  \centerline{\includegraphics[width=\linewidth]{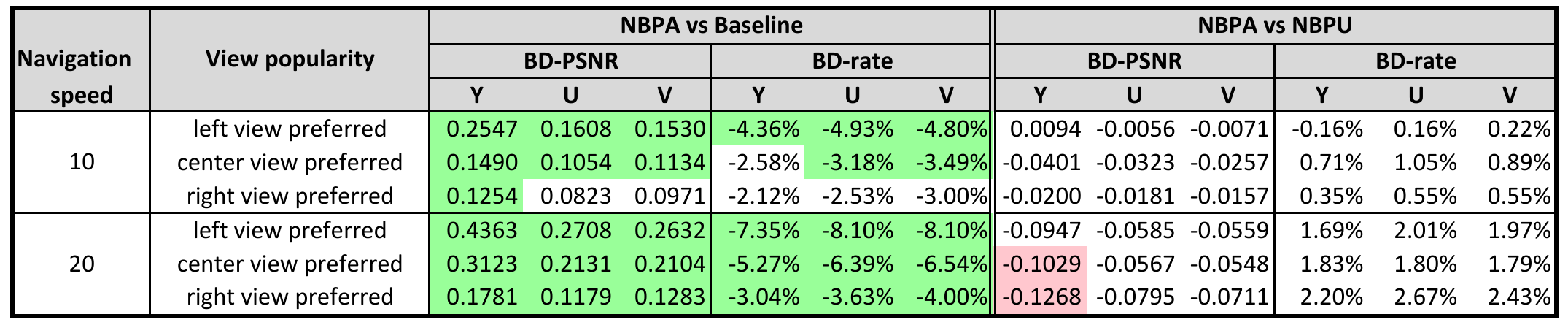}}
  % \centerline{ (a) center view preferred }
\end{minipage}
%\vspace{10pt}
\end{table*}

\begin{figure}[tb]
\vspace{-1pt}
 \centering
\begin{minipage}[htb]{.485\linewidth}
  \centering
  \centerline{\includegraphics[width=\linewidth]{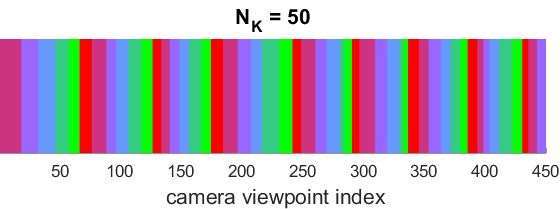}}
 % \vspace{1.5cm}
  \centerline{ (a) $f_e = 30$ } %\medskip
\end{minipage} %\hspace{10pt}
\begin{minipage}[htb]{.485\linewidth}
  \centering
  \centerline{\includegraphics[width=\linewidth]{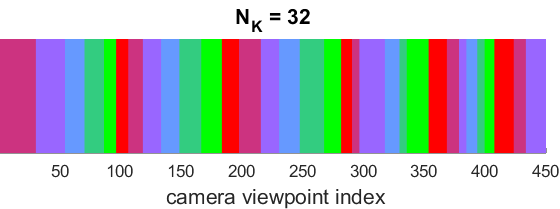}}
 % \vspace{1.5cm}
  \centerline{ (b) $f_e = 120$ } %\medskip
\end{minipage}
\caption{ Partitions of navigation segments for different $ f_e $ (NBPU with $\Delta = 10$). }
\label{fig:expr_part_fe}
%\vspace{-5pt}
\end{figure}

\begin{figure}[tb]
 \centering
\begin{minipage}[htb]{.75\linewidth}
  \centering
  \centerline{\includegraphics[width=\linewidth]{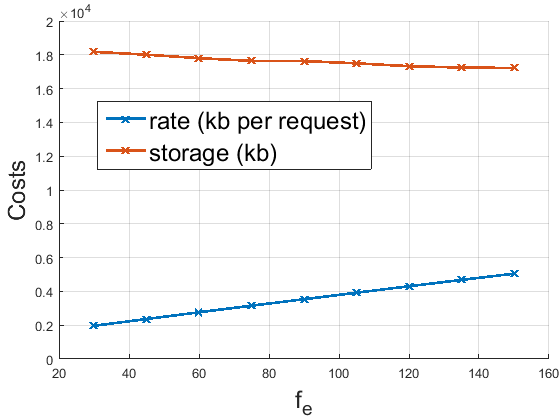}}
 % \vspace{1.5cm}
  \centerline{  } %\medskip
\end{minipage}
\caption{ Evolution of rate and storage costs versus $ f_e $ (NBPU with $\Delta = 10$). }
\label{fig:expr_rs_fe}
\vspace{-5pt}
\end{figure}

\emph{\textbf{The request interval $f_e$}} changes the size of the navigation ball as indicated in Eq. (\ref{eq:opt_t}), which further influences the partition of segments.
Fig. \ref{fig:expr_part_fe} illustrates the partition pattern of NBPU for different values of $f_e$. Similarly to the navigation speed $\Delta$, a larger $f_e$ leads to a larger size of navigation balls, which further results in wider navigation segments. 
Fig. \ref{fig:expr_rs_fe} plots the rate and storage costs versus $f_e$. Similar trend of rate and storage costs is observed due to the growing size of navigation segments as $f_e$ increases.
The choice of $f_e$ in practical navigation systems depends on the storage and bandwidth limitations but also on the system delay. As long as the storage and bandwidth capacities allow, a larger $f_e$ is preferred because it leads to wider navigation segments and increases the efficiency of compression and transmission. On the other hand, more data is transmitted at each request for the larger $f_e$, which however increases the system delay for transmitting and processing the data. Therefore, $f_e$ can not be arbitrarily large in order to enable low-delay navigation. 


\subsection{Complete System Evaluation}
\label{subsec:eval_part+strm}
We now evaluate the performance of the complete system, i.e. a joint evaluation of partition and allocation of the navigation segments, where we further apply the allocation solution in Eq. (\ref{eq:nv_comple_alloc_sol}) for each data request based on the different partitioning methods discussed above.

\begin{figure}[tb]
 \centering
\begin{minipage}[htb]{.85\linewidth}
  \centering
  \centerline{\includegraphics[width=\linewidth]{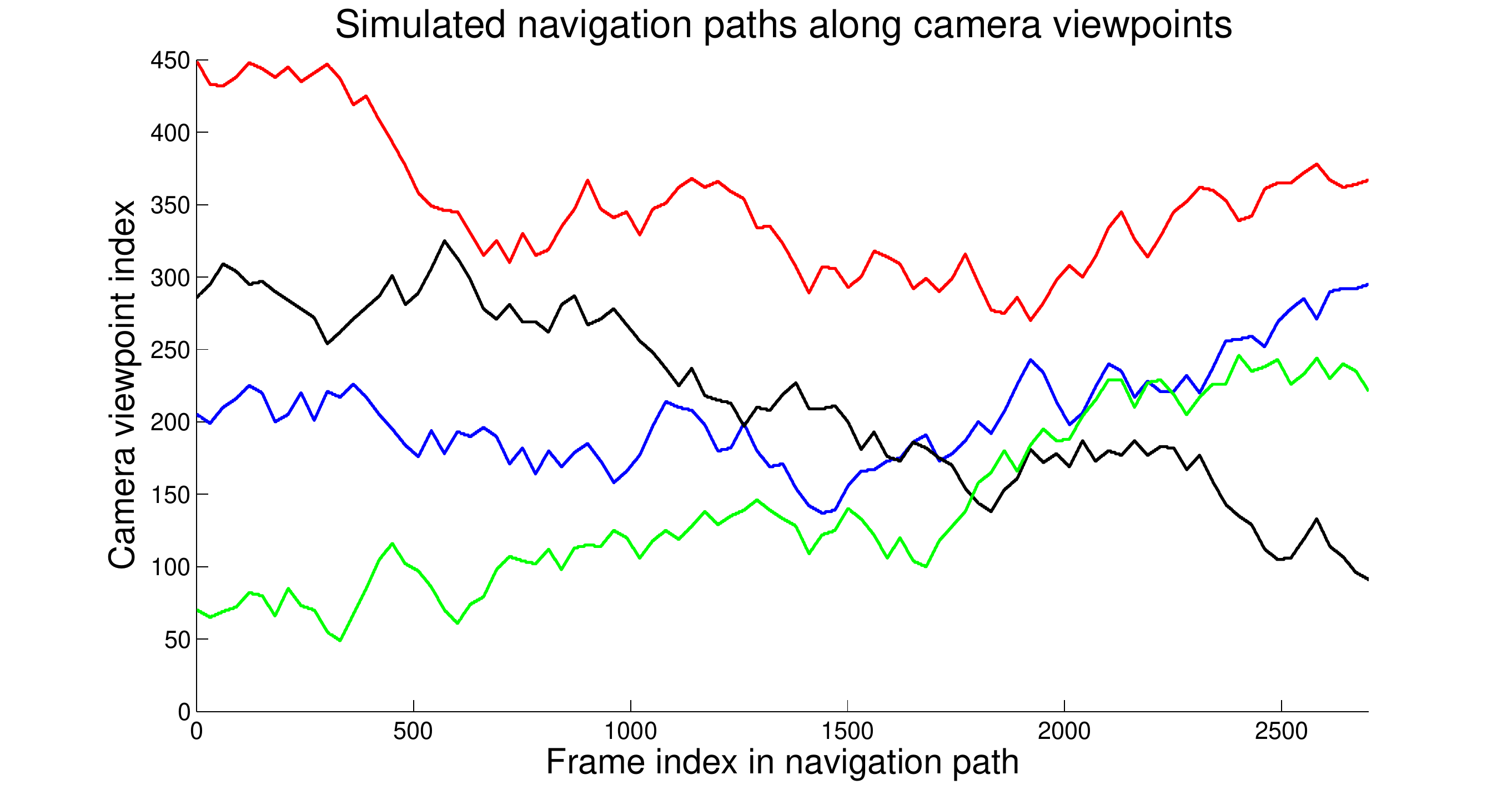}}
  % \vspace{1.5cm}
  % \centerline{  } %\medskip
\end{minipage}
\caption{Simulated navigation paths along camera viewpoints (pre-defined view popularity: center view preferred, navigation speed: $\Delta = 10$). The colors represent different navigation paths.}
\label{fig:expr_sim_navpath_cam}
%\vspace{10pt}
\end{figure}

\begin{flushleft}
\textbf{Navigation paths along real views}
\vspace{-5pt}
\end{flushleft}
We first carry out the experiments on navigation paths composed of real camera viewpoints, where the ground-truth images are always available and we can compute the distortion directly. 
In our experiment, we generate simulated navigation paths from a pre-defined view popularity distribution. In particular, the first camera viewpoint in the navigation path is generated based on this distribution. In order to choose the next viewpoint, we build a navigation ball with navigation speed $\Delta$ centred at the previous viewpoint, and then randomly pick up a viewpoint from the normalized view popularity distribution within the navigation ball. This process is repeated until the last viewpoint in the navigation path is reached. Then interpolation is applied between consecutive viewpoints according to the frame rate.
Fig. \ref{fig:expr_sim_navpath_cam} shows the simulated navigation paths along camera viewpoints in our experiments.
We further fix $ t_S = t^\star $ in the allocation solution Eq. (\ref{eq:nv_comple_alloc_sol}) to avoid the rendering of camera views caused by insufficient data, and therefore the distortion is uniquely affected by quantization. In this case, the allocation solution is indeed $ S_0 $.
We then adjust the QP value in order to derive the rate-distortion (RD) curves. Four QP values $\{25, 30, 35, 40\}$ for I-frame are tested, where we rerun the partitioning algorithm for each QP value and apply the allocation algorithm afterwards.
Table \ref{tab:expr_part+strm_bd_cam} summarizes the RD performances of different partitioning methods, which is averaged over $100$ simulated navigation paths. The Bjonteggard metric \cite{bjontegaard2001bmetric} is adopted for RD comparison.

\begin{figure*}[tb]
 \centering
\begin{minipage}[htb]{.4\linewidth}
  \centering
  \centerline{\includegraphics[width=\linewidth]{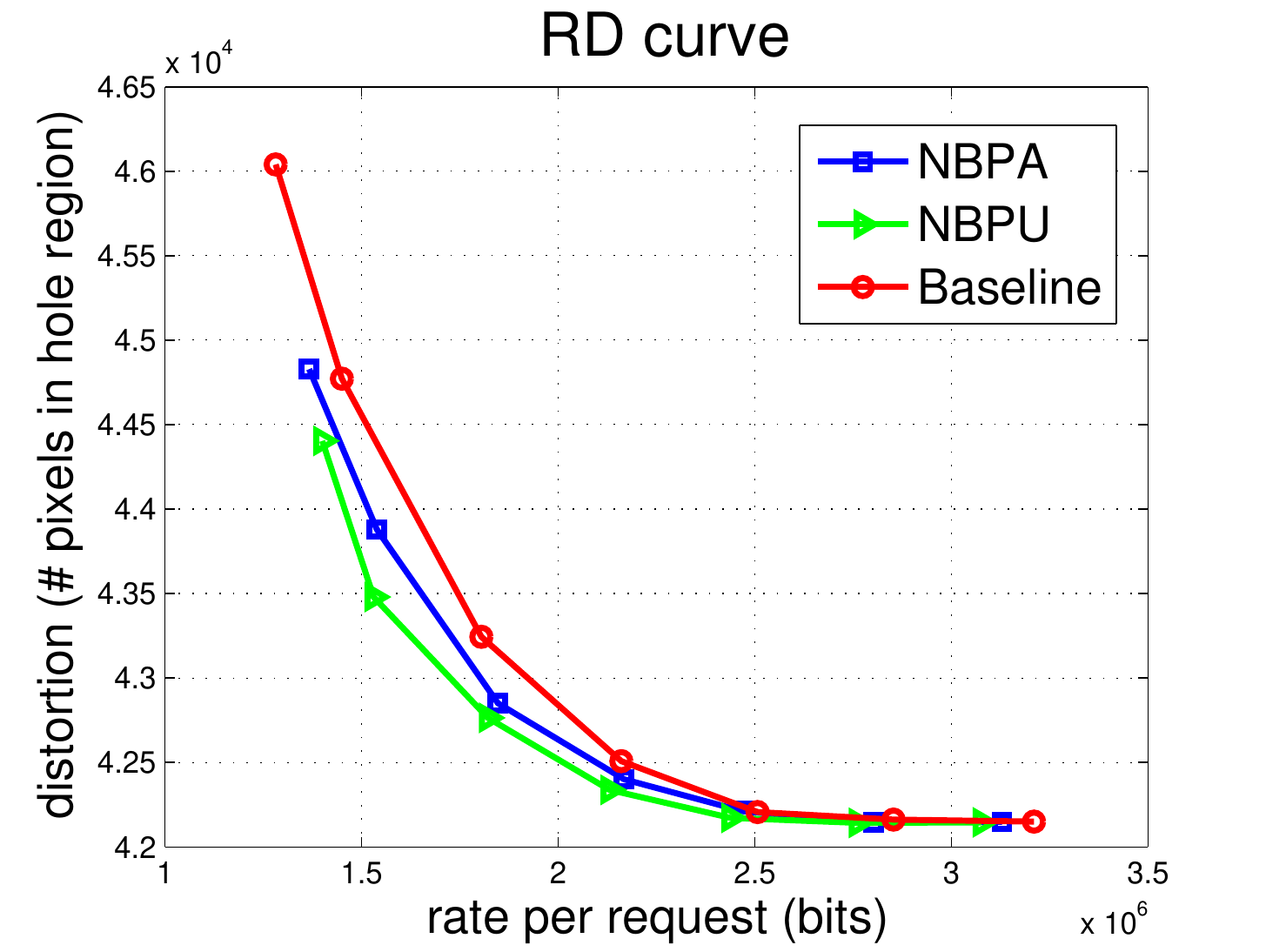}}
  % \vspace{1.5cm}
  % \centerline{ (a) rate-distortion curve } %\medskip
\end{minipage}
\begin{minipage}[htb]{.4\linewidth}
  \centering
  \centerline{\includegraphics[width=\linewidth]{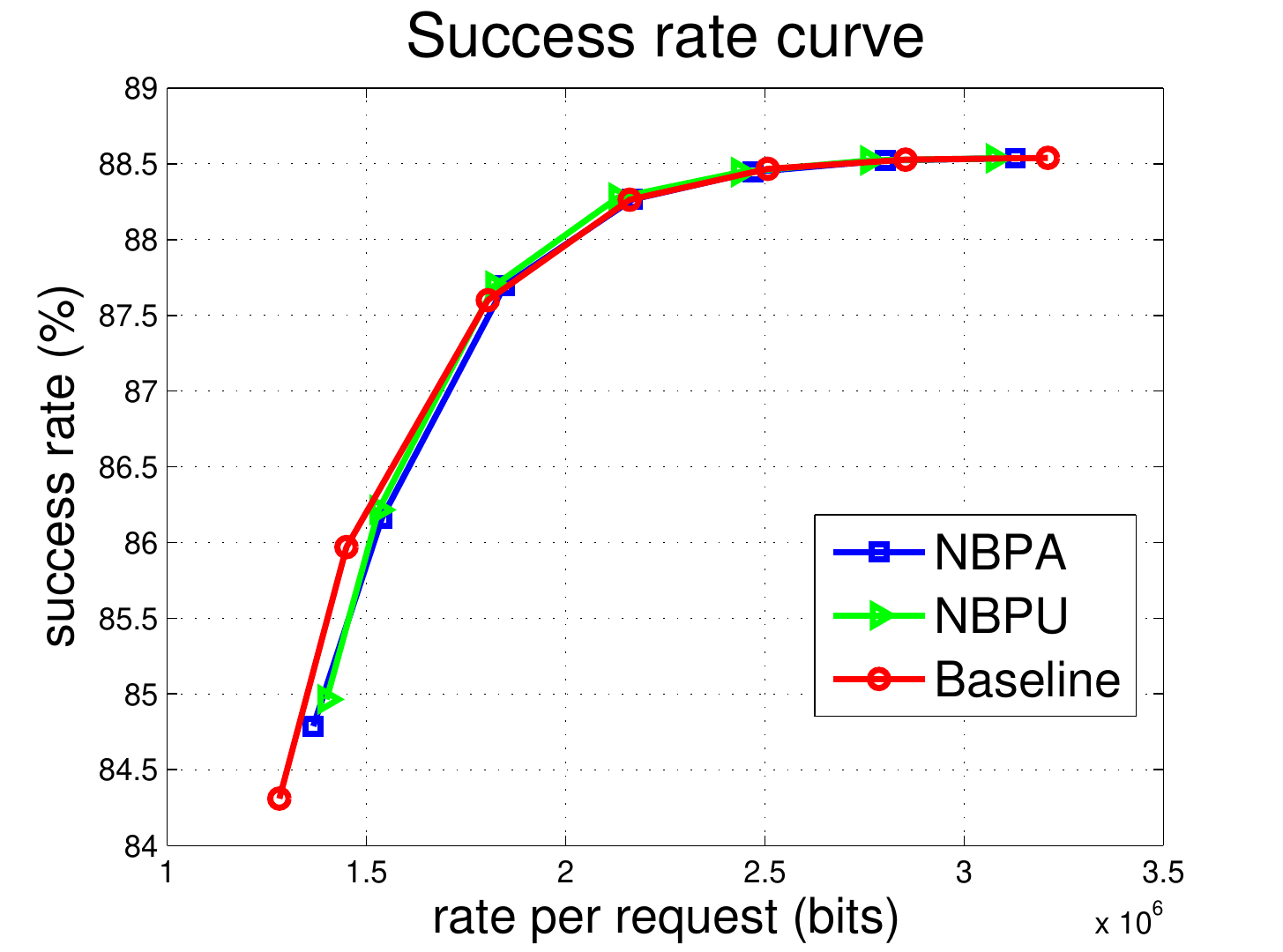}}
  % \vspace{1.5cm}
  % \centerline{ (b) success rate curve } %\medskip
\end{minipage}
\vspace{5pt}
\centerline{ (a) center view preferred, $\Delta = 10$ } \\
\begin{minipage}[htb]{.4\linewidth}
  \centering
  \centerline{\includegraphics[width=\linewidth]{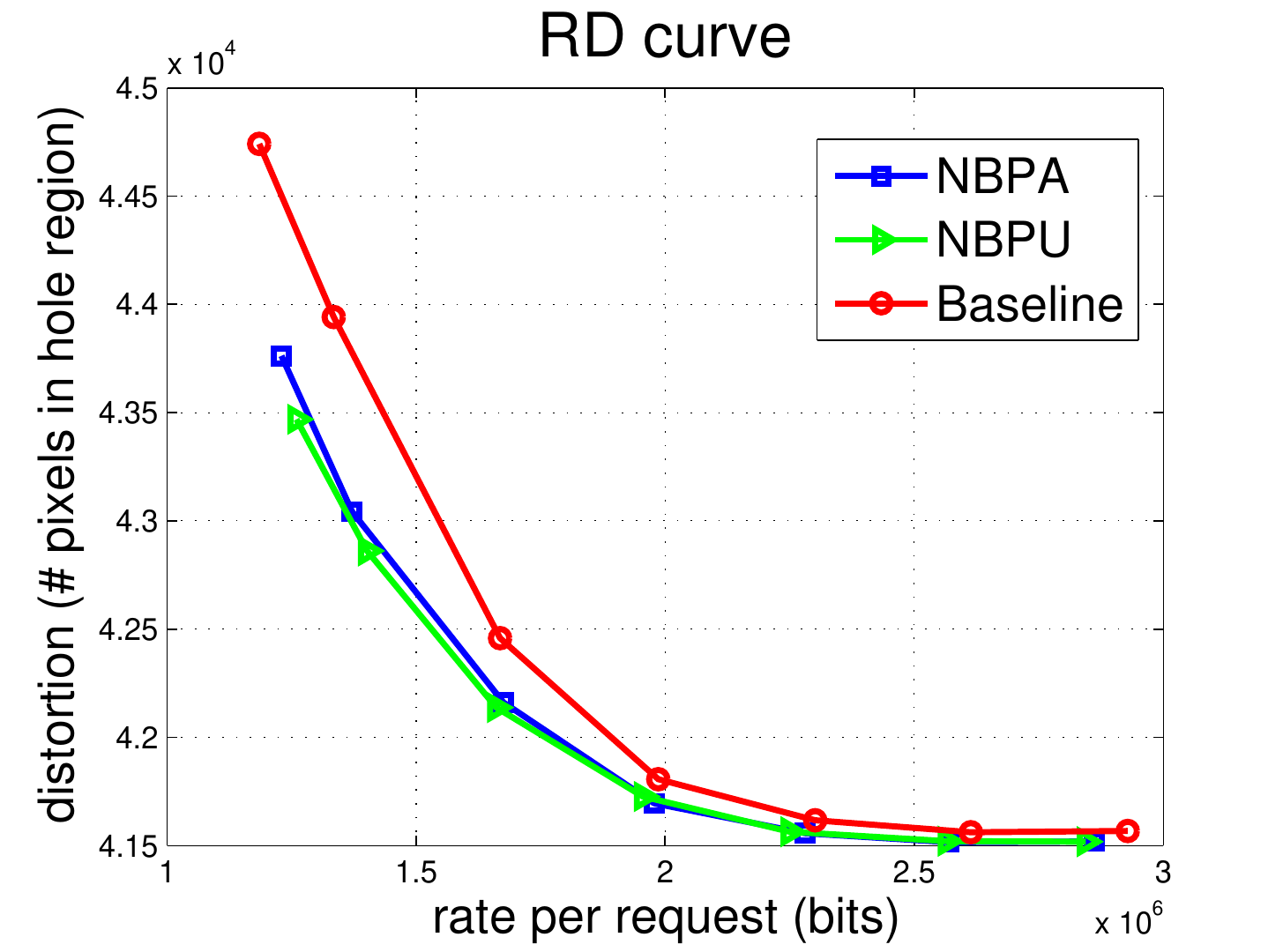}}
  % \vspace{1.5cm}
  %\centerline{ (a) rate-distortion curve } %\medskip
\end{minipage}
\begin{minipage}[htb]{.4\linewidth}
  \centering
  \centerline{\includegraphics[width=\linewidth]{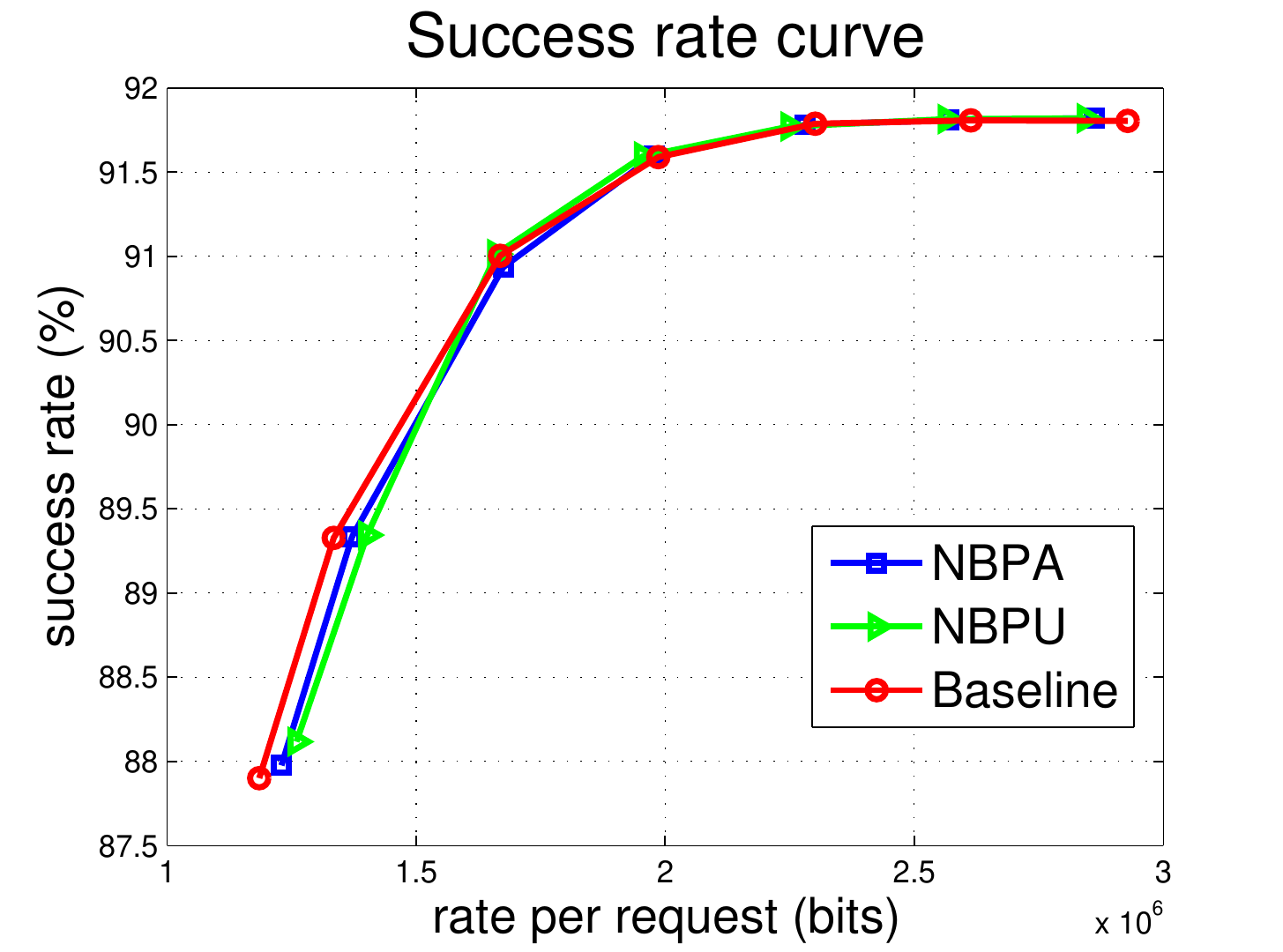}}
  % \vspace{1.5cm}
  %\centerline{ (b) success rate curve } %\medskip
\end{minipage}
\centerline{ (b) right view preferred, $\Delta = 10$ }
\caption{ RD performances of different methods averaged over $100$ simulated navigation paths along virtual viewpoints. }
\label{fig:expr_part+strm}
%\vspace{10pt}
\end{figure*}

It is observed that the proposed NBPA achieves better RD performance than the baseline method in all configurations, and the gain is larger as the navigation speed increases.
This is mainly due to the great rate reduction achieved by the proposed method as discussed before.
On the other hand, NBPA and NBPU have quite similar RD performances.
If we compare Fig. \ref{fig:expr_part_speed} and Fig. \ref{fig:expr_part_viewpop}, it is clearly seen that, under the same $\Delta$, the segments are quite similar in popular views for different popularity distributions, while the main differences lie in the unpopular views.
Since the popular views are required more frequently, it makes a primary contribution to the final RD performance. Therefore the similar segment patterns lead to the close RD performances between NBPA and NBPU.
It should be reminded that, although the gain in RD performance is limited, considering the view popularity brings benefit to the system in terms of lower resource consumptions as demonstrated previously.

\begin{figure}[tb]
 \centering
\begin{minipage}[htb]{.98\linewidth}
  \centering
  \centerline{\includegraphics[width=\linewidth]{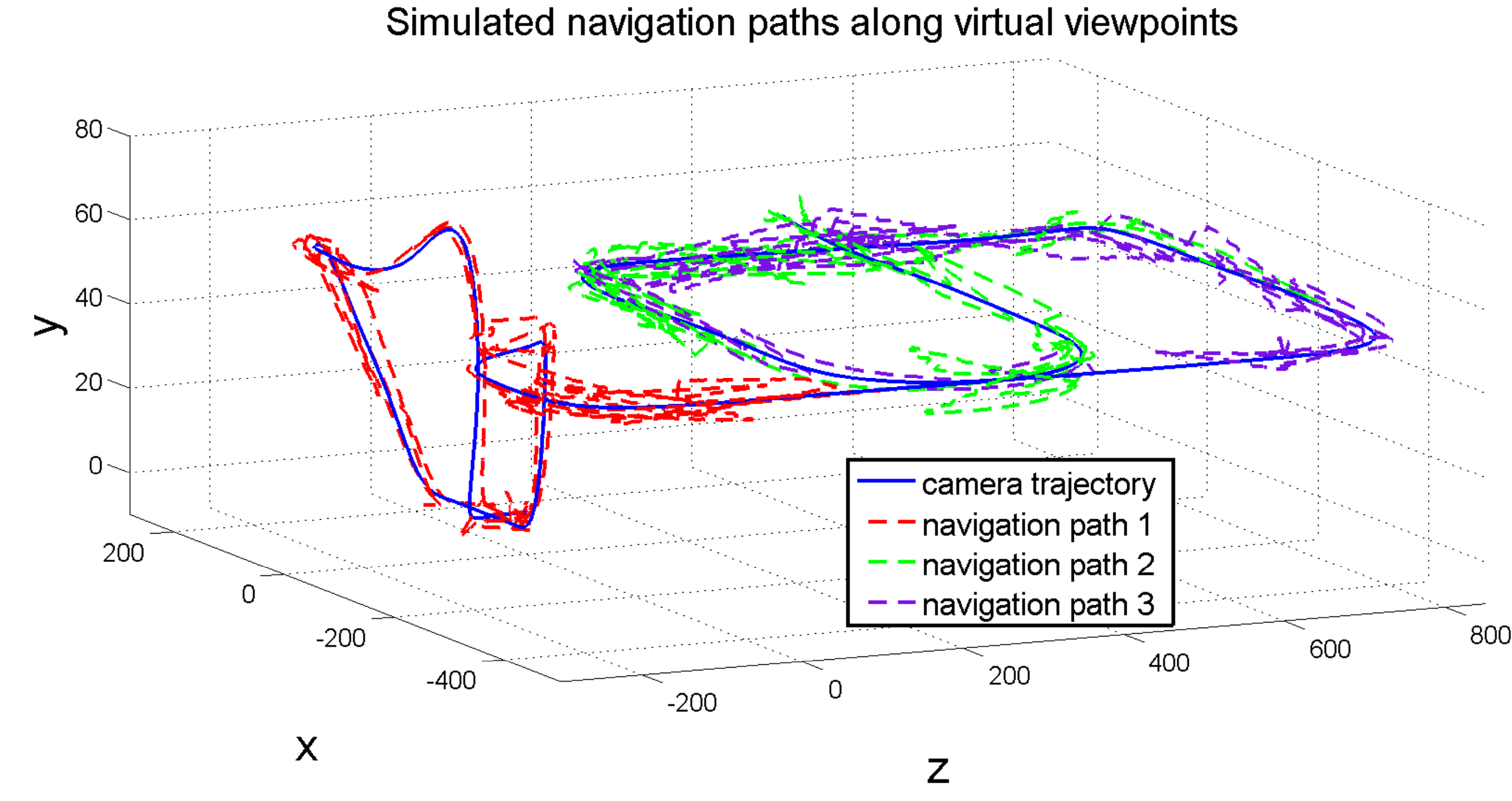}}
  % \vspace{1.5cm}
  % \centerline{  } %\medskip
\end{minipage}
\caption{ Simulated navigation paths along virtual viewpoints plotted by their positions in 3-D space (center view preferred, $\Delta = 10$). The solid curve denotes the camera trajectory, while the dash curves denote different simulated navigation paths.}
\label{fig:expr_sim_navpath}
%\vspace{10pt}
\end{figure}

\begin{flushleft}
\textbf{Navigation paths along virtual views}
\vspace{-5pt}
\end{flushleft}
We next conduct experiments on navigation paths composed of virtual viewpoints which surrounds the real camera viewpoints.
We generate them by adding a 6-D (position plus orientation) random shift onto the previously derived navigation paths composed of camera viewpoints. Fig. \ref{fig:expr_sim_navpath} shows the resulting navigation paths.
Different from the camera viewpoints, the virtual viewpoints require view synthesis and the distortion is in general difficult to compute due to the unavailability of the ground-truth images. Instead, we estimated the view synthesis distortion using Eq. (\ref{eq:model_dist_dibr}).
We further fix $\text{QP}=25$ so that the distortion only comes from inpainting, and we adjust $ t_S $ in Eq. (\ref{eq:nv_comple_alloc_sol}) in order to obtain a series of RD points, based on which we plot the RD curves.

Fig. \ref{fig:expr_part+strm} illustrates the RD curves of different partitioning methods.
Here the distortion is represented by the number of pixels in the hole regions, because it is proportional to the view synthesis distortion as we assume a constant inpainting distortion in Eq. (\ref{eq:model_dist_dibr}).
We further label the view synthesis as \emph{successful} when the size of the hole regions is less than a threshold (half the image size in our experiment). Otherwise our distortion model is ineffective because the inpainting distortion drastically increases and the virtual views are heavily distorted. We therefore discard the unsuccessful rendering with extremely large hole regions and evaluate the RD performance only for successful ones. We further plot the success rate curves on the right side of Fig. \ref{fig:expr_part+strm} for reference.
It is observed that all methods provide similar success rate curves. It indicates that similar percentages of successful rendering are evaluated for different methods and therefore the RD comparison between them is fair. 
The proposed NBPA clearly outperforms the baseline partitioning method in different configurations of navigation speed and view popularity. Similarly to the experiments on navigation paths along camera views, the RD gain mainly comes from a lower transmission rate provided by the proposed partitioning methods.
We also note that NBPA and NBPU have very close RD performances, and sometimes NBPU is even better than NBPA. This is because the segment partition is optimized only for the fixed allocation solution $ \mathcal{S}_0 $. However, in this experiment, we adopt the complementary allocation solutions in order to derive different rate-distortion combinations. There is no guarantee that NBPA always has better RD performance than NBPU in this case. 
On the other hand, as we investigate the trade-offs between different navigation costs of the system, the consideration of the better solution should be evaluated in all different aspects. Although NBPA is less efficient than NBPU in terms of RD performance in some cases, it is validated in previous experiments that NBPA always achieves lower rate and storage costs than NBPU in different situations.

The above evaluations of the complete system demonstrate the effectiveness of the proposed navigation segment representation for a practical navigation system with low resource consumptions and high navigation quality.

\section{Conclusion}
\label{sec:concl}
In this paper, we study the optimal navigation segment representation for an end-to-end interactive multiview navigation system. 
Experimental results verify the effectiveness of the proposed data representation for practical navigation systems. Our method provides lower resource consumption, higher navigation quality and higher adaptation to different navigation parameters when compared to the baseline representation method.
The idea of navigation segment representation can be further extended to more challenging navigation scenarios like the dynamic environment or more complex camera arrangements like the 2-D manifold camera arrays. These problems will be considered in our future work.

\appendix[Estimation of the size of hole regions in single view DIBR]
In virtual view rendering, the size of the hole regions $ \Omega $ can be theoretically derived from the scene geometry. However this approach is time-consuming because it requires to perform actual view rendering.
In order to derive a fast approach for practical usage, we consider to estimate the upper bound of $ \Omega $ as follows.
%Throughout our estimation process, we follow a particular rule that the estimated $ \Omega $ is always no less than a real value. In other words, we estimate the upper bound of $ \Omega $.

As we assume a single reference view for rendering, we suppose the view change between the target view and the reference view is a 6-D vector $ [ \Delta x, \Delta y, \Delta z, \Delta \theta, \Delta \phi, \Delta \psi ]^T \in \mathbb{R}^6 $, which denotes the change in both position and orientation.
We first study the hole regions induced by each component. Let $ \Omega(\Delta x) $ be the hole regions induced only by $ \Delta x $. The disparity value for any pixel location is determined by the equation
\begin{equation} \label{eq:app1}
d(u,v) = f \Delta x / z(u,v),
\end{equation}
where $ d(u,v) $ is the disparity at $ (u,v) $ location. $ z $ is the depth and $ f $ is the focal length.
Then $ \Omega(\Delta x) $ can be estimated by
\begin{equation}
\Omega(\Delta x) \leq f |\Delta x| ( E_V(\frac{1}{z_{\min}}-\frac{1}{z_{\max}}) ) + f |\Delta x| H \frac{1}{z_{\max}},
\end{equation}
where $ E_V $ denotes the edge along vertical direction and $ H $ is the image height. Both $ E_V $ and $ H $ are represented in pixels. $ z_{\min}$ and $ z_{\max} $ denote the minimum and maximum depth values within the image respectively. On the right side of the above inequality, the first term denotes the disoccluded pixels and the second term is the new appearing pixels due to the view change. 
%The inequality is due to the fact that we use the minimum and maximum depth in the expression. 
Note that $ E_V \leq H $. By replacing  $ E_V $ with $H$, we could further derive
\begin{equation}
\Omega(\Delta x) \leq  \frac{fH |\Delta x| }{z_{\min}}.
\end{equation}
Similarly, we could obtain,
\begin{equation}
\Omega(\Delta y) \leq  \frac{fW |\Delta y| }{z_{\min}},
\end{equation}
where $W$ is the image width.
The influence of $ \Delta z $ is similar to zoom in / out. When $ \Delta z \leq 0 $, $ \Omega(\Delta z) = 0 $, because it is a zoom-in action and we assume all pixel values can be obtained by interpolation. When $ \Delta z > 0 $, it is a zoom-out action and we assume the holes are mainly due to the new appearing pixels at the image borders. Using Eq. (\ref{eq:app1}), it can be derived that the image borders are changed by
\begin{equation*}
\begin{aligned}
W - W' \leq W - \frac{z_{\min} W}{z_{\min} + \Delta z} = \frac{\Delta z W}{z_{\min} + \Delta z} \leq \frac{\Delta z W}{z_{\min}},
\end{aligned}
\end{equation*}
and similarly, \[ H - H' \leq \frac{\Delta z H}{z_{\min}}. \]
Therefore,
\begin{equation}
\Omega(\Delta z) \leq H (W-W') + W (H-H') \leq \frac{2HW\Delta z}{z_{\min}}, \Delta z > 0.
\end{equation}
The influences of $ \Delta \theta $ and $ \Delta \phi $ are similar to pitch and yaw respectively. These actions do not introduce disocclusion and the holes uniquely come from the new appearing pixels. We can estimate the new appearing pixels from the angle change as follows
\begin{equation}
\begin{aligned}
\Omega(\Delta \theta) = f W |\Delta \theta|, -\pi \leq \Delta \theta < \pi, \\
\Omega(\Delta \phi) = f H |\Delta \phi|, -\pi \leq \Delta \phi < \pi.
\end{aligned}
\end{equation}
The influence of $ \Delta \psi $ is similar to rotation, which also introduces no disoccluded pixels. The new appearing pixels induced by $ \Delta \psi $ can be computed from the geometry of the image as follows
\begingroup\makeatletter\def\f@size{9}\check@mathfonts
\begin{equation}
\Omega(\Delta \psi) = \left\lbrace
\begin{array}{l}
\frac{\tan (\Delta \psi)}{4} \left( (H^2+W^2) (1+\tan^2(\frac{\Delta \psi}{2}) \right. \\ \left. \quad - 4WH\tan(\frac{\Delta \psi}{2}) \right), \quad 0 \leq \Delta \psi < 2 \tan^{-1}(\frac{H}{W}) \vspace{4pt} \\ 
WH - \frac{H^2}{\sin (\Delta \psi)}, \quad 2 \tan^{-1}(\frac{H}{W}) \leq \Delta \psi \leq \frac{\pi}{2} \vspace{4pt} \\
\Omega(\pi - \Delta \psi), \quad \frac{\pi}{2} < \Delta \psi \leq \pi.
\end{array} \right.
\end{equation}
\endgroup

The overall size of hole regions induced by all the components $ [ \Delta x, \Delta y, \Delta z, \Delta \theta, \Delta \phi, \Delta \psi ]^T $ is difficult to estimate, because the individual components would influence each other and the joint effect is not a simple addition. However we can estimate the upper bound by aggregating the individual areas of hole regions induced by each component, i.e.
\begin{equation}
\Omega \leq \min \{ \sum\limits_{\omega \in \{ x,y,z,\theta,\phi,\psi \}} \Omega(\Delta \omega), ~~W H \}.
\end{equation}
This is because each time a new component is added, the newly induced hole regions might have intersections with the existing hole regions, and therefore the overall size of hole regions is always smaller than a plain addition of the size of hole regions induced by each component. Also note that $ \Omega $ is bounded by $ WH$, which is the size of the whole image area.

%% use section* for acknowledgment
%\section*{Acknowledgment}

%% Can use something like this to put references on a page
%% by themselves when using endfloat and the captionsoff option.
%\ifCLASSOPTIONcaptionsoff
%  \newpage
%\fi


\bibliographystyle{IEEEtran}
\bibliography{IEEEabrv,navi_3d}


%\begin{IEEEbiography}{Rui Ma}
%\end{IEEEbiography}
%\begin{IEEEbiographynophoto}{Thomas Maugey}
%% insert where needed to balance the two columns on the last page with
%%\newpage
%
%\begin{IEEEbiographynophoto}{Pascal Frossard}


%\vfill


\end{document}